\documentclass[prb,aps,amssymb,twocolumn,nofootinbib,superscriptaddress]{revtex4-2}
\usepackage{amsmath}
\usepackage{amsthm}
\usepackage{listings}
\usepackage{braket}
\usepackage[dvips]{graphicx}
\usepackage{hyperref}
\usepackage{psfrag}
\usepackage{comment}
\usepackage{bm}
\usepackage{color}
\usepackage{lineno}

\newcommand{\beq}{\begin{equation}}
	\newcommand{\eneq}{\end{equation}}
\newcommand{\bs}[1]{\boldsymbol{#1}}


\begin{document}
	\title{Rotoinversion-symmetric bulk-hinge correspondence and its applications to higher-order Weyl semimetals}
	
	\author{Yutaro Tanaka}
	\affiliation{
		Department of Physics, Tokyo Institute of Technology, 2-12-1 Ookayama, Meguro-ku, Tokyo 152-8551, Japan
	}
	\author{Ryo Takahashi}
	\affiliation{
		Department of Physics, Tokyo Institute of Technology, 2-12-1 Ookayama, Meguro-ku, Tokyo 152-8551, Japan
	}
	\author{Ryo Okugawa}
	\affiliation{
		Graduate School of Information Sciences, Tohoku University, Sendai 980-8579, Japan
	}
	\author{Shuichi Murakami}
	\affiliation{
		Department of Physics, Tokyo Institute of Technology, 2-12-1 Ookayama, Meguro-ku, Tokyo 152-8551, Japan
	}
	\affiliation{
		TIES, Tokyo Institute of Technology, 2-12-1 Ookayama, Meguro-ku, Tokyo 152-8551, Japan
	}
	
	\begin{abstract}
		We give a bulk-hinge correspondence for higher-order topological phases protected by  rotoinversion $C_{4}\mathcal{I}$ symmetry in magnetic systems. Our approach allows us to show the emergence of the chiral hinge modes only from the information of the $C_{4}\mathcal{I}$ eigenvalues at the high-symmetry points in the Brillouin zone. In addition, based on the bulk-hinge correspondence, we propose a class of higher-order Weyl semimetals (HOWSMs) being Weyl semimetals with hinge modes and Fermi-arc surface states. The HOWSM is characterized by topological invariants for three-dimensional higher-order topological insulators, and the topological invariants are determined by the $C_{4}\mathcal{I}$ symmetry eigenvalues at the high-symmetry points. This HOWSM has chiral hinge modes as a direct consequence of the three-dimensional higher-order topology in the bulk.		
	\end{abstract}
	
	\maketitle

	\section{introduction}
	
	Higher-order topological insulators (HOTIs) have invoked paradigm shifts in condensed matter physics \cite{PhysRevLett.108.126807,PhysRevLett.110.046404,benalcazar2017quantized, PhysRevB.96.245115,PhysRevLett.119.246402, PhysRevLett.119.246401,schindler2018higher, fang2017rotation}. 
	Two-dimensional (2D) HOTIs have corner modes localized at the corners of the systems, and three-dimensional (3D) HOTIs have corner modes or hinge modes localized at the intersections between the surfaces.
	Conventional $n$-dimensional topological insulators \cite{RevModPhys.82.3045, RevModPhys.83.1057}  have ($n-1$)-dimensional gapless states, namely, the gapless modes have codimension $d_c =1$. In this case, the topological phases are classified as first-order topological phases.
	On the other hand, $n$-dimensional HOTIs have ($n-2$)- or ($n-3$)-dimensional gapless states. 
	The former has codimension $d_c=2$, and the latter has codimension $d_c=3$, and therefore the former and the latter are second-order and third-order topological phases, respectively. 
	Moreover, the coexistence of boundary modes with $d_c=m-1$ and those with $d_c=m$ have been studied in recent years and is referred to as hybrid-order topological phase \cite{zhang2020symmetry, PhysRevB.102.041122, PhysRevB.102.125126, PhysRevLett.126.156801}. 
	
	Weyl semimetals (WSM) \cite{murakami2007phase, PhysRevB.83.205101, PhysRevB.84.075129,  PhysRevLett.107.127205,doi:10.1126/sciadv.1602680} and Dirac semimetals \cite{PhysRevB.85.195320, PhysRevLett.108.140405, PhysRevLett.112.036403} are topological phases with topologically protected gapless points in the bulk. 
	WSMs have topological Fermi-arc surface states \cite{PhysRevB.86.195102, PhysRevLett.107.186806, PhysRevB.87.245112, PhysRevB.89.235315}, and those with the boundary modes with codimension $d_c =1$ are classified as the first-order topological phases. Recently, the concept of the higher-order topological phase (HOTP) is generalized to higher-order Dirac semimetals  \cite{PhysRevLett.120.026801, PhysRevB.98.241103, PhysRevB.99.041301,wieder2020strong, PhysRevLett.124.156601, PhysRevB.101.241104, PhysRevResearch.2.043197} and higher-order Weyl semimetals (HOWSM) \cite{PhysRevResearch.1.032048, PhysRevLett.125.266804,PhysRevLett.125.146401,fleury2021sound,luo2021observation, wei2021higher,  PhysRevB.104.014203, PhysRevB.103.184510}. The HOWSM has both Fermi-arc surface states and hinge modes, and therefore it is classified as the hybrid-order topological phase. 
	
	Among various classes of HOTPs \cite{PhysRevB.97.205135, PhysRevB.97.241405, schindler2018higherbismuth, PhysRevB.98.205147, serra2018observationnature555, peterson2018quantizedNature7695, imhof2018topolectricalnatphys, PhysRevLett.123.016806, wang2018higher, sheng2019two, agarwala2019higher, chen2019higher, PhysRevB.98.035147, okugawa2019second, ghosh2019engineering, hirayama2020higher, chen2020universal}, some classes of 3D HOTIs have chiral hinge modes \cite{PhysRevLett.108.126807,PhysRevLett.110.046404,PhysRevB.96.245115,schindler2018higher} akin to the edge modes of 2D quantum Hall effect \cite{PhysRevLett.45.494}. In addition, the HOTIs with chiral hinge modes protected by spatial symmetries are characterized by symmetry-based indicators \cite{PhysRevX.7.041069, po2017symmetry, bradlyn2017topological, watanabe2018structure, song2018quantitative, PhysRevB.98.115150}.
	For example, HOTIs with chiral hinge modes protected by inversion ($\mathcal{I}$) symmetry \cite{PhysRevB.97.205136, PhysRevB.98.205129, PhysRevLett.122.256402, PhysRevResearch.2.043274} can be characterized by a $\mathbb{Z}_{4}$ index in terms of the parity eigenvalues at high-symmetry points in the Brillouin zone (BZ) \cite{PhysRevX.8.031070, PhysRevB.98.115150, PhysRevB.101.115120, PhysRevResearch.2.013300}. 
	According to the classification of topological phases (key space groups) in Ref.~\cite{PhysRevB.98.115150}, the chiral hinge modes detected by symmetry-based indicators are essentially limited to those protected by $\mathcal{I}$ symmetry and those protected by rotoinversion $C_{4}\mathcal{I}$ symmetry, where $C_4$ is a four-fold rotation symmetry.  
	The chiral hinge modes protected by $\mathcal{I}$ symmetry have been explored in detail in Refs.~\cite{PhysRevB.97.205136, PhysRevB.98.205129, PhysRevB.98.115150, PhysRevB.101.115120, PhysRevResearch.2.013300, PhysRevResearch.2.043274},
	and the bulk-hinge correspondence for the HOTPs has been shown in Refs.~\cite{PhysRevB.101.115120, PhysRevResearch.2.013300}. 
	Similarly, $C_{4}\mathcal{I}$-symmetric HOTPs have been studied in Refs.~\cite{schindler2018higher, PhysRevB.97.155305, PhysRevB.98.081110}. Nevertheless, they are based on $\boldsymbol{k}\cdot \boldsymbol{p}$ surface theory or Wannier approach, and these approaches cannot be applied to some systems as we discuss in Appendix~\ref{subsec:Comparison_to_pre}.
	Furthermore, the conventional discussions are insufficient in the following two points. 
	First, the indicator does not explicitly appear in the discussions, and the role of the indicator is not clear in the bulk-hinge correspondence.
	Second, the discussions so far make use of certain properties of the models other than the indicator.
	Since the topological phase is characterized only by the indicator, its bulk-hinge correspondence should be explained by the indicator alone in principle.
	Thus, a general proof of the bulk-hinge correspondence characterized by the indicator has not been given.

\begin{figure}
		\includegraphics[width=1\columnwidth]{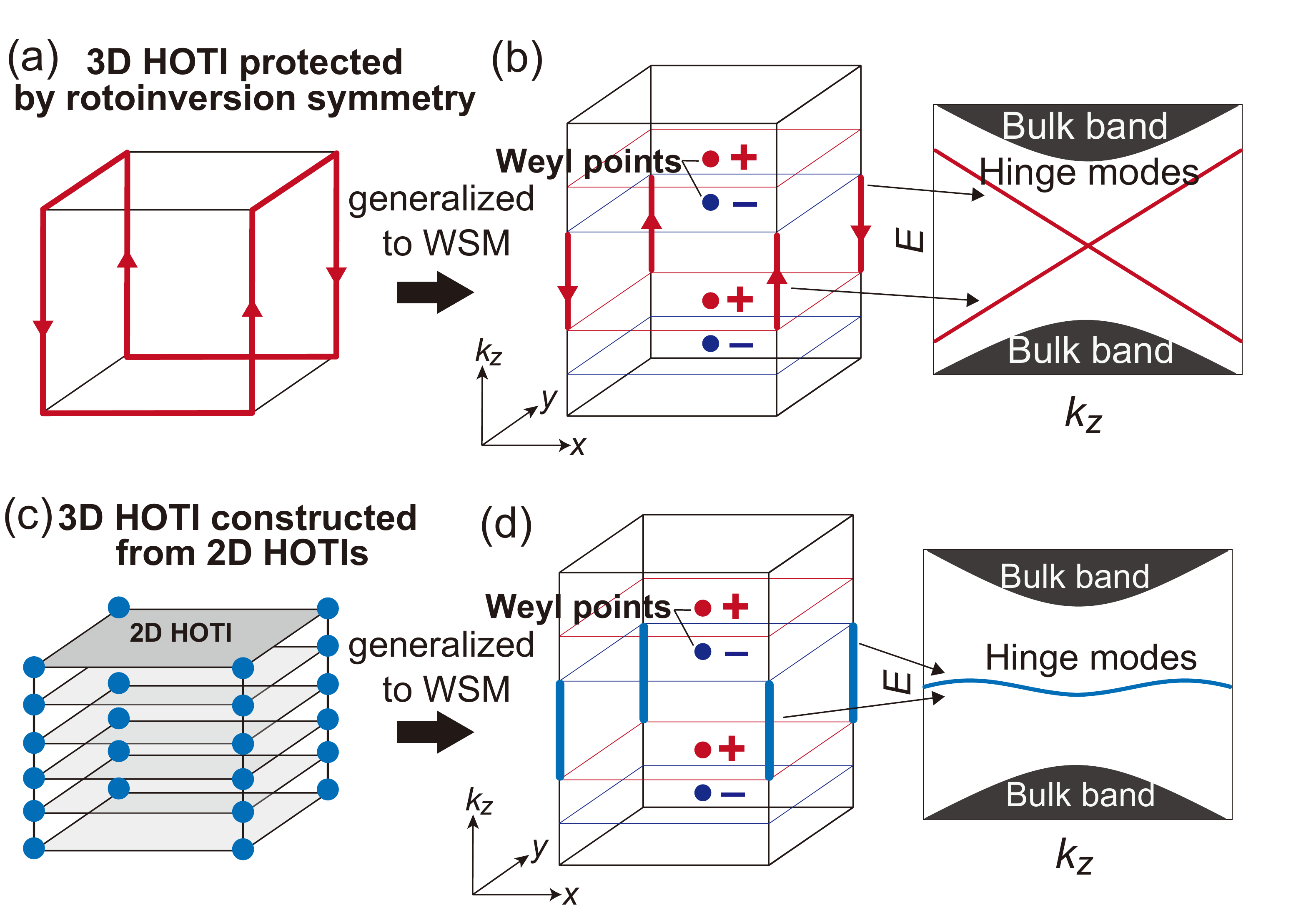}
		\caption{Schematic drawings of HOTIs and HOWSMs. (a) 3D HOTIs protected by the rotoinversion symmetry. (b) HOWSMs with the chiral hinge modes originating from the 3D higher-order topology protected by the rotoinversion symmetry. Red and blue points represent monopoles and anti-monopoles of Berry curvature,  respectively. In this paper, we focus on this HOWSM with the hinge modes characterized by the topological invariants for 3D HOTIs. 
		(c) 3D HOTIs constructed by stacking 2D HOTIs with corner modes. (d) HOWSMs studied in the previous works, where the hinge modes are not chiral. 
		}
		\label{fig:concept_WSM}
	\end{figure}

	In this paper, we show the bulk-hinge correspondence for HOTPs protected by $C_{4}\mathcal{I}$ symmetry. 
	Our approach in this paper allows us to show the emergence of the chiral hinge modes only from the information of the $C_{4}\mathcal{I}$ symmetry eigenvalues at the high-symmetry points. 
	In addition, we propose a class of the HOWSMs having chiral hinge modes protected by $C_{4}\mathcal{I}$ symmetry based on our theory of bulk-hinge correspondence. 
	This HOWSM as a direct consequence of the  3D higher-order topology in the bulk is regarded as a  generalization of the 3D HOTIs protected by $C_{4}\mathcal{I}$ symmetry [Figs.~\ref{fig:concept_WSM}(a) and~\ref{fig:concept_WSM}(b)]. 
	Unlike the HOWSMs in the previous works \cite{PhysRevLett.125.266804,PhysRevLett.125.146401}, which are characterized by topological invariants for the 2D HOTI [Figs.~\ref{fig:concept_WSM}(c) and~\ref{fig:concept_WSM}(d)], the HOWSM in this paper has the chiral hinge modes characterized by the topological invariants for 3D HOTIs.
	Because our HOWSM has the Fermi-arc surface states and the chiral hinge modes,  it can be an ideal platform to study the interplay between first-order and second-order topology as the hybrid-order topological phase. 
	
	This paper is organized as follows. In Sec.~\ref{sec:HOWSM_model}, we propose the HOWSM with the chiral hinge modes protected by $C_{4}\mathcal{I}$ symmetry in terms of a tight-binding model. 
	In Sec.~\ref{sec:bulk_hinge_correspondence}, we show that the emergence of chiral hinge modes  in insulators and WSMs with $C_{4} \mathcal{I}$ symmetry can be determined by the information of the $C_{4}\mathcal{I}$ symmetry eigenvalues. It is a proof of bulk-hinge correspondence for the HOTPs protected by $C_{4}\mathcal{I}$ symmetry. Conclusion and discussion are given in Sec.~\ref{sec:conclusion_and_discussion}.
	
	\section{Higher-order Weyl semimetals with chiral hinge modes protected by rotoinversion symmetry}\label{sec:HOWSM_model}
	In this section, we propose the concept of HOWSMs protected by $C_{4}\mathcal{I}$ symmetry based on a tight-binding model and demonstrate the bulk-hinge correspondence by numerical calculations. The proof of the bulk-hinge correspondence is given in Sec.~\ref{sec:bulk_hinge_correspondence}.
	
	\begin{figure}
		\includegraphics[width=1\columnwidth]{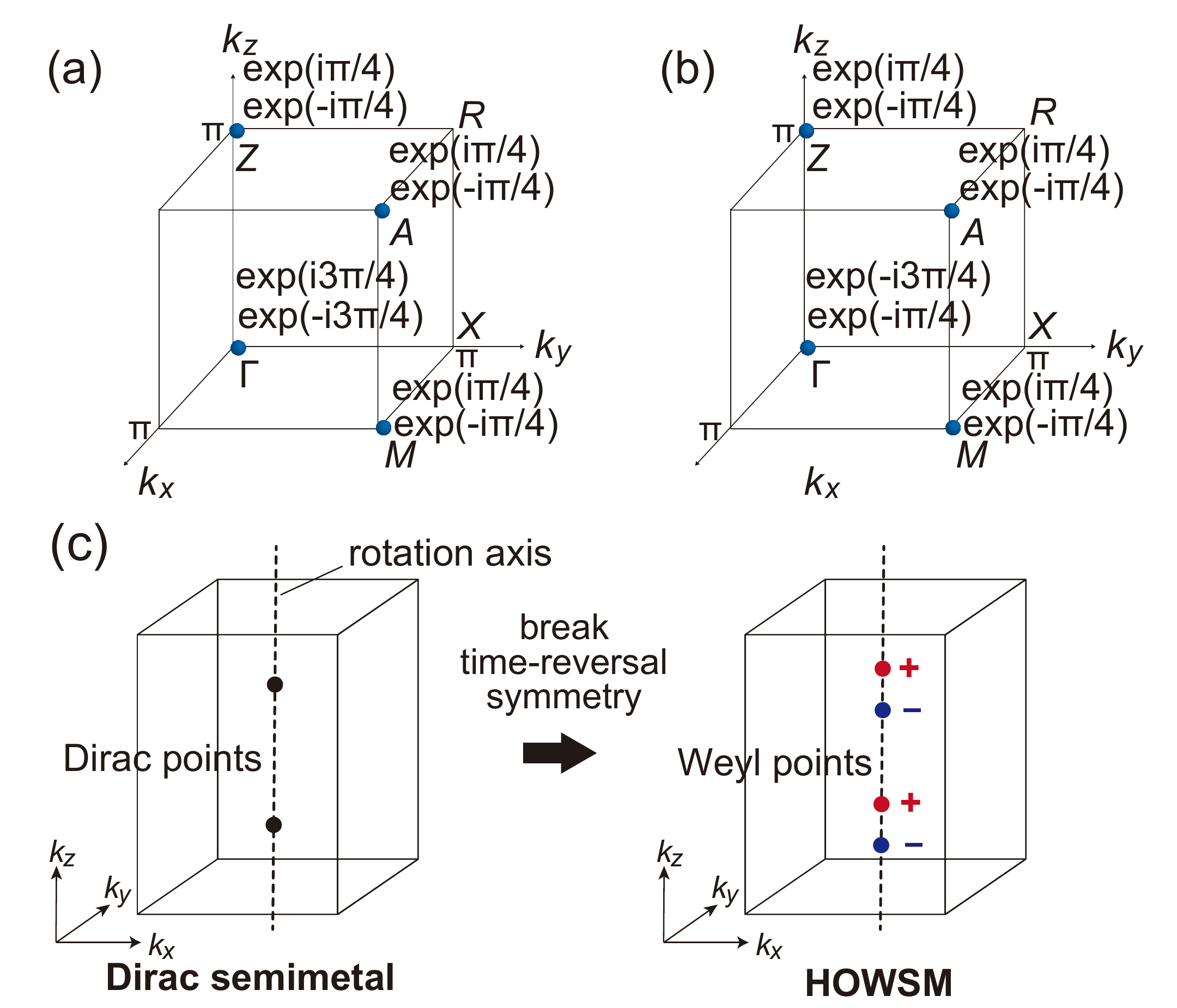}
		\caption{
			(a,b) 1/8 of the BZ and examples of the $C_{4z}\mathcal{I}$ symmetry eigenvalues at the high-symmetry points. These $C_{4z}\mathcal{I}$ symmetry eigenvalues at the high-symmetry points satisfy $\chi^{(+)}_{C_{4z}\mathcal{I}}=\chi^{(-)}_{C_{4z}\mathcal{I}}= 1$ in (a) and  $\chi^{(+)}_{C_{4z}\mathcal{I}}=1, \chi^{(-)}_{C_{4z}\mathcal{I}}= 0$ in (b).
			(c) Phase transition from a $C_{4}$-symmetric Dirac semimetal to a HOWSM with  $\chi^{(+)}_{C_{4z}\mathcal{I}}=\chi^{(-)}_{C_{4z}\mathcal{I}}= 1$.
			The HOWSM with $C_{4z}\mathcal{I}$ symmetry is obtained by adding a perturbation breaking $\mathcal{T}$ symmetry to the 3D Dirac semimetal. Because of $C_{4z}\mathcal{I}$ symmetry, the two Dirac points split to become the two pairs of the Weyl points on the rotation axis. 
		} 
		\label{fig:model_bulk_spectral}
	\end{figure}
	
	\subsection{HOWSM protected by $C_{4z}\mathcal{I}$ symmetry}\label{subsec:symmetry-based_indicators_c4I}
	\subsubsection{Topological invariants}
	Here, we discuss topological invariants for the HOTPs protected by $C_{4z}\mathcal{I}$, where $C_{4 z}$ is a four-fold rotational symmetry about the $z$-axis. 
	The operation $C_{4z}\mathcal{I}$ changes wave vectors as  $(k_{x},k_{y},k_{z})\rightarrow (k_{y},-k_{x},-k_{z})$. Therefore, this system has four $C_{4z}\mathcal{I}$-invariant points in the $k$-space: $K_{4}=\{ \Gamma=(0,0,0)$, $M=(\pi,\pi,0)$, $Z=(0, 0, \pi)$, $A=(\pi,\pi,\pi)\}$.
	Figure \ref{fig:model_bulk_spectral}(a) shows 1/8 of the 3D BZ and the high-symmetry points. 
	To capture the HOTPs protected by $C_{4z} \mathcal{I}$ symmetry, let us introduce the following indices in terms of the numbers of occupied states with symmetry eigenvalues at $K_{4}$: 
	\begin{align}\label{eq:s4symmetric_SI}
		\chi^{(\pm)}_{C_{4z}\mathcal{I}} \equiv \frac{1}{2}\biggl[&n_{\pm \frac{\pi}{4}}(Z)-n_{\mp  \frac{3\pi}{4}}(Z)
		+n_{\pm \frac{\pi}{4}}(A) \nonumber \\
		-&n_{\mp \frac{3\pi}{4}}(A)-n_{\pm \frac{\pi}{4}}(\Gamma)+n_{\mp \frac{3\pi}{4}}(\Gamma)\nonumber \\
		-&n_{\pm \frac{\pi}{4}}(M)+n_{\mp \frac{3\pi}{4}}(M)
		\biggr]\ \ {\rm mod\ 2}, 
	\end{align}
	where $n_{+ \frac{\pi}{4}}(\boldsymbol{k})$, $n_{- \frac{\pi}{4}}(\boldsymbol{k})$, $n_{+ \frac{3 \pi}{4}}(\boldsymbol{k})$, and $n_{- \frac{3 \pi}{4}}(\boldsymbol{k})$ are the numbers of occupied states with $C_{4z}\mathcal{I}$ eigenvalues, $e^{+ i \frac{ \pi}{4}}$, $e^{- i \frac{ \pi}{4}}$, $e^{+ i \frac{ 3\pi}{4}}$, and $e^{- i \frac{ 3\pi}{4}}$ at the $C_{4z} \mathcal{I}$-invariant wavevector $\boldsymbol{k}$, respectively. 
	Here we assume that the Chern numbers at $k_{z}=0$ and $k_{z}=\pi$  are zero.
	This ensures that $\chi^{(\pm)}_{C_{4z}\mathcal{I}}$  are integers.
	When the Chern numbers are nonzero, the hinge modes will be hidden by chiral surface states due to a nonzero Chern number. In this paper, to study the hinge modes, we restrict ourselves to systems with the Chern numbers at $k_{z}=0$ and $k_{z}=\pi$ being zero. 
	The topological invariants $\chi^{(\pm)}_{C_{4z}\mathcal{I}}$ can be defined  both for insulators and for WSMs when the bulk and the surfaces of the WSMs are gapped at $k_{z}=0$ and $k_{z}=\pi$. Figures \ref{fig:model_bulk_spectral}(a) and \ref{fig:model_bulk_spectral}(b) show two examples of the $C_{4z}\mathcal{I}$ symmetry eigenvalues at the high-symmetry points satisfying $\chi^{(+)}_{C_{4z}\mathcal{I}}=\chi^{(-)}_{C_{4z}\mathcal{I}}= 1$ and $\chi^{(+)}_{C_{4z}\mathcal{I}}=1, \chi^{(-)}_{C_{4z}\mathcal{I}}= 0$, respectively.
	In Sec.~\ref{sec:bulk_hinge_correspondence}, we will show that when $\chi^{(+)}_{C_{4z}\mathcal{I}} \equiv \chi^{(-)}_{C_{4z}\mathcal{I}} \equiv1$ mod 2, chiral hinge modes appear in insulators or WSMs.
	
	\subsubsection{Phase transition for the HOWSM}
	Now we discuss how to construct a HOWSM protected by $C_{4z}\mathcal{I}$ symmetry.
	The HOWSM can be realized by adding a  perturbation breaking time-reversal ($\mathcal{T}$) symmetry to a 3D non-magnetic Dirac semimetal with $C_{4z}$ symmetry.
	We start with a 3D Dirac semimetal with two four-fold degenerate points (Dirac points) on the rotation axis [Fig.~\ref{fig:model_bulk_spectral}(c)].
	Then, we perturb it to split each Dirac point into a pair of Weyl points on the rotation axis while preserving  $C_{4z}\mathcal{I}$ symmetry.
	The  topological phase transition for the HOWSM can be induced by magnetic doping or an external magnetic field \cite{PhysRevB.85.195320, du2015dirac} without breaking $C_{4z}\mathcal{I}$ symmetry in the Dirac semimetals.
	
	\subsubsection{Tight-binding model of the HOWSM}
	According  to the above discussion, we introduce a spinful tight-binding model of a HOWSM with 
	$\chi^{(+)}_{C_{4z}\mathcal{I}} \equiv 1$ and $\chi^{(-)}_{C_{4z}\mathcal{I}} \equiv 1$ mod 2. 
	This tight-binding model is defined on the simple tetragonal lattice, and each site has four orbitals. 
	In the following, we set the lattice constants to 1.
	The Hamiltonian is  given by
	\begin{align}\label{Eq:Model_WS_hinge_rotoin}
		\mathcal{H}(\boldsymbol{k})=&\mathcal{H}_{\rm DSM}(\boldsymbol{k})+\mathcal{H}_{C_{4z}\mathcal{I}}(\boldsymbol{k})+\mathcal{H}_z(\boldsymbol{k}),
	\end{align}
	with
	\begin{align}
		\mathcal{H}_{\rm DSM}(\boldsymbol{k})\equiv &\biggl(-m+c\sum_{j=x,y,z}\cos k_j \biggr)\sigma_{0} \tau_{z}\nonumber \\
		&-v\biggl(  \sum_{j=x,y}  \sin k_j \sigma_j\biggr) \tau_{x}, \label{eq:model_for_Dirac_semimetals} \\
		\mathcal{H}_{C_{4z}\mathcal{I}}(\boldsymbol{k}) \equiv &
		v_{s}( \cos  k_x- \cos k_y )\sigma_{0} \tau_{x}\nonumber \\
		&+v_{t}( \cos k_x-\cos k_y )\sigma_{0} \tau_{y}, \label{eq:Hamiltonian_z} \\
		\mathcal{H}_z(\boldsymbol{k})= &v_z \sin k_z \sigma_z \tau_{x}+B_{z}\sigma_{z}\tau_{0},
	\end{align}
	where $\sigma_{j}$, $\tau_{j}$ ($j=x,y,z$) are the Pauli matrices corresponding to the spin and the orbital degrees of freedom, respectively, and $\sigma_{0}$, $\tau_{0}$ are the $2\times 2$ identity matrices. $\mathcal{H}_{\rm DSM}(\boldsymbol{k})$ is a model for a 3D Dirac semimetal with $\mathcal{T}$ symmetry, where $\mathcal{T}$ can be expressed as $\mathcal{T}=-i \sigma_{y}\tau_{0}K$ with $K$ being the complex conjugation. On the other hand, the second term of $ \mathcal{H}_{C_{4z}\mathcal{I}}(\boldsymbol{k})$ breaks $\mathcal{T}$ symmetry.  
	In addition, $\mathcal{H}_z(\boldsymbol{k})$ makes the Dirac points of $\mathcal{H}_{\rm DSM} (\boldsymbol{k})$ split into Weyl points.
	
	\begin{figure}
		\includegraphics[width=1\columnwidth]{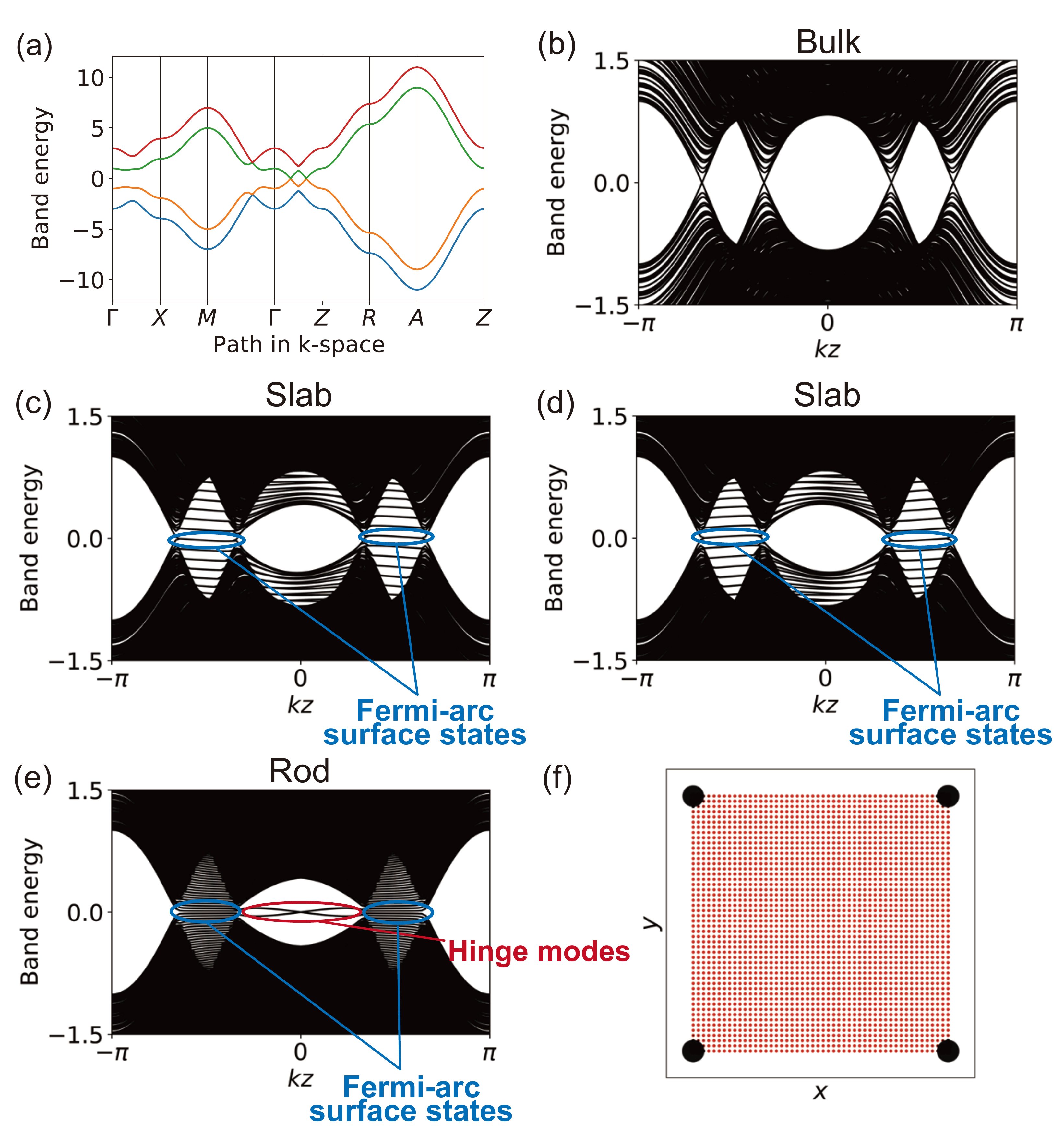}
		\caption{
			(a) Bulk band structure of the tight-binding model of Eq.~(\ref{Eq:Model_WS_hinge_rotoin}). 
			(b-e) The band structures with  the system size being $L_x \times L_y =50\times 50$ in the $x$- and the $y$-directions.
			The band structures with (b) the periodic boundary conditions in the $x$- and the $y$-directions, (c)  the open boundary condition in the $x$-direction and the periodic boundary condition in the $y$-direction, (d)  the periodic boundary condition in the $x$-direction and the open boundary condition in the $y$-direction, (e)  the open boundary conditions in the $x$- and the $y$-directions. (f) The real-space distribution of zero-energy modes of (e).} 
		\label{fig:rotoin_weyl_band_rodgeom}
	\end{figure}
	
	The overall Hamiltonian has $C_{4z} \mathcal{I}$ symmetry
	\begin{align}
		C_{4z} \mathcal{I} \mathcal{H}(k_{x},k_y, k_z) (C_{4z} \mathcal{I})^{-1} 
		=\mathcal{H}(k_{y},-k_x, -k_z),
	\end{align}
	where $C_{4z} \mathcal{I}=e^{-i\frac{\pi}{4}\sigma_{z}}\tau_{z}$. This Hamiltonian does not have 
	$C_{4z} \mathcal{T}$ symmetry because the first term of 
	$\mathcal{H}_{C_{4z}\mathcal{I}}(\boldsymbol{k})$ breaks it. Therefore, the topological phases of this Hamiltonian cannot be characterized by a topological invariant protected by $C_{4z} \mathcal{T}$ symmetry \cite{schindler2018higher}. Instead, we need to use the topological invariant $\chi_{C_{4z}\mathcal{I}}^{(\pm)}$ to analyze the topological phases in this model. 
	In the following, we set the parameters as $m=4$, $c=2$, $v=1$, $v_{z}=0.2$, $v_s = 0.4$, $v_t =1$, and $B_{z}=1$, and we set the Fermi energy to be zero. 
	In this model with these parameters,  the $C_{4z}\mathcal{I}$ eigenvalues of occupied states  at the high-symmetry points are shown in Fig.~\ref{fig:model_bulk_spectral}(a), and therefore the numbers of $C_{4z}\mathcal{I}$ eigenvalues at high-symmetry points satisfy $\chi_{C_{4z}\mathcal{I}}^{(+)}=\chi_{C_{4z}\mathcal{I}}^{(-)}=1$. 
	In the following, we perform calculations of the tight-binding model using the PythTB package \cite{PythTB}.
	Figure \ref{fig:rotoin_weyl_band_rodgeom}(a) shows a bulk band structure of the tight-binding model [Eq.~(\ref{Eq:Model_WS_hinge_rotoin})]. In Fig.~\ref{fig:rotoin_weyl_band_rodgeom}(a), Weyl points appear between $\Gamma$ and $Z$ in the $k$-space. 
	
	Next, to see behaviors of the hinge modes and the surface states of our model, we calculate band structures in the bulk, in the slab and in the rod geometries, and compare them. 
	Figures~\ref{fig:rotoin_weyl_band_rodgeom}(b-e) show the band structures with these different geometries. 
	The system has surfaces in both Fig.~\ref{fig:rotoin_weyl_band_rodgeom}(c) and Fig.~\ref{fig:rotoin_weyl_band_rodgeom}(d), and the Fermi-arc surface states appear in these cases.
	In Fig.~\ref{fig:rotoin_weyl_band_rodgeom}(e), the system has surfaces and hinges, resulting in the  emergence of the hinge modes.
	While the Fermi-arc surface states exist here, the surface states have a mini gap  in Fig.~\ref{fig:rotoin_weyl_band_rodgeom}(e) because of the finite-size effects. 
	
	From the model calculations, we have found several features of the HOWSM protected by $C_{4z}\mathcal{I}$ symmetry. The Fermi-arc surface states appear when the system has the open bondary condition in one direction, and the chiral hinge modes appear only when the system has the open boundary conditions in the two directions. The chiral hinge modes of the HOWSM are localized at the four corners in the real space [see Fig.~\ref{fig:rotoin_weyl_band_rodgeom}(f)] because of $C_{4z}\mathcal{I}$ symmetry. 
	
	As a comparison with this $C_{4z}\mathcal{I}$-symmetric HOWSM,  we calculate a tight-binding model of a HOWSM protected by $\mathcal{I}$ symmetry in Appendix \ref{ap:HOWSM_inversion}. 
	The calculations in Appendix \ref{ap:HOWSM_inversion} show that the chiral hinge modes of the $\mathcal{I}$-symmetric HOWSM appear at two corners facing in opposite directions in the real space. Therefore, the localizations of the hinge modes at the four corners are the characteristic of the  $C_{4z}\mathcal{I}$-symmetric HOWSM.
    As we discuss in Appendix~\ref{appendix:key_spa_gr},  the chiral hinge modes are limited to $\mathcal{I}$-symmetric ones and  $C_{4z}\mathcal{I}$-symmetric ones in type-I magnetic space groups.
	Thus, we exhaust all possible phases of HOWSMs with chiral hinge modes in type-I magnetic space groups. 
	
	\subsection{Topological invariants for insulators with $C_{4z}\mathcal{I}$ symmetry}
	So far we mainly discussed HOWSMs with $C_{4z}\mathcal{I}$ symmetry. In this subsection, we discuss the topological invariants for insulators with $C_{4z}\mathcal{I}$ symmetry. For insulators, we can greatly  simplify  our indices $\chi^{(\pm)}_{C_{4z}\mathcal{I}}$ and connect them to the symmetry-based indicators in insulators \cite{PhysRevX.7.041069, po2017symmetry, bradlyn2017topological, watanabe2018structure,PhysRevB.98.115150, elcoro2021magnetic} by considering the two-fold rotation ($C_{2z}$) symmetry about the $z$-axis because $C_{2z}=(C_{4z}\mathcal{I})^2 $.
	Let  $n^{(+\frac{\pi}{2})}_{C_{2z}}(\boldsymbol{k})$ and $n^{(-\frac{\pi}{2})}_{C_{2z}}(\boldsymbol{k})$ denote the number of occupied states with the $+i$ and $-i$ eigenvalues of the $C_{2z}$ symmetry at high-symmetry points, respectively.  
	$C_{2z}$ symmetry is preserved on the high-symmetry lines $\Gamma$-$Z$ and $M$-$A$, and therefore in insulators, the numbers of occupied bands with the $+i$ and $-i$ eigenvalues  of the $C_{2z}$ symmetry does not change along these lines:
	\begin{align}
		&n^{(+\frac{\pi}{2})}_{C_{2z}}(\Gamma)=n^{(+\frac{\pi}{2})}_{C_{2z}}(Z),\ n^{(-\frac{\pi}{2})}_{C_{2z}}(\Gamma)=n^{(-\frac{\pi}{2})}_{C_{2z}}(Z), \label{eq:compatibility_c2_1} \\
		&n^{(+\frac{\pi}{2})}_{C_{2z}}(M)=n^{(+\frac{\pi}{2})}_{C_{2z}}(A),\ n^{(-\frac{\pi}{2})}_{C_{2z}}(M)=n^{(-\frac{\pi}{2})}_{C_{2z}}(A).\label{eq:compatibility_c2_2}
	\end{align}
	Such restrictions for the numbers of occupied states with the symmetry eigenvalues at high-symmetry points are referred to as compatibility relations. 
	The compatibility relations hold in any band insulators. 
	
	The states with the $e^{i\frac{\pi}{4}}$ and $e^{-i\frac{3\pi}{4}}$ eigenvalues of $C_{4z} \mathcal{I}$ symmetry have the $+i$ eigenvalue of $C_{2z}$ symmetry. Thus, Eq.~(\ref{eq:compatibility_c2_1}) and Eq.~(\ref{eq:compatibility_c2_2}) can be expressed as
	\begin{align}
		&n_{+\frac{\pi}{4}} (Z)+ n_{-\frac{3\pi}{4}} (Z) = n_{+\frac{\pi}{4}} (\Gamma)+ n_{-\frac{3\pi}{4}} (\Gamma), \label{eq:compatibility_c4i} \\
		&n_{+\frac{\pi}{4}} (A)+ n_{-\frac{3\pi}{4}} (A) = n_{+\frac{\pi}{4}} (M)+ n_{-\frac{3\pi}{4}} (M). \label{eq:compatibility_c4i_2}
	\end{align}
	By combining Eqs.~(\ref{eq:s4symmetric_SI}), (\ref{eq:compatibility_c4i}), and (\ref{eq:compatibility_c4i_2}), we can rewrite our index $\chi^{(+)}_{C_{4z} \mathcal{I}}$  as
	\begin{align}\label{eq:simple_index_1}
		\chi^{(+)}_{C_{4z} \mathcal{I}}& \equiv n_{+\frac{\pi}{4}}(Z)-n_{+\frac{\pi}{4}}(\Gamma)+n_{+\frac{\pi}{4}}(A)-n_{+\frac{\pi}{4}}(M)\nonumber \\
		& \equiv \sum_{\Gamma_i \in K_4} n_{+\frac{\pi}{4}} (\Gamma_i)\ \ {\rm mod}\ 2.
	\end{align}
	Similarly, the compatibility relations for the $-i$ eigenvalue of $C_{2z}$ symmetry allow us to rewrite our index $\chi^{(-)}_{C_{4z} \mathcal{I}}$ as 
	\begin{align}\label{eq:simple_index_2}
		\chi^{(-)}_{C_{4z} \mathcal{I}} \equiv \sum_{\Gamma_i \in K_4} n_{-\frac{\pi}{4}} (\Gamma_i)\ \ {\rm mod}\ 2.
	\end{align}
	In addition, as we show in Appendix \ref{ap:symm_ind_chern}, the difference between the Chern numbers on $k_{z}=0$ and $k_{z}=\pi$ is given by
	\begin{equation}\label{eq:difference_between_Chern}
		{\rm Ch}|_{k_z = 0}-{\rm Ch}|_{k_z = \pi}\equiv 2( \chi^{(+)}_{C_{4z} \mathcal{I}} +  \chi^{(-)}_{C_{4z} \mathcal{I}})\ \ {\rm mod}\ 4.
	\end{equation}
	Therefore, when one of our indices $ \chi^{(\pm)}_{C_{4z} \mathcal{I}}$ is an odd number and the other is an even number, the value of Eq.~(\ref{eq:difference_between_Chern}) is equal to 2 mod 4, and the system has a nontrivial Chern number on the $k_{z}=0$ or  $k_{z}=\pi$. 
	
	The space group generated by $C_{4z}\mathcal{I}$ is $\#$ 81 ($P\bar{4}$). Because we do not assume $\mathcal{T}$ symmetry, it is the type-I magnetic space group $\#$81.33 generated by  $C_{4z}\mathcal{I}$.
	Its symmetry-based indicators are given by $\mathbb{Z}_2 \times \mathbb{Z}_2 \times \mathbb{Z}_4$ \cite{watanabe2018structure}. The $\mathbb{Z}_4$ index corresponds to a weak index being equal to the Chern number ${\rm Ch}|_{k_z =0}$ modulo 4.
	The first $\mathbb{Z}_2$ index is a strong index to detect whether  the number of Weyl points between $0\leq k_z \leq \pi$ is $4n$ or $4n+2$ ($n=$ integer).
	It is equal to the difference of  the Chern numbers between the $k_z = 0$ and $k_z = \pi$ \cite{PhysRevB.98.115150, elcoro2021magnetic, peng2021topological}.
	Therefore,  this $\mathbb{Z}_2$ index is equal to the parity of $\chi^{(+)}_{C_{4z} \mathcal{I}} +  \chi^{(-)}_{C_{4z} \mathcal{I}}$ from Eq.~(\ref{eq:difference_between_Chern}). 
	In this paper, we focus on the system with the Chern numbers on the $k_z = 0$ and $k_z = \pi$ being zero. 
	This assumption leads to  $\chi^{(+)}_{C_{4z} \mathcal{I}} +  \chi^{(-)}_{C_{4z} \mathcal{I}}\equiv 0$ mod 2.
	This allows two possibilities; $\chi^{(+)}_{C_{4z} \mathcal{I}} \equiv  \chi^{(-)}_{C_{4z} \mathcal{I}} \equiv 0$ or 1 modulo 2. They are distinguished in terms of  the second $\mathbb{Z}_2$ index.  Namely, when  $\chi^{(+)}_{C_{4z} \mathcal{I}} \equiv \chi^{(-)}_{C_{4z} \mathcal{I}} \equiv 1$ mod 2, the second $\mathbb{Z}_2$ is equal to $\mathbb{Z}_2=1$, and it is the HOTPs protected by $C_{4z}\mathcal{I}$ symmetry.  In this way, our indices match with the symmetry-based indicators in the previous works.
	
	In the previous work \cite{PhysRevB.98.115150,peng2021topological}, a quantity $\mu_{4}$ defined by
	\begin{equation}
		\mu_{4}=\frac{1}{\sqrt{2}}\sum_{\Gamma_{i}\in K_{4}} \sum_{\alpha}e^{i \alpha}n_{\alpha}(\Gamma_{i}),
	\end{equation}
	is proposed as a symmetry-based indicator in the magnetic space group $\#81.33$, where $\alpha$ runs over $\alpha=\pm \pi/4, \pm 3\pi/4$.
	According to Refs.~\cite{PhysRevB.98.115150,peng2021topological}, $\mu_{4}=2$ or $\mu_{4}=2i$ correspond to nontrivial phases.  
	Furthermore, in the previous work~\cite{elcoro2021magnetic}, the symmetry-based indicator is given by
	\begin{equation}
		z_2 \equiv \frac{1}{2}\sum_{\Gamma_{i}\in K_{4}}  \Bigl( n_{-\frac{\pi}{4}} (\Gamma_i) - n_{+\frac{3\pi}{4}} (\Gamma_i) \Bigr)\ \  {\rm mod\ 2},
	\end{equation} 
	 in the magnetic space group $\#81.33$. According to Ref.~\cite{elcoro2021magnetic}, $z_2=1$ corresponds to nontrivial phases. 
	When the Chern numbers at $k_{z}=0$ and $\pi$ are zero, both $\mu_4= 2, 2i$ and $z_2 =1$ are equivalent to $\chi_{C_{4z} \mathcal{I}}^{(+)} \equiv \chi_{C_{4z} \mathcal{I}}^{(-)} \equiv 1$ mod 2. 
	While the indices $\chi_{C_{4z} \mathcal{I}}^{(\pm)}$ detect the same HOTP as the symmetry-based indicators $\mu_4$ and  $z_2$,  we propose the indices  $\chi_{C_{4z} \mathcal{I}}^{(\pm)}$ in this paper to be the useful indices to detect the HOTP protected by $C_{4z}\mathcal{I}$ symmetry from the discussions of spectral flows in Sec.~\ref{sec:bulk_hinge_correspondence}.
	In contrast, the formula of the symmetry-based indicator $\mu_{4}$ and $z_2$ may not have a direct physical interpretation.

	\section{Bulk-hinge correspondence for higher-order topological phases}\label{sec:bulk_hinge_correspondence}
	In this section, we show the bulk-hinge correspondence for HOTPs in insulators and WSMs with $C_{4z}\mathcal{I}$ symmetry only from the $C_{4z}\mathcal{I}$ eigenvalues at the high-symmetry points in the BZ. From this bulk-hinge correspondence, we can confirm that the chiral hinge modes of our model in the previous section originate from the nontrivial indices $\chi_{C_{4z} \mathcal{I}}^{(\pm)}$ introduced in the previous section. 
	In the following discussion, we show that when our indices $\chi_{C_{4z}\mathcal{I}}^{(\pm)}$ [see Eq.(\ref{eq:s4symmetric_SI})] satisfy $\chi^{(+)}_{C_{4z}\mathcal{I}}\equiv \chi^{(-)}_{C_{4z}\mathcal{I}} \equiv1$ mod 2, the chiral hinge modes appear.
	
	\subsection{Setup of the problem}
	Here, we consider a general 3D spinful electronic system with $C_{4z}\mathcal{I}$ symmetry.
	In addition, we assume that the system is an insulator and a WSM with the bulk being gapped at $k_{z}=0$ and $k_{z}=\pi$. 
	We do not assume $\mathcal{T}$ symmetry.
	In the following discussion, we assume that the surfaces are gapped at $k_{z}=0$ and $k_{z}=\pi$ in the system with finite sizes in the $x$- and $y$-directions. 
	Here, we consider a large but finite system in the $x$- and $y$-directions with the system size $L_{x}\times L_{y}=(2M+1)\times (2M+1)$ ($M$: an integer), where the length of the system is measured in the unit of the lattice constant. The lattice sites $(x,y)$ with $x, y \in \mathbb{Z}$ satisfy $-M\leq x \leq M$ and $-M\leq y \leq M$. Furthermore, we set the system size $L_z$ along the $z$ direction as $L_{z} \rightarrow \infty$ and
	electronic states are labeled by the wavenumber $k_{z}$ along the $z$-direction.
	
	\begin{figure}
		\includegraphics[width=1\columnwidth]{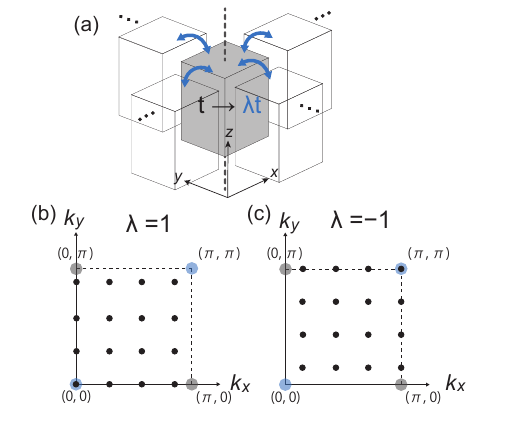}
		\caption{Cutting procedures and boundary conditions. (a) Geometry of the system with boundary conditions controlled by a cutting parameter $\lambda$ with the finite system size $L_x \times L_y$ in the $x$- and the $y$-directions. (b) Possible wave vectors $(k_{x},k_{y})$ with $\lambda=1$. Black dots represent the possible wave vectors. $\lambda=1$ leads to the periodic boundary conditions in the $x$- and $y$-directions and the the wave vectors $k_{i}=\tfrac{2\pi}{L_{i}}m_{i}$ ($-M\leq m_{i} \leq M$) ($i=x,y$). (c) Possible wave vectors $(k_{x},k_{y})$ with $\lambda=-1$. $\lambda=-1$ leads to the anti-periodic boundary conditions in the $x$- and $y$-directions and the the wave vectors $k_{i}=\tfrac{2\pi}{L_{i}}m_{i}+\tfrac{\pi}{L_{i}}$ ($-M\leq m_{i} \leq M$) ($i=x,y$).} 
		\label{fig:s4_eigenvalues_cutting}
	\end{figure}

	Next, we introduce hoppings between sites $x=-M$ and $x=M$ in the similar way as in the bulk but  all the hopping amplitudes between $x=-M$ and $x=M$ are multiplied by a real parameter $\lambda$. Similarly, we introduce the hoppings between the sites $y=-M$ and $y=M$ multiplied by $\lambda$. For example, when the bulk Hamiltonian has a hopping parameter $t$ in the $x$- ($y$-) direction, the hoppings between  $x=-M$ and $x=M$ ($y=-M$ and $y=M$)  are given by $\lambda t$ [Fig.~\ref{fig:s4_eigenvalues_cutting}(a)].
	The system has two parameters, $k_{z}$ and $\lambda$, and we write the Hamiltonian of the system as $\mathcal{H}(k_{z}, \lambda)$. 
	The Hamiltonian $\mathcal{H}(k_{z}, \lambda)$ has the periodic boundary conditions in the $x$- and $y$-directions when $\lambda=1$, while it has the open boundary conditions in the $x$- and $y$-directions when $\lambda=0$.
	
	The Hamiltonian $\mathcal{H}(k_{z}, \lambda)$ has $C_{2z}$ symmetry because $(C_{4z}\mathcal{I})^{2}=C_{2z}$: 
	\begin{equation}
		C_{2z}\mathcal{H}(k_{z}, \lambda)C_{2z}^{-1}=\mathcal{H}(k_{z}, \lambda).
	\end{equation}
	Therefore, under an appropriate unitary transformation $U$, $\mathcal{H}(k_{z}, \lambda)$ can be transformed into the following direct sum of $(+)$ and $(-)$  sectors labeled by the $+i$  and $-i$ eigenvalues of $C_{2z}$, respectively:
	\begin{align}
		&U\mathcal{H}(k_{z}, \lambda)U^{\dagger}\nonumber \\
		=&\tilde{\mathcal{H}}(k_{z}, \lambda)=
		\begin{pmatrix}
			\mathcal{H}^{(+)}(k_{z},\lambda) & 0 \\
			0 & \mathcal{H}^{(-)}(k_{z},\lambda)
		\end{pmatrix}. 
	\end{align}
	Then we define the total numbers of occupied states with $C_{4z}\mathcal{I}$ eigenvalues $e^{i\frac{\pi}{4}}$, $e^{-i\frac{\pi}{4}}$, $e^{i\frac{ 3\pi}{4}}$, and $e^{-i\frac{ 3\pi}{4}}$ as $N_{+ \frac{\pi}{4}}(k_{z},\lambda)$, $N_{- \frac{\pi}{4}}(k_{z},\lambda)$, $N_{+\frac{3\pi}{4}}(k_{z}, \lambda)$, and $N_{-\frac{3\pi}{4}}(k_{z}, \lambda)$, respectively. 
	In addition, for the following discussions, we introduce the following integers from the numbers of occupied states with $C_{4z}\mathcal{I}$ eigenvalues:
	\begin{equation}
		\mathcal{N}^{(\pm)}_{C_{4z}\mathcal{I}} (k_{z})|_{\lambda} \equiv N_{\pm \frac{\pi}{4}} (k_{z},\lambda) - N_{\mp  \frac{3\pi}{4}}(k_{z},\lambda).
	\end{equation}
	The square of $e^{i\frac{\pi}{4}}$ and the square of $e^{-i\frac{ 3\pi}{4}}$ are $+i$, and the square of $e^{-i\frac{\pi}{4}}$ and the square of $e^{i\frac{ 3\pi}{4}}$ are $-i$. Thus, ($\pm$) labels of $\mathcal{N}^{(\pm)}_{C_{4z}\mathcal{I}} (k_{z})|_{\lambda}$ corresponds to the ($\pm$) sectors of the Hamiltonian $\mathcal{H}^{(+)}(k_{z},\lambda)$ and $\mathcal{H}^{(-)}(k_{z},\lambda)$.
	
	\subsection{Cutting procedures and spectral flows}\label{subsec:cuttign_procedure_spectral_flows}
	Here, we show the bulk-hinge correspondence with $C_{4z}\mathcal{I}$ symmetry by considering changes in the energy spectra through changing the cutting parameter.
	When the cutting parameter is $\lambda=1$, the wave vectors in the $x$- and $y$-directions are
	$
	k_{i}=\frac{2\pi}{L_i}m_{i}\ \ (-M\leq m_{i} \leq M)\ \ (i=x,y),
	$
	because of the periodic boundary condition in the $x$- and $y$-directions.
	On the other hand, in the case with $\lambda=-1$, the wave vectors are given by
	$
	k_{i}=\frac{2\pi}{L_i}m_{i}+\frac{\pi}{L_i}\ \ (-M\leq m_{i} \leq M)\ \ (i=x,y)
	$
	because of the antiperiodic boundary condition.
	The proof of this is given in Appendix~\ref{app:wave_vector_APBC}.
	We note that here $k_x$ and $k_y$ are well-defined only for $\lambda=\pm 1$ because of the periodic boundary conditions ($\lambda=1$) and the anti-periodic boundary conditions  ($\lambda=-1$) [see Figs.~\ref{fig:s4_eigenvalues_cutting}(b) and \ref{fig:s4_eigenvalues_cutting}(c)]. 
	For both the cases with $\lambda=1$ and $\lambda=-1$, each wave function $\psi_{m}$ at non-$C_{4 z}\mathcal{I}$ symmetric momenta $\bs{k}$ with $k_{z}=0$ or $\pi$ can always be mixed with the wave functions at $\hat{C}_{4z}\hat{\mathcal{I}}\bs{k}$, $(\hat{C}_{4 z}\hat{\mathcal{I}})^{2}\bs{k}$ and $(\hat{C}_{4z}\hat{\mathcal{I}})^{3}\bs{k}$ to construct four eigenstates of $C_{4z}\mathcal{I}$:
	\begin{align}
		&\ket{\phi_{\alpha}}\equiv \ket{\psi_{m}(\bs{k})} +e^{-i\alpha} \hat{C}_{4z}\hat{\mathcal{I}} \ket{\psi_{m}(\bs{k})} \nonumber \\
		&+e^{-2i\alpha}(\hat{C}_{4z}\hat{\mathcal{I}})^{2} \ket{\psi_{m}(\bs{k})} +e^{-3i\alpha} (\hat{C}_{4z}\hat{\mathcal{I}})^{3}\ket{\psi_{m}(\bs{k})},
	\end{align}
	\begin{align}
		\hat{C}_{4z} \hat{\mathcal{I}} \ket{\phi_{\alpha}}=e^{i\alpha} \ket{\phi_{\alpha}},
	\end{align}
	where $\alpha=\pm \frac{\pi}{4}, \pm \frac{3\pi}{4}$.
	Therefore, states at non-${C}_{4z} \mathcal{I}$ symmetric points  do not contribute to $\mathcal{N}^{(\pm)}_{{C}_{4z} \mathcal{I}}$. 
	
	On the other hand, the states at $C_{4z}\mathcal{I}$-symmetric momenta contribute to $\mathcal{N}^{(\pm)}_{C_{4z}\mathcal{I}}(k_{z})|_{\lambda}$. 
	When $\lambda=1$, the wave vector $(k_{x}, k_{y})$ can take a value $(0,0)$, but not $(\pi, \pi)$ [Fig.~\ref{fig:s4_eigenvalues_cutting}(b)]. 
	When $\lambda=-1$, the wave vector $(k_{x},k_{y})$ can take a value $(\pi,\pi)$, but not (0,0) [Fig.~\ref{fig:s4_eigenvalues_cutting}(c)].
	Therefore, $\mathcal{N}^{(\pm)}_{C_{4z}\mathcal{I}}(k_{z})|_{\lambda=\pm 1}$ is determined by the number of occupied states with the $C_{4z}\mathcal{I}$ eigenvalues at $\bs{k}$:
	\begin{align}\label{eq:NC4I_with_lam1_and_lam-1}
		\mathcal{N}^{(\pm)}_{C_{4z \mathcal{I}}}(0)|_{\lambda=1}=n_{\pm \frac{\pi}{4}}(\Gamma)-n_{\mp \frac{3\pi}{4}}(\Gamma),\nonumber \\
		\mathcal{N}^{(\pm)}_{C_{4z \mathcal{I}}}(0)|_{\lambda=-1}=n_{\pm \frac{\pi}{4}}(M)-n_{\mp \frac{3\pi}{4}}(M),\nonumber \\
		\mathcal{N}^{(\pm)}_{C_{4z \mathcal{I}}}(\pi)|_{\lambda=1}=n_{\pm \frac{\pi}{4}}(Z)-n_{\mp \frac{3\pi}{4}}(Z),\nonumber \\
		\mathcal{N}^{(\pm)}_{C_{4z \mathcal{I}}}(\pi)|_{\lambda=-1}=n_{\pm \frac{\pi}{4}}(A)-n_{\mp \frac{3\pi}{4}}(A),
	\end{align}
	where $\Gamma$, $M$, $Z$, and $A$ are the high-symmetry points in the BZ defined in Sec.~\ref{subsec:symmetry-based_indicators_c4I} [see Fig.~\ref{fig:model_bulk_spectral}(a)]. 
	
\begin{figure}
		\includegraphics[width=1\columnwidth]{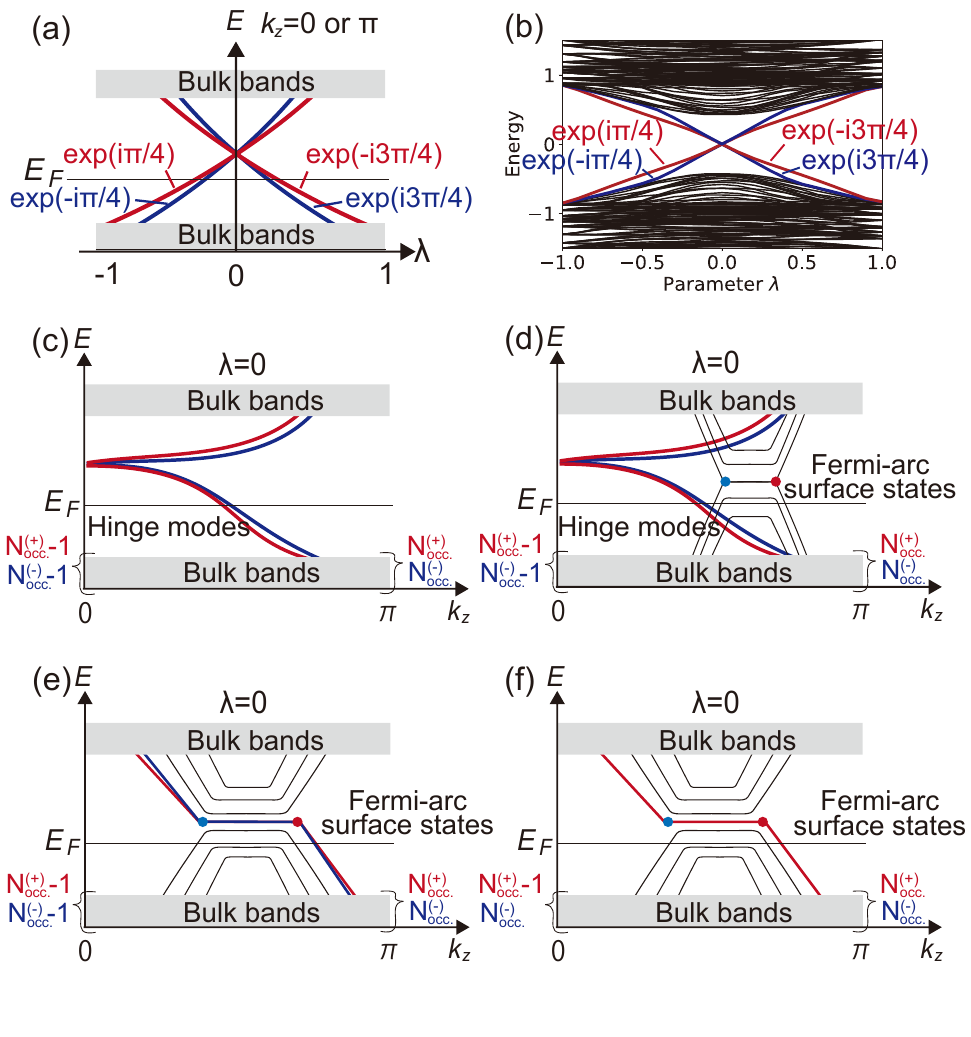}
		\caption{
			Spectral flows in the band gaps and the emergence of hinge modes.
			The spectral flows in the sectors labeled by the $(+)$ and $(-)$ eigenvalues of $C_{2z}$ are red and blue lines, respectively. 
			(a) The changes of the energy spectra in changing $\lambda$ from $\lambda=-1$ to $\lambda=1$ with $k_z =0$ or $\pi$. 
			(b) Spectral flows of the tight-binding model of Eq.~(\ref{Eq:Model_WS_hinge_rotoin}) in the process of changing $\lambda$ from $\lambda=-1$ to $\lambda=1$. 
			(c) Emergence of chiral hinge modes between $k_{z}=0$ and $k_{z}=\pi$ in insulators. 
			The numbers of occupied bands in the $(\pm)$ sectors labeled by the $C_{2z}$ eigenvalues satisfy $\nu^{(\pm)}(k_{z}=0)=N^{(\pm)}_{occ.}-1$ and $\nu^{(\pm)}(k_{z}=\pi)=N^{(\pm)}_{occ.}$. 
			(d) Emergence of chiral hinge modes between $k_{z}=0$ and $k_{z}=\pi$ in WSMs with $\chi^{(+)}_{C_{4z}\mathcal{I}}\equiv \chi^{(-)}_{C_{4z}\mathcal{I}} \equiv 1$ mod 2. 
			(e) Emergence of Fermi-arc surface states in WSMs with $\chi^{(+)}_{C_{4z}\mathcal{I}}\equiv \chi^{(-)}_{C_{4z}\mathcal{I}} \equiv 1$ mod 2. 
			(f) Emergence of Fermi-arc surface states with $\chi^{(+)}_{C_{4z}\mathcal{I}}\equiv 1$ mod 2 and  $\chi^{(-)}_{C_{4z}\mathcal{I}} \equiv 0$ mod 2.
		}
		\label{fig:spectralflow}
	\end{figure}

	Next, we discuss changes of the band structures while continuously changing the parameter $\lambda$ from $\lambda=-1$ to $\lambda=1$. We call these changes of the band energy in this process ``spectral flows''. When the bulk and the surfaces are gapped at $k_{z}=0$ and $\pi$ and $M$ is sufficiently large, the spectral flows in the band gap have the following important properties: (i) The spectral flows in the band gap are symmetric under $\lambda \leftrightarrow -\lambda$, namely, $E(k_{z}, \lambda)=E(k_{z}, -\lambda)$ [Fig.~\ref{fig:spectralflow}(a)]. (ii) When a state with the energy $E(k_{z},\lambda)$ has the $e^{i\alpha}$ eigenvalue of $C_{4z}\mathcal{I}$ ($\alpha=\pm \pi/4$, $\pm 3 \pi/4$), a state with the energy $E(k_{z},-\lambda)$ has the $-e^{i\alpha}$ eigenvalue [Fig.~\ref{fig:spectralflow}(a)]. The proofs of these  properties (i) and (ii) of the spectral flows are given in Appendix~\ref{app:property_1_spectral} and Appendix~\ref{app:property_2_spectral}, respectively. 
	From the  properties (i) and (ii), we find that when $N_{+\frac{\pi}{4}}(k_{z},\lambda)$ decreases by an integer $N$ in changing $\lambda$ from $\lambda=-1$ to $\lambda=0$, $N_{-\frac{3\pi}{4}}(k_{z},\lambda)$ increases by $N$ in  changing $\lambda$ from $\lambda=0$ to $\lambda=1$. Similarly, when $N_{-\frac{\pi}{4}}(k_{z},\lambda)$ decrease by $N$ in changing $\lambda$ from $\lambda=-1$ to $\lambda=0$, $N_{+\frac{3\pi}{4}}(k_{z},\lambda)$ increase by $N$ in  changing $\lambda$ from $\lambda=0$ to $\lambda=1$. 
	In this way, the properties (i) and (ii) restrict changes of $N_{+\frac{\pi}{4}}(k_{z},\lambda)$, $N_{-\frac{\pi}{4}}(k_{z},\lambda)$,  $N_{+\frac{3\pi}{4}}(k_{z},\lambda)$, and $N_{-\frac{3\pi}{4}}(k_{z},\lambda)$ through the cutting procedure.
	Therefore, the changes of $\mathcal{N}^{(\pm)}_{C_{4z} \mathcal{I}}(k_{z})|_\lambda$  from $\lambda=-1$ to $\lambda=0$ are equal to the changes of 
	$\mathcal{N}^{(\pm)}_{C_{4z} \mathcal{I}}(k_{z})|_\lambda$  from $\lambda=0$ to $\lambda=1$:
	\begin{align}\label{eq:symmetry_spectral_flow}
		&\mathcal{N}^{(\pm)}_{C_{4z}\mathcal{I}}(k_z)|_{\lambda=1}-\mathcal{N}^{(\pm)}_{C_{4z}\mathcal{I}}(k_z)|_{\lambda=0}\nonumber \\
		=&\mathcal{N}^{(\pm)}_{C_{4z}\mathcal{I}}(k_z)|_{\lambda=0}-\mathcal{N}^{(\pm)}_{C_{4z}\mathcal{I}}(k_z)|_{\lambda=-1}.
	\end{align}
	By combining Eq.~(\ref{eq:NC4I_with_lam1_and_lam-1}) and Eq.~(\ref{eq:symmetry_spectral_flow}), we obtain the following equations:
	\begin{align}
		\mathcal{N}^{(\pm)}_{C_{4z}\mathcal{I}}(0)|_{\lambda=0}=&\frac{1}{2}\left( n_{\pm \frac{\pi}{4}}(\Gamma)-n_{\mp \frac{ 3 \pi}{4}}(\Gamma)  \right) \nonumber \\
		&+\frac{1}{2}\left( n_{\pm \frac{\pi}{4}}(M)-n_{\mp \frac{ 3 \pi}{4}}(M)  \right), \\
		\mathcal{N}^{(\pm)}_{C_{4z}\mathcal{I}}(\pi)|_{\lambda=0}=&\frac{1}{2}\left( n_{\pm \frac{\pi}{4}}(Z)-n_{\mp \frac{ 3 \pi}{4}}(Z)  \right) \nonumber \\
		&+\frac{1}{2}\left( n_{\pm \frac{\pi}{4}}(A)-n_{\mp \frac{ 3 \pi}{4}}(A)  \right). 
	\end{align}
	
	In addition, to confirm that our theory are consistent with the previous section, we calculate the spectral flows of our model [Eq.~(\ref{Eq:Model_WS_hinge_rotoin})] in the cutting procedure with $L_{x}\times L_{y}=15 \times 15$, and the periodic boundary condition in the $z$-direction. Figure~\ref{fig:spectralflow}(b) shows the spectral flows through changing the cutting parameter $\lambda$ from $\lambda=-1$ to $\lambda=1$ with $k_{z}=0$.
	In changing $\lambda$ from $\lambda=-1$ to $\lambda=1$,
	occupied states with the eigenvalues $e^{i\frac{\pi}{4}}$ and $e^{-i\frac{\pi}{4}}$ move to conduction bands, while unoccupied states with the eigenvalues $e^{i\frac{3\pi}{4}}$ and $e^{-i\frac{3\pi}{4}}$ move to valence bands, which perfectly agrees with the above discussion [Fig.~\ref{fig:spectralflow}(a)].

	To see behaviors of hinge modes, we introduce $\nu^{(\pm)}(k_{z})$ as the total numbers of occupied states  of $\mathcal{H}^{(\pm)}(k_{z}, \lambda)$ with $\lambda=0$, where $(\pm)$ refers to the sectors with $C_{2z}=\pm i$. They can be expressed as
	\begin{align}
		\nu^{(\pm)}(k_{z}) &= N_{ \pm \frac{\pi}{4}}(k_{z}, \lambda=0)+N_{\mp \frac{3\pi}{4}}(k_{z}, \lambda=0) \nonumber \\
		&\equiv \mathcal{N}^{(\pm)}_{C_{4z}\mathcal{I}}(k_{z})|_{\lambda=0}\ \  {\rm mod\ 2},
	\end{align}
	with $k_{z}=0$ or $\pi$. Therefore, the differences between the total numbers of occupied states with $k_{z}=0$  and that with  $k_{z}=\pi$ are given by
	\begin{align}\label{eq:mainresult}
		\nu^{(\pm)}(\pi)-\nu^{(\pm)}(0) &\equiv \mathcal{N}^{(\pm)}_{C_{4z}\mathcal{I}}(\pi)|_{\lambda=0} - \mathcal{N}^{(\pm)}_{C_{4z}\mathcal{I}}(0)|_{\lambda=0} \nonumber \\
		& \equiv\chi^{(\pm)}_{C_{4z}\mathcal{I}} \ \ {\rm mod\ 2}.
	\end{align}
	This equation means that the parity of the difference between $\nu^{\pm}(\pi)$ and $\nu^{\pm}(0)$ is determined by the numbers of occupied states with $C_{4z} \mathcal{I}$ eigenvalues at high-symmetry points. Therefore, when $\chi^{(+)}_{C_{4z}\mathcal{I}}\equiv 1$ mod 2, gapless states of $\mathcal{H}^{(+)}(k_{z},\lambda=0)$ appear, and  the number of times the  gapless states cross the Fermi level between $k_z = 0$ and $k_z = \pi$ is odd.
	Similarly, when $\chi^{(-)}_{C_{4z}\mathcal{I}}\equiv 1$ mod 2, 
	this holds in the $C_{2z}=-i$ sector.
	
	Figure \ref{fig:spectralflow}(c) shows an example when $\chi^{(+)}_{C_{4z}\mathcal{I}}\equiv \chi^{(-)}_{C_{4z}\mathcal{I}}\equiv 1$ mod 2. In Fig.~\ref{fig:spectralflow}(c), the numbers of occupied bands satisfy $\nu^{(\pm)}(0)=N^{(\pm)}_{occ.}-1$ and $\nu^{(\pm)}(\pi)=N^{(\pm)}_{occ.}$, where  $N^{(\pm)}_{occ.}$ is an integer.
	Therefore, the states in the $C_{2z}=+ i$ sector  cross the Fermi level an odd number of times and so do the states in the $C_{2z}=-i$ sector. 
	As mentioned at the beginning of this section, we assume that the surfaces are gapped at  $k_{z}=0$ and $k_{z}=\pi$. Therefore, these states crossing the Fermi level are localized at the hinges in the real space, namely, these gapless states are hinge modes. 
	Therefore, the hinge modes always appear in insulators with $C_{4z}\mathcal{I}$ symmetry when $\chi^{(+)}_{C_{4z}\mathcal{I}}\equiv \chi^{(-)}_{C_{4z}\mathcal{I}} \equiv 1$ mod 2. 
	
		Under the condition that the bulk and the surfaces are  gapped at $k_{z}=0$ and $k_{z}=\pi$, 
		this scenario can be applied not only to insulators but also to WSMs. [Fig.~\ref{fig:spectralflow}(d)]. 
		In Fig.~\ref{fig:spectralflow}(d), the chiral hinge modes contribute to  $\nu^{(\pm)}(\pi)-\nu^{(\pm)}(0)$, but the Fermi-arc surface states do not contribute to that. 
		The model of the HOWSM in Sec.~\ref{sec:HOWSM_model} corresponds to this case. 
		On the other hand, another case of HOWSM is possible, where the Fermi-arc surface states account for the nonzero $\nu^{(\pm)}(\pi)-\nu^{(\pm)}(0)$, and the chiral hinge modes do not appear [Fig.~\ref{fig:spectralflow}(e)]. When $\chi^{(+)}_{C_{4z}\mathcal{I}}\equiv 1$ mod 2 and  $\chi^{(-)}_{C_{4z}\mathcal{I}} \equiv 0$ mod 2,  it leads to $\nu^{(+)}(\pi)-\nu^{(+)}(0)\equiv 1$ mod 2 and  $\nu^{(-)}(\pi)-\nu^{(-)}(0)\equiv 0$ mod 2 [Fig.~\ref{fig:spectralflow}(f)]. 
		The nonzero value of $\nu^{(+)}(\pi)-\nu^{(+)}(0)$ means that there should be gapless modes.
		Nonetheless, in this case,  $\nu^{(+)}(\pi)-\nu^{(+)}(0)$ and $\nu^{(-)}(\pi)-\nu^{(-)}(0)$ are not equal. Therefore, these gapless modes are not hinge modes because a hinge mode at a given $k_z$ $(\neq 0,\pi)$ lying at two hinges related by $C_{2z}$ symmetry, leads to two modes, one with $C_{2z}=+i$ and the other with $C_{2z}=-i$. 
		In this way, from Eq.~(\ref{eq:mainresult}), we can determine whether the gapless modes cross the Fermi level an odd number of times or an even number of times although whether the gapless modes are the hinge modes or the  Fermi-arc surface states are not determined only by  Eq.~(\ref{eq:mainresult}).
			
\section{Conclusion and discussion}\label{sec:conclusion_and_discussion}
	In summary, we show that when the topological invariants $\chi_{C_{4z}\mathcal{I}}^{(\pm)}$ are nontrivial, the  chiral hinge modes appear, and thereby show general bulk-hinge correspondence for HOTPs.
	Our approach gives direct evidence to show the bulk-hinge correspondence only from the information of symmetry eigenvalues without relying on Dirac surface theory or on Wannier representations used in the previous works.  
	In addition, we proposed a class of  HOWSMs having the chiral hinge modes protected by rotoinversion $C_{4}\mathcal{I}$ symmetry in this paper. 
	Unlike the HOWSMs in the previous works, which are characterized by topological invariants for the 2D HOTI  and are realized by stacking the 2D HOTIs, the HOWSM in this paper has the chiral hinge modes characterized by the topological invariants for 3D HOTIs.  
	
	Lastly, we discuss the possibilities of the material realizations of our HOWSMs with chiral hinge modes. 
	Our model $\mathcal{H}_{\rm DSM}(\boldsymbol{k})$ has similar energy spectra to a Dirac semimetal ${\rm Cd}_3 {\rm As}_2$ \cite{PhysRevB.88.125427, yi2014evidence, doi:10.1021/ic403163d} [see Appendix \ref{appendix:Cd3As2}]. 
	By adding perturbations  to this model, we construct both the model of the HOWSMs protected by $C_4 \mathcal{I}$ symmetry in Sec.\ref{sec:HOWSM_model}  and the model  protected by $\mathcal{I}$ symmetry in Appendix \ref{ap:HOWSM_inversion}.
	In addition, the recent experiment shows that  ${\rm Cd}_3 {\rm As}_2$ has hinge modes \cite{PhysRevLett.124.156601}. In terms of this experimental result and our theoretical results, we propose that our HOWSMs may be realized by adding perturbations such as magnetic doping or an external magnetic field to  ${\rm Cd}_3 {\rm As}_2$ \cite{PhysRevB.88.125427, liu2014stable} without breaking $C_4 \mathcal{I}$ symmetry or  $\mathcal{I}$ symmetry.
	
	\begin{acknowledgments}
		This work was supported by  JSPS KAKENHI Grant No.~JP18H03678 and No.~JP21J22264, and  by JSPS Grant-in-Aid for Scientific Research
		on Innovative Areas “Discrete Geometric Analysis for Materials
		Design” Grant No.~JP17H06469 and No.~JP20H04633.
	\end{acknowledgments}
	
	\appendix
	\section{Comparison to the previous works}\label{subsec:Comparison_to_pre}
		Here, we discuss the relationship between this work and the previous works. Although $C_{4z}\mathcal{I}$-symmetric bulk-hinge correspondence has been studied in the previous works \cite{schindler2018higher, PhysRevB.97.155305, PhysRevB.98.081110}, they are merely indirect evidences to show that this bulk-hinge correspondence is reasonable. Discussions on the bulk-hinge correspondence with $C_{4z}\mathcal{I}$ symmetry in the previous works are classified into the following two types: (i) $\boldsymbol{k}\cdot \boldsymbol{p}$ surface theory \cite{schindler2018higher} and (ii) Wannier approach \cite{PhysRevB.98.081110}. 
		In the approach (i), one starts from the surface Dirac Hamiltonian with a Dirac mass term, and show that the Dirac mass term has a different sign between the two surfaces sharing the same hinge, resulting in the emergence of the chiral hinge modes.
		In the approach (ii), one calculates the Wannier centers for two pseudo-2D systems corresponding to two subspaces at $k_{z}=0$ and $k_{z}=\pi$. When excess corner charges obtained from the Wannier centers in the two subspaces are different, chiral hinge modes appear as a pumping of the corner charges. 
		However, both of these approaches study only the special cases of $C_{4z}\mathcal{I}$-symmetric systems.
		Whether they hold in general remains unclear, except in cases that are adiabatically connected to those cases. The approach (i) cannot be applied to cases where the surface cannot be represented by Dirac Hamiltonian. The approach (ii) is based on Wannier representations and assumes that electronic states are described  in terms of  localized Wannier functions. However, some electronic states, such as those with fragile topology  \cite{PhysRevLett.121.126402},  cannot be described in terms of localized Wannier states.
		On the other hand, our approach allows us to show emergence of the chiral hinge modes only from the $C_{4z}\mathcal{I}$ symmetry eigenvalues without relying on Dirac surface Hamiltonian or on Wannier representations.
	
		Next, we discuss relationship between our topological invariants $\chi_{C_{4z}\mathcal{I}}^{(\pm)}$ and a topological invariant $\nu_{c}$ in Ref.~\cite{schindler2018higher}. 
		In the previous work \cite{schindler2018higher}, the authors restrict themselves to systems with $\mathcal{I} \mathcal{T}$ symmetry in order to define $\nu_{c}$. 
		Because  $[\hat{C}_{4z}\hat{\mathcal{I}}, \hat{\mathcal{I}}\hat{\mathcal{T}}]=0$ in $\mathcal{I} \mathcal{T}$-symmetric systems and $\hat{\mathcal{I}}\hat{\mathcal{T}}$ is anti-unitary, the eigenvalues of $C_{4z}\mathcal{I}$ at the high-symmetry points can be obtained by pairs $\{ e^{i\frac{\pi}{4}} \xi_{n, \boldsymbol{k}}, e^{-i\frac{\pi}{4}} \xi_{n, \boldsymbol{k}} \}$, where $\xi_{n,\boldsymbol{k}}=\pm 1$, $n=1, \cdots, N/2$, and $N$ is the number of occupied bands. The topological invariant $\nu_{c}$ \cite{schindler2018higher} is given by 
		\begin{equation}
			(-1)^{\nu_c}=\prod_{n=1}^{N/2}\prod_{\boldsymbol{k}\in K_4} \xi_{n, \boldsymbol{k}}. 
		\end{equation}
		 This topological invariant is well-defined only in systems with both $C_{4z}\mathcal{I}$ symmetry and $\mathcal{I} \mathcal{T}$ symmetry. On the other hand, our topological invariants $\chi_{C_{4z}\mathcal{I}}^{(\pm)}$ can be applied to general $C_{4z}\mathcal{I}$-symmetric systems. 
	
		In the following, we show that when a given system has $\mathcal{I}\mathcal{T}$ symmetry, the topologically nontrivial case in our theory, $\chi_{C_{4z}\mathcal{I}}^{(+)}\equiv \chi_{C_{4z}\mathcal{I}}^{(-)} \equiv 1$ mod 2 is equivalent to the case with $\nu_{c}\equiv 1$ mod 2. The topological invariant $\nu_{c}$ can be rewritten as 
		\begin{equation}
			\nu_{c} \equiv \frac{1}{2}\sum_{\boldsymbol{k}\in K_{4}} n_{ \xi_{\boldsymbol{k}},-}(\boldsymbol{k})\ \ {\rm mod\ 2},
		\end{equation}
		where $n_{ \xi_{\boldsymbol{k}},-}(\boldsymbol{k})$ is the number of occupied states with $\xi_{n,\boldsymbol{k}}=-1$, and  $n_{ \xi_{\boldsymbol{k}},-}(\boldsymbol{k})$ is an even number because the eigenvalues of $C_{4z}\mathcal{I}$ at $\boldsymbol{k} \in K_{4}$ are given by a pair $\{ e^{i\frac{\pi}{4}} \xi_{n, \boldsymbol{k}}, e^{-i\frac{\pi}{4}} \xi_{n, \boldsymbol{k}} \}$. 
		Now that we are considering an insulator, our topological invariants $\chi_{C_{4z}\mathcal{I}}^{(\pm)}$ are given by Eqs.~(\ref{eq:simple_index_1}) and (\ref{eq:simple_index_2}). 
		In addition, the pair $\{ e^{i\frac{\pi}{4}} \xi_{n, \boldsymbol{k}}, e^{-i\frac{\pi}{4}} \xi_{n, \boldsymbol{k}} \}$ leads to $n_{+\frac{\pi}{4}} (\boldsymbol{k})=n_{-\frac{\pi}{4}} (\boldsymbol{k})$, and therefore we obtain
		\begin{align}\label{appendix:nuc_chi}
			\nu_{c} & \equiv \frac{1}{2}\sum_{\boldsymbol{k}\in K_{4}} \biggl(n_{+\frac{3\pi}{4}} (\boldsymbol{k})+ n_{-\frac{3\pi}{4}}(\boldsymbol{k}) \biggr) \nonumber \\
			&=\frac{1}{2}\sum_{\boldsymbol{k}\in K_{4}} \biggl( \nu - n_{+\frac{\pi}{4}} (\boldsymbol{k})- n_{-\frac{\pi}{4}} (\boldsymbol{k}) \biggr) \nonumber \\
			&=\frac{1}{2}\sum_{\boldsymbol{k}\in K_{4}} \biggl( \nu - 2n_{+\frac{\pi}{4}} (\boldsymbol{k}) \biggr) \nonumber \\
			&\equiv \chi_{C_{4z}\mathcal{I}}^{(+)}\ \ {\rm mod\ 2},
		\end{align}
		where $\nu$ is the number of occupied bands in the bulk.
		Furthermore, by using $n_{+\frac{\pi}{4}} (\boldsymbol{k})=n_{-\frac{\pi}{4}} (\boldsymbol{k})$, we get
		\begin{equation}\label{appendix:chi_equiv_pm}
			\chi_{C_{4z}\mathcal{I}}^{(+)} \equiv \chi_{C_{4z}\mathcal{I}}^{(-)}\ \ {\rm mod\ 2}. 
		\end{equation}
		By combining Eq.~(\ref{appendix:nuc_chi}) and Eq.~(\ref{appendix:chi_equiv_pm}), we find that $\chi_{C_{4z}\mathcal{I}}^{(+)}\equiv \chi_{C_{4z}\mathcal{I}}^{(-)} \equiv 1$ mod 2 is equivalent to $\nu_{c}\equiv 1$ mod 2.

	\section{Higher-order Weyl semimetals protected by inversion symmetry}\label{ap:HOWSM_inversion}
\begin{figure}
		\includegraphics[width=1\columnwidth]{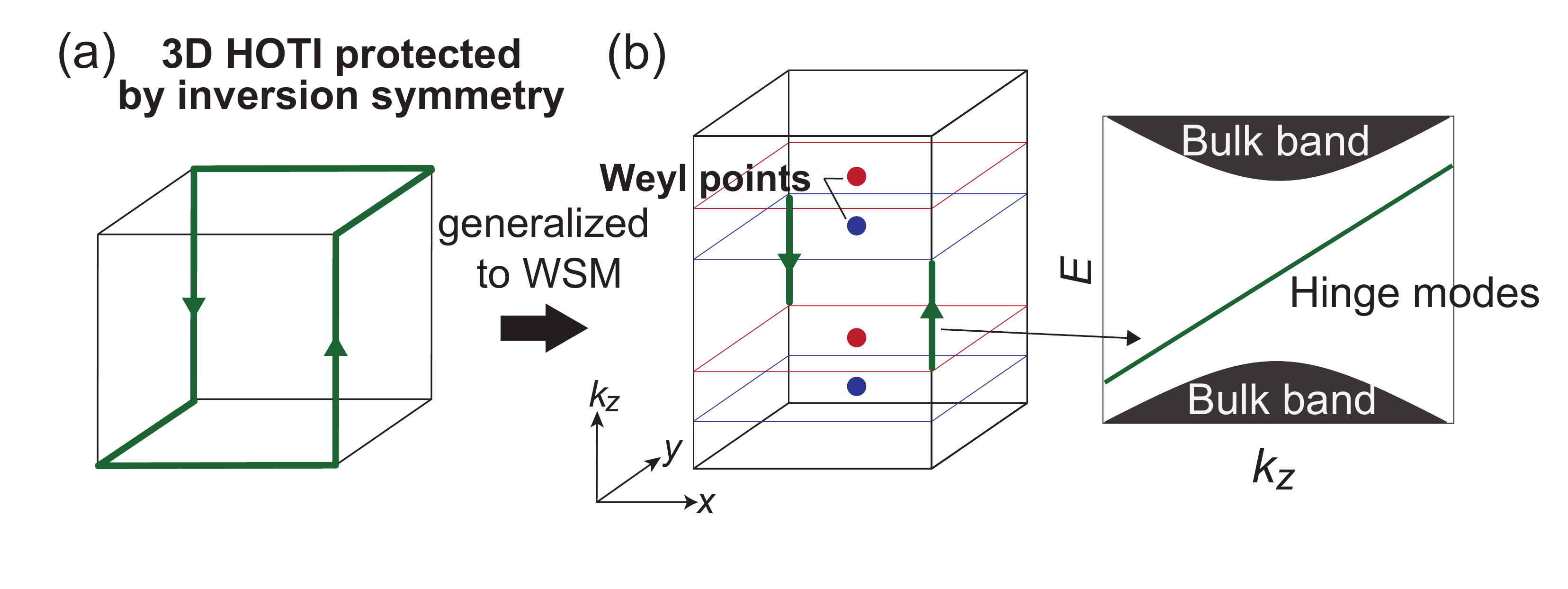}
		\caption{(a) 3D HOTIs with the chiral hinge modes protected by $\mathcal{I}$ symmetry. (b) HOWSMs with the chiral hinge modes protected by $\mathcal{I}$ symmetry. They are characterized by the same topological invariants as that of the $\mathcal{I}$-symmetric HOTIs.
		}
		\label{fig:concept_inversion_WSM}
	\end{figure}

\begin{figure}
		\includegraphics[width=1\columnwidth]{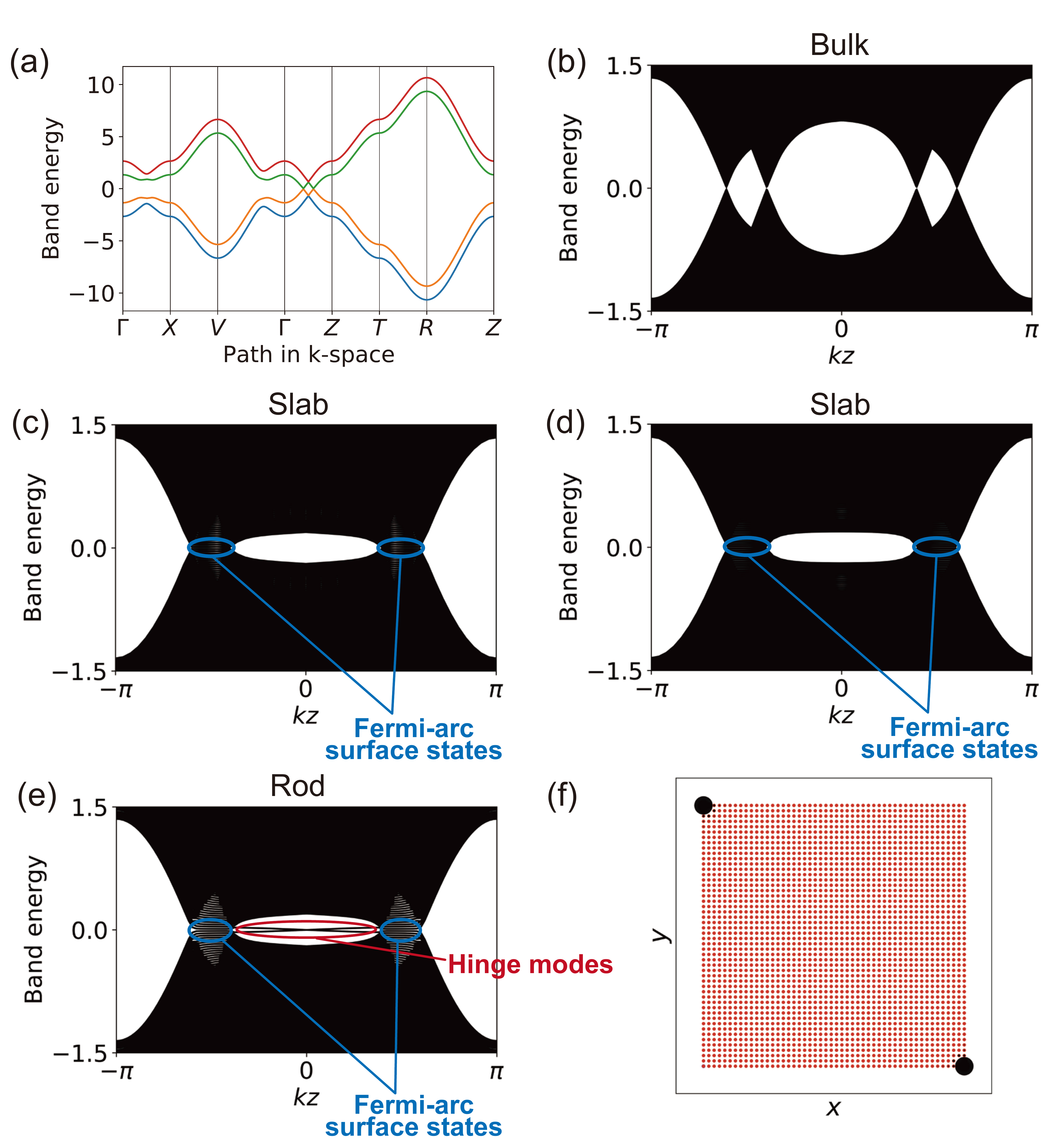}
		\caption{HOWSM with chiral hinge modes protected by $\mathcal{I}$ symmetry. (a) Bulk band structure of the tight-binding model of Eq.~(\ref{eq:model_for_HOWSM_inversion}) along the high-symmetry lines. (b) The band structure with the  periodic boundary conditions in the $x$- and the $y$-directions and with various values of $k_{x}$ and $k_{y}$. (c) The band structure with the open boundary condition in the $x$-direction and the periodic boundary condition in the $y$-direction and with various values of $k_y$, where the system size is $L_x= 200$ in the $x$-direction. (d) The band structure with the open boundary condition in the $y$-direction and the periodic boundary condition in the $x$-direction and with various values of $k_x$, where the system size is $L_y= 200$ in the $y$-direction. (e)  The band structure with the open boundary conditions in the $x$- and the $y$-directions, where the system size is $L_{x}\times L_y =50\times 50$. (f) The real-space distribution of zero-energy modes of (e). 
		} 
		\label{fig:inversion_weyl_band_rodgeom}
	\end{figure}

	In the main text, we discussed a HOWSM protected by $C_{4z}\mathcal{I}$ symmetry. For comparison, 
	in this appendix, we discuss a HOWSM with chiral hinge modes protected by $\mathcal{I}$ symmetry in class A, one of the Altland-Zirnbauer symmetry classes \cite{PhysRevB.55.1142}.
	This HOWSM is regarded as a  generalization of the 3D HOTIs protected by $\mathcal{I}$ symmetry [Figs.~\ref{fig:concept_inversion_WSM}(a) and~\ref{fig:concept_inversion_WSM}(b)]. 
	Similarly to the HOWSM protected by $C_{4z}\mathcal{I}$ symmetry  in Sec.~\ref{sec:HOWSM_model}, we start from Eq.~(\ref{eq:model_for_Dirac_semimetals}) 
	of the Hamiltonian for a 3D Dirac semimetal. 
	This model is defined on a primitive orthorhombic lattice.
	In the following, we set the lattice constants to 1.
	We add perturbations
	\begin{equation}
		\mathcal{H}_{\mathcal{I}}=\biggl(  \sum_{j=x,y,z}B_{j}  \sigma_j\biggr)\tau_0+v_{f}\sin k_{z}\sigma_x \tau_y, 
	\end{equation}
    to this Dirac semimetal. These perturbations preserve $\mathcal{I}$ symmetry, and the overall Hamiltonian can be expressed as
	\begin{equation}\label{eq:model_for_HOWSM_inversion}
		\mathcal{H}' (\boldsymbol{k}) = \mathcal{H}_{\rm DSM}(\boldsymbol{k})+ 	\mathcal{H}_{\mathcal{I}}.
	\end{equation}
	In the following, we set the parameters of the model $\mathcal{H}' (\boldsymbol{k})$ as $m=4$, $c=2$, $v=1$, $v_{f}=0.05$, $B_{x}=B_{y}=0.3$, and $B_{z}=0.5$, and we set the Fermi level to be zero. 
	The high-symmetry points in the Brillouin zone are given by $\Gamma=(0,0,0)$, $X=(\pi,0,0)$, $V=(\pi,\pi,0)$, $Z=(0,0,\pi)$, $T=(0,\pi, \pi)$, and $R=(\pi,\pi,\pi)$. 
	Figure \ref{fig:inversion_weyl_band_rodgeom}(a) shows band structures of the bulk Hamiltonian $\mathcal{H}' (\boldsymbol{k})$ along the high-symmetry lines.
	
	This model has $\mathcal{I}$ symmetry,
	\begin{equation}
		\mathcal{I} \mathcal{H}' (\boldsymbol{k}) \mathcal{I}^{-1} =\mathcal{H}' (-\boldsymbol{k}),
	\end{equation}
	and therefore the parity eigenvalues of occupied states are well-defined at the $\mathcal{I}$-symmetric momenta in the  $k$-space. 
	These $\mathcal{I}$-symmetric momenta  can be written as $\Gamma_{j=(n_{x},n_{y},n_{z})}=(n_{x}\boldsymbol{b}_{x}+n_{y}\boldsymbol{b}_{y}+n_{z}\boldsymbol{b}_{z})/2$ with 
	three integers $n_{l}=0,1$ defined modulo 2, where $\boldsymbol{b}_{l}$ represents primitive reciprocal lattice vectors.
	Here we consider the space group $\#$2 ($P\bar{1}$) generated by $\mathcal{I}$. Because we do not assume $\mathcal{T}$ symmetry, it is the type-I magnetic space group $\# 2.4$ generated by $\mathcal{I}$.
	Its symmetry-based indicators are given by $X_{\rm BS}=\mathbb{Z}_{2}\times \mathbb{Z}_{2}\times \mathbb{Z}_{2}\times \mathbb{Z}_{4}$  \cite{watanabe2018structure}. 
	The three $\mathbb{Z}_{2}$ factors are weak topological indices and  can be written as 
	\begin{equation}\label{weakz2index_inversion}
		\nu_{a}\equiv \sum_{\Gamma_{j}\land n_{a}=1}n_{-}(\Gamma_{j})\ \ \ {\rm mod}\ 2\ \ (a=x,y,z),
	\end{equation}
	where $n_{-}(\Gamma_{i})$ denotes the numbers of odd-parity occupied states at $\Gamma_{j}$, and the summation is taken over the $\mathcal{I}$-symmetric momenta on the plane $n_{a}=1$. These weak indices $\nu_{x}$, $\nu_{y}$, and $\nu_{z}$ correspond to Chern numbers on the $k_{x}=0$, $k_{y}=0$, and $k_{z}=0$ planes, respectively.
	When these indices are nontrivial, the gapless surface states appear and mask possible hinge modes. Therefore, we assume these indices are trivial. 
	On the other hand, the $\mathbb{Z}_{4}$ index of $X_{\rm BS}$  is given by an integer $\mu_1$  \cite{PhysRevB.98.115150} written as 
	\begin{align}\label{z4index_inversion}
		\mu_{1}\equiv 
		&\frac{1}{2}\sum_{\Gamma_{j}}\Bigl(n_{+}(\Gamma_{j})-n_{-}(\Gamma_{j})\Bigr) \ \ \ {\rm mod}\ 4
	\end{align}
	where $n_{+}(\Gamma_{j})$ is the number of even-parity occupied states at the $\mathcal{I}$-symmetric momenta $\Gamma_{j}$. 
	In the previous work \cite{PhysRevB.101.115120}, the authors showed that  when $\mu_1 =2$ and  $\nu_{a}=0$ ($a=x,y,z$),  hinge modes always appear in systems with $\mathcal{I}$ symmetry. 
	Although this bulk-hinge correspondence for HOTPs focuses on insulators, this theory can be also applied to WSMs when Fermi-arc surface states do not cover the hinge modes.
	The parity eigenvalues of occupied states  at the $\mathcal{I}$-symmetric momenta in our model satisfy  $\nu_{x}=\nu_{y}=\nu_z= 0$ and $\mu_{1}=2$.

	To see behaviors of the hinge modes and the surface states of our model, we calculate band structures in the bulk, in the slab and in the rod geometries, and compare them. 
	Figures~\ref{fig:inversion_weyl_band_rodgeom}(b-e) show the band structures with these different geometries. 
	These model calculations were performed using the PythTB package \cite{PythTB}. 
	The system has surfaces in both Fig.~\ref{fig:inversion_weyl_band_rodgeom}(c) and Fig.~\ref{fig:inversion_weyl_band_rodgeom}(d), and the Fermi-arc surface states appear in these cases.
	In Fig.~\ref{fig:inversion_weyl_band_rodgeom}(e), the system has surfaces and hinges, resulting in the  emergence of the hinge modes.
	The  zero-energy modes are localized at the corners in the real space [Fig.~\ref{fig:inversion_weyl_band_rodgeom}(f)], and here the surface states have a mini gap because of the finite size effects.  
	The hinge modes are localized at the two corners facing in opposite directions in the real space, unlike the  $C_{4z}\mathcal{I}$-symmetric HOWSM where the hinge modes appear at four corners in Sec.~\ref{sec:HOWSM_model}. 
	
	\section{Classification of topological phases based on symmetry-based indicators}\label{appendix:key_spa_gr}
		In this appendix, we review the classifications of topological phases in type-I magnetic space groups based on   Ref.~\cite{PhysRevB.98.115150} to see which magnetic space groups allow HOTPs with chiral hinge modes. According to Ref.~\cite{PhysRevB.98.115150}, it is sufficient to discuss the following  seven key space groups, $P\bar{1}$, $Pn$, $Pn/m$, $P\bar{4}$, $Pmmm$ (spinful), $P4/mmm$ (spinful), and $Pcc2$ (spinless), where  the strong indices of the symmetry-based indicators are given by $\mu_1$ ($P\bar{1}$),  a mirror Chern number ($Pn/m$), Chern number  ($P\bar{4}$), $\mu_4$ ($P\bar{4}$), $\kappa_1$ ($Pmmm$), $\Delta=\kappa_{1}-2\kappa_4$ ($P4/mmm$), and $\mu_{2}$ ($Pcc2$).  
		$\kappa_1$ is the indicator to detect HOTPs with helical hinge modes. A mirror Chern number and $\Delta$ are the indicators to detect topological crystalline insulator phases protected by mirror symmetries leading to gapless surface states on mirror-invariant surfaces. Similarly, $\mu_2$ is the indicator for topological crystalline insulator protected by   glide symmetries leading to gapless surface states on glide-invariant surfaces. 
		 Thus, only $\mu_1$ and $\mu_4$ can capture HOTPs with chiral hinge modes, and therefore the chiral hinge modes are limited to  ones protected by $\mathcal{I}$ symmetry and ones protected by $C_{4}\mathcal{I}$ symmetry.
	
	\section{Symmetry-based indicators and the difference between the Chern numbers on the $k_{z}=0$ and $k_{z}=\pi$  planes}\label{ap:symm_ind_chern}
	In this appendix, we show Eq.~(\ref{eq:difference_between_Chern}) in Sec.~\ref{sec:HOWSM_model}. 
	Here we consider a general spinful 3D insulator with $C_{4z}\mathcal{I}$ symmetry, and the high symmetry points in the $k$-space are given by 
	$\Gamma=(0,0,0)$, $M=(\pi,\pi,0)$, $Z=(0, 0, \pi)$, $A=(\pi,\pi,\pi)$, $X=({0,\pi,0})$, and $R=(0,\pi,\pi)$ [see Fig.~\ref{fig:model_bulk_spectral}(c)].
	On the $k_z=0$ and $k_z = \pi$ planes, the $C_{4z}\mathcal{I}$ operation  changes   the wavevectors as $(k_{x},k_y,k_z) \rightarrow (k_y, -k_x, k_z)$. 
	Similar to the formula of the Chern number expressed in terms of rotation eigenvalues in Ref.~\cite{PhysRevB.86.115112}, we can easily derive the formula of the Chern number in terms of the $C_{4z}\mathcal{I}$ and $C_{2z}$  symmetry eigenvalues as
	\begin{align}
		&i^{{\rm Ch}|_{k_z=0}}=\prod_{n:occ.} - \xi_n (\Gamma) \xi_n (M) \zeta_n (Y),\label{eq:ch_kz=0_eigenvalues} \\
		&i^{{\rm Ch}|_{k_z=\pi}}=\prod_{n:occ.}-\xi_n (Z) \xi_n (A) \zeta_n (R),\label{eq:ch_kz=pi_eigenvalues}
	\end{align}
	where $\xi_{n}(\boldsymbol{k})$ and $\zeta_{n}(\boldsymbol{k})$ are eigenvalues of $C_{4z}\mathcal{I}$ and $C_{2z}$ at $C_{4z}\mathcal{I}$-invariant momenta $\boldsymbol{k}$, respectively. Here, the following equations hold:
	\begin{align}
		&\prod_{n:occ.} \xi_n (\boldsymbol{k}) = \prod_{\alpha}{\exp}\biggl[ i \alpha n_{\alpha}(\boldsymbol{k}) 
		\biggr],\\
		&\prod_{n:occ.} \zeta_n (\boldsymbol{k}) = \prod_{\beta}{\exp}\biggl[ i\beta n_{C_{2z}}^{(\beta)}(\boldsymbol{k}) 
		\biggr],
	\end{align}
	where $\alpha$ runs over $ \alpha=\pm \pi/4$,  $\pm 3\pi/4$, and $\beta$ runs over $\beta=\pm \pi/2$, and 
	$n_{\alpha}(\boldsymbol{k})$ denotes the numbers of occupied states with the $C_{4z}\mathcal{I}$ eigenvalues $e^{i\alpha}$. 
	Similarly, $n_{C_{2z}}^{(\beta)}(\boldsymbol{k})$ is the number of occupied states with the $C_{2z}$ eigenvalues $e^{i\beta}$.
	By combining Eq.~(\ref{eq:ch_kz=0_eigenvalues}) and Eq.~(\ref{eq:ch_kz=pi_eigenvalues}), we obtain the following equation:
	\begin{align}\label{eq:differece_equation_1}
		&i^{({\rm Ch}|_{k_z=\pi}-{\rm Ch}|_{k_z=0})}\nonumber \\
		=&\prod_{\alpha} {\exp} \biggl[
		i\alpha \Bigl( n_{\alpha}(Z)+n_{\alpha}(A)
		-n_{\alpha}(\Gamma)-n_{\alpha}(M)
		\Bigr) \biggr] \nonumber \\
		&\times \prod_{\beta} {\exp} \biggl[
		i\beta \Bigl( n^{(\beta)}_{C_{2z}}(R)-n^{(\beta)}_{C_{2z}}(X)
		\Bigr) \biggr].
	\end{align}
	
	By using a compatibility relations of $C_{2z}$ symmetry along the high-symmetry line $X$-$R$
	\begin{align}
		n_{C_{2z}}^{(\beta)}(X)=n_{C_{2z}}^{(\beta)}(R), 
	\end{align}
	we rewrite Eq.~(\ref{eq:differece_equation_1}) as 
	\begin{widetext}
		\begin{align}\label{eq:difference/equation_2}
			&i^{({\rm Ch}|_{k_z=\pi}-{\rm Ch}|_{k_z=0})}
			=\prod_{\alpha} {\exp} \biggl[
			i\alpha \Bigl( n_{\alpha}(Z)+n_{\alpha}(A) -n_{\alpha}(\Gamma)-n_{\alpha}(M)
			\Bigr) \biggr], \nonumber \\
			=&{\exp} \biggl[
			i\frac{\pi}{4} \sum_{\boldsymbol{k} \in Z, A} \Bigl( n_{+\frac{\pi}{4}}(\boldsymbol{k})+3n_{+\frac{3\pi}{4}}(\boldsymbol{k})
			-n_{-\frac{\pi}{4}}(\boldsymbol{k})
			-3n_{-\frac{3\pi}{4}}(\boldsymbol{k})
			\Bigr) -i\frac{\pi}{4} \sum_{\boldsymbol{k} \in \Gamma, M} \Bigl( n_{+\frac{\pi}{4}}(\boldsymbol{k})+3n_{+\frac{3\pi}{4}}(\boldsymbol{k})
			-n_{-\frac{\pi}{4}}(\boldsymbol{k})
			-3n_{-\frac{3\pi}{4}}(\boldsymbol{k})
			\Bigr)
			\biggr]\nonumber \\
			=&{\exp} \biggl[
			i\frac{\pi}{4} \sum_{\boldsymbol{k} \in Z, A} \Bigl( n_{+\frac{\pi}{4}}(\boldsymbol{k})+n_{-\frac{3\pi}{4}}(\boldsymbol{k})
			-n_{+\frac{3\pi}{4}}(\boldsymbol{k})
			-n_{-\frac{\pi}{4}}(\boldsymbol{k})
			\Bigr) -i\frac{\pi}{4} \sum_{\boldsymbol{k} \in \Gamma, M} \Bigl( n_{+\frac{\pi}{4}}(\boldsymbol{k})+n_{-\frac{3\pi}{4}}(\boldsymbol{k})
			-n_{+\frac{3\pi}{4}}(\boldsymbol{k})
			-n_{-\frac{\pi}{4}}(\boldsymbol{k})
			\Bigr)
			\biggr] \nonumber \\
			&\times {\exp} \biggl[
			i\pi \sum_{\boldsymbol{k} \in Z, A} \Bigl( n_{+\frac{3 \pi}{4}}(\boldsymbol{k})-n_{-\frac{3\pi}{4}}(\boldsymbol{k})
			\Bigr) -i\pi \sum_{\boldsymbol{k} \in \Gamma, M}  \Bigl( n_{+\frac{3 \pi}{4}}(\boldsymbol{k})-n_{-\frac{3\pi}{4}}(\boldsymbol{k})
			\Bigr)
			\biggr],\nonumber \\
			=&{\exp} \biggl[
			i\pi \sum_{\boldsymbol{k} \in Z, A} \Bigl( n_{+\frac{3 \pi}{4}}(\boldsymbol{k})-n_{-\frac{3\pi}{4}}(\boldsymbol{k})
			\Bigr)-i\pi \sum_{\boldsymbol{k} \in \Gamma, M}  \Bigl( n_{+\frac{3 \pi}{4}}(\boldsymbol{k})-n_{-\frac{3\pi}{4}}(\boldsymbol{k})
			\Bigr)
			\biggr],
		\end{align}
	\end{widetext}
	where the last equality follows from the compatibility relations of the $C_{2z}$ symmetry eigenvalues along the high-symmetry lines $\Gamma$-$Z$ and $M$-$A$
	\begin{align}
		n_{+\frac{\pi}{4}}(\Gamma)+n_{-\frac{3\pi}{4}}(\Gamma)=
		n_{+\frac{\pi}{4}}(Z)+n_{-\frac{3\pi}{4}}(Z),\nonumber \\
		n_{-\frac{\pi}{4}}(\Gamma)+n_{+\frac{3\pi}{4}}(\Gamma)=
		n_{-\frac{\pi}{4}}(Z)+n_{+\frac{3\pi}{4}}(Z), \nonumber \\
		n_{+\frac{\pi}{4}}(M)+n_{-\frac{3\pi}{4}}(M)=
		n_{+\frac{\pi}{4}}(A)+n_{-\frac{3\pi}{4}}(A),\nonumber \\
		n_{-\frac{\pi}{4}}(M)+n_{+\frac{3\pi}{4}}(M)=
		n_{-\frac{\pi}{4}}(A)+n_{+\frac{3\pi}{4}}(A)\nonumber.
	\end{align}
	Noting that $\exp [-i2\pi (n_{+\frac{3\pi}{4}}(\boldsymbol{k})-n_{-\frac{3\pi}{4}}(\boldsymbol{k}))]=1$, we obtain the following equation:
	\begin{equation}\label{eq:difference_Chern_is_Delta_Chern}
		i^{({\rm Ch}|_{k_z=\pi}-{\rm Ch}|_{k_z=0})}
		=\exp [ i\pi \Delta' ], 
	\end{equation}
	with
	\begin{equation}
		\Delta' \equiv \sum_{\boldsymbol{k} \in K_{4}} \Bigl( n_{+\frac{3\pi}{4}} (\boldsymbol{k}) - n_{-\frac{3\pi}{4}} (\boldsymbol{k}) \Bigr)\  {\rm mod}\ 2,
	\end{equation}
	where  $K_{4}$ are the $C_{4z}\mathcal{I}$-symmetric momenta $K_{4}= \{ \Gamma$, $M$, $Z$, $A$\}.
	Furthermore, $\Delta' $ can be expressed as
	\begin{align}
		\Delta' & \equiv  \sum_{\boldsymbol{k} \in K_{4}}  \Bigl(\nu - n_{+\frac{\pi}{4}} (\boldsymbol{k}) - n_{-\frac{\pi}{4}} (\boldsymbol{k}) \Bigr)\nonumber \\
		& \equiv  \sum_{\boldsymbol{k} \in K_{4}}  \Bigl( n_{+\frac{\pi}{4}} (\boldsymbol{k}) + n_{-\frac{\pi}{4}} (\boldsymbol{k}) \Bigr)
		\ {\rm mod}\ 2,
	\end{align}
	where $\nu$ is a number of occupied bands in the bulk. 
	Therefore, by using the topological indices $ \chi^{(+)}_{C_{4}\mathcal{I}}$ and $ \chi^{(+)}_{C_{4}\mathcal{I}}$  [see Eq.~(\ref{eq:simple_index_1}), and Eq.~(\ref{eq:simple_index_2}) in Sec.~\ref{sec:HOWSM_model}] , we  can rewrite $	\Delta'$ as 
	\begin{equation}\label{eq:Delta_Chern_is_Chiplus_Chiminus}
		\Delta'  \equiv \chi^{(+)}_{C_{4}\mathcal{I}}+ \chi^{(-)}_{C_{4}\mathcal{I}}\ {\rm mod}\ 2.
	\end{equation}
	By combining Eq.~(\ref{eq:difference_Chern_is_Delta_Chern}) and Eq.~(\ref{eq:Delta_Chern_is_Chiplus_Chiminus}), we get  Eq.~(\ref{eq:difference_between_Chern}).
	
	\section{Properties of spectral flows with rotoinversion eigenvalues}
	In this appendix, we show the following three properties of the spectral flows in the cutting procedures that we use in Sec.~\ref{subsec:cuttign_procedure_spectral_flows}. (i) When the cutting parameter is $\lambda = -1$, the wave vector in the $x$- and $y$-direction are $k_i = \tfrac{2\pi}{L_{i}}m_i+\tfrac{\pi}{L_i}$ ($-M\leq m_i \leq M$) ($i=x,y$). 
	(ii) The spectral flows in the band gap are symmetric under $\lambda \leftrightarrow -\lambda$, namely, $E(k_{z},\lambda) = E(k_{z}, -\lambda)$. 
	(iii) When a state with the energy $E(k_{z},\lambda)$ has the $e^{i\alpha}$ eigenvalue, a state with the energy $E(k_{z},-\lambda)$ has the $-e^{i\alpha}$ eigenvalue ($\alpha$= $\pi/4$, $-\pi/4$, $3\pi/4$, and $- 3\pi/4$). We assume $M$ to be sufficiently large.
	
	The Hamiltonian considered in Sec.~\ref{subsec:cuttign_procedure_spectral_flows} can be expressed as
	\begin{align}
		&\mathcal{H}(k_{z},\lambda)
		=\sum_{x=-M}^{M}\sum_{y=-M}^{M} H_{0}(k_{z}) \otimes \ket{x,y}\bra{x,y}\nonumber \\
		&+\sum_{x=-M}^{M-1}\sum_{y=-M}^{M} \Bigl[H_{x}(k_z) \otimes \ket{x+1,y}\bra{x,y}+\rm{H.c.}\Bigr] \nonumber \\
		&+\sum_{y=-M}^{M} \Bigl[ \lambda H_{x}(k_z) \otimes \ket{-M,y}\bra{M,y} +\rm{H.c.} \Bigr] \nonumber \\
		&+\sum_{x=-M}^{M}\sum_{y=-M}^{M-1} \Bigl[H_{y} (k_z) \otimes \ket{x,y+1} \bra{x,y}+\rm{H.c.}\Bigr] \nonumber \\
		&+\sum_{x=-M}^{M} \Bigl[ \lambda H_{y} (k_z)\otimes \ket{x,-M}\bra{x,M} +\rm{H.c.} \Bigr],
	\end{align}
	where $H_{0}(k_{z})$, $H_{x}(k_{z})$ and  $H_{y}(k_{z})$ are $n\times n$ matrices, with $n$ being the number of internal degrees of freedom per site. For simplicity, we consider a case where the hopping is limited up to the nearest neighbor unit cells. 
	The following discussions can be easily extended to the case with $m$-th nearest neighbor hoppings as long as $m\ll M$.
	
	\subsection{Proof of $k_i = (2m_i+1)\pi/L_{i}$  when $\lambda = -1$}\label{app:wave_vector_APBC}
	Here, we show that by changing the cutting parameter from $\lambda=1$ to $\lambda=-1$, the wave vectors are shifted as $k_{i}=2 m_i \pi/L_i \rightarrow (2 m_i+1)\pi/L_i$. 
	In fact, this is obvious from the fact that $\lambda=1$ and $\lambda=-1$ correspond to periodic and antiperiodic boundary conditions, respectively. Here, we show this explicitly for the purpose of using the results to prove other properties in the following subsections.
	Here, the translation operators $T_{x}$ and $T_{y}$ are defined as 
	\begin{align}
		T_{x}\equiv &\sum_{x=-M}^{M-1} \sum_{y=-M}^{M} \Bigl[ \ket{x+1,y}\bra{x,y}\Bigr] \nonumber \\
		&+ \sum_{y=-M}^{M} \Bigl[\ket{-M,y}\bra{M,y}\Bigr], \nonumber \\
		T_{y}\equiv & \sum_{x=-M}^{M} \sum_{y=-M}^{M-1} \Bigl[ \ket{x,y+1}\bra{x,y}\Bigr]\nonumber \\
		&+\sum_{x=-M}^{M} \Bigl[ \ket{x,-M}\bra{x,M}\Bigr].
	\end{align}
	Therefore, when $\lambda=1$, the Hamiltonian can be expressed as
	\begin{align}
		\mathcal{H}(k_{z},\lambda=1)
		=&\sum_{x=-M}^{M}\sum_{y=-M}^{M} H_{0}(k_{z}) \otimes \ket{x,y}\bra{x,y}\nonumber \\
		+&\Bigr[ H_{x}\otimes T_x+H_{y}\otimes T_y +\rm{H.c.}\Bigr].
	\end{align}
	In addition, we introduce two unitary transformation
	\begin{equation}
		U_{x}=\exp \left( i\frac{\pi}{L_{x}}\hat{x} \right),\  U_{y}=\exp \left( i\frac{\pi}{L_{y}}\hat{y} \right).
	\end{equation}
	By using these unitary transformation, we can obtain the following equation:
	\begin{align}\label{app:eq:unitary_transform}
		&U_{x}U_{y}\mathcal{H}(k_{z},\lambda) (U_{x}U_{y})^{-1} \nonumber \\
		=&\sum_{x=-M}^{M}\sum_{y=-M}^{M} H_{0}(k_{z}) \otimes \ket{x,y}\bra{x,y}\nonumber \\
		&+\sum_{x=-M}^{M-1}\sum_{y=-M}^{M} \Bigl[ e^{i\frac{\pi}{L_{x}}} H_{x}\otimes \ket{x+1,y}\bra{x,y}+\rm{H.c.}\Bigr] \nonumber \\
		&+\sum_{y=-M}^{M} \Bigl[ -\lambda e^{i\frac{\pi}{L_{x}}} H_{x}\otimes \ket{-M,y}\bra{M,y} +\rm{H.c.} \Bigr] \nonumber \\
		&+\sum_{x=-M}^{M}\sum_{y=-M}^{M-1} \Bigl[ e^{i\frac{\pi}{L_{y}}} H_{y}\otimes \ket{x,y+1} \bra{x,y}+\rm{H.c.}\Bigr] \nonumber \\
		&+\sum_{x=-M}^{M} \Bigl[ -\lambda e^{i\frac{\pi}{L_{x}}} H_{y}\otimes \ket{x,-M}\bra{x,M} +\rm{H.c.} \Bigr].
	\end{align}
	When $\lambda=-1$, this equation can be expressed as
	\begin{align}
		&U_{x}U_{y}\mathcal{H}(k_{z},\lambda=-1) (U_{x}U_{y})^{-1} \nonumber \\
		=&\sum_{x=-M}^{M}\sum_{y=-M}^{M} H_{0}(k_{z}) \otimes \ket{x,y}\bra{x,y}\nonumber \\
		&+\Bigr[ H_{x}\otimes \tilde{T}_x+H_{y}\otimes \tilde{T}_y +\rm{H.c.}\Bigr],
	\end{align}
	where  $\tilde{T}_x\equiv e^{i\frac{\pi}{L_x}}T_x$ and $\tilde{T}_y \equiv e^{i\frac{\pi}{L_y}}T_y$.
	The eigenvalues of $T_{i}$ and $\tilde{T}_{i}$ ($i=x,y$) are defined by $e^{i k_{i}}$ and $e^{i \tilde{k}_{i}}$, respectively, and these wave vectors are give by
	\begin{equation}
		k_{i}=\frac{2\pi}{L_i}m_i\ \ (-M\leq m_i \leq M),
	\end{equation}
	\begin{equation}
		\tilde{k}_{i}=\frac{2\pi}{L_i}m_i+ \frac{\pi}{L_i}\ \ (-M\leq m_i \leq M),
	\end{equation}
	with $i=x,y$. From this, we conclude that  $U_{x}U_{y}\mathcal{H}(k_{z},\lambda=-1) (U_{x}U_{y})^{-1}$ is unitary equivalent to $\mathcal{H}(k_z, \lambda=1)$ with the wave vector being shifted as $k_i \rightarrow k_i+\frac{\pi}{L_i}$.
	In other words, by changing $\lambda$ from $-1$ to 1, the wave vectors are shifted as 
	$k_{i}=2 m_i \pi/L_i \rightarrow (2 m_i+1)\pi/L_i$. 
	
	\subsection{Proof of $E(k_{z},\lambda) = E(k_{z},-\lambda)$ for spectral flows}\label{app:property_1_spectral}
	Here, we show that the spectral flows in the band gap are symmetric under $\lambda \leftrightarrow -\lambda$ when $L_{x}$ $(=L_y)$ is sufficiently large. 
	By approximating $e^{i\frac{\pi}{L_{x}}}\approx 1$ and $e^{i\frac{\pi}{L_{y}}}\approx 1$, from Eq.~(\ref{app:eq:unitary_transform}), we obtain the following equation:
	\begin{equation}
		U_{x}U_{y}\mathcal{H}(k_{z},\lambda) (U_{x}U_{y})^{-1} \approx  \mathcal{H} (k_{z}, -\lambda),
	\end{equation}
	where $L_{x}$ and $L_{y}$ are sufficiently large.
	Therefore, under the unitary transformations $U_x$ and $U_y$, $\mathcal{H}(k_{z},\lambda)$ can be transformed into  $\mathcal{H} (k_{z}, -\lambda)$, and these two Hamiltonian has the same energy: $E(k_{z},\lambda)=E(k_{z},-\lambda)$. In other words, the spectral flows are symmetric under $\lambda \leftrightarrow -\lambda$ when $L_{x}$ and $L_{y}$ are sufficiently large. 
	
	\subsection{Spectral flows with the opposite rotoinversion eigenvalues}\label{app:property_2_spectral}
	Next, we give a proof that when a state with the energy $E(k_{z},\lambda)$ has a $e^{i\alpha}$ eigenvalue of $C_{4z}\mathcal{I}$ ($\alpha=\pm \pi/4$, $\pm 3 \pi/4$), a state with the energy $E(k_{z},-\lambda)$ has the $-e^{i\alpha}$ eigenvalue. 
	In the previous subsection, we showed that $U_{x}U_{y}\mathcal{H}(k_{z},\lambda) (U_{x}U_{y})^{-1} \approx  \mathcal{H} (k_{z}, -\lambda)$ when $L_x$ and $L_y$ are sufficiently large.
	Therefore, the following equation holds:
	\begin{equation}
		\mathcal{H}(k_{z},-\lambda) U_x U_y \ket{\psi_{\alpha}(k_z, \lambda)} = E   U_x U_y \ket{\psi_{\alpha}(k_z, \lambda)},
	\end{equation}
	when $\mathcal{H}(k_{z},\lambda)$ has an eigenstate $\ket{\psi_{\alpha}(k_z, \lambda)}$ with the $e^{i \alpha}$ eigenvalue ($\alpha=\pm \pi/4$, $\pm 3 \pi/4$) of $C_{4z}\mathcal{I}$.
	From this, we obtain the following equation:
	\begin{align}
		C_{4z} \mathcal{I} U_x U_y \ket{\psi_{\alpha}(k_z, \lambda)} 
		=&U_y U_x^{\dagger} C_{4z} \mathcal{I} \ket{\psi_{\alpha}(k_z, \lambda)} \nonumber \\
		=& (U_x^{\dagger})^{2} U_x U_y e^{i\alpha} \ket{\psi_{\alpha}(k_z, \lambda)}\nonumber \\
		=& e^{i\frac{2\pi }{L_x}\hat{x}} U_x U_y e^{i\alpha} \ket{\psi_{\alpha}(k_z, \lambda)}\nonumber \\
		\simeq&-e^{i\alpha} U_x U_y \ket{\psi_{\alpha}(k_z, \lambda)},
	\end{align}
	where $x\simeq \pm M$. Therefore, the localized states $\ket{\psi_{\alpha}(k_z, \lambda)}$ and $U_x U_y \ket{\psi_{\alpha}(k_z, \lambda)}$ at the boundary  $x\simeq \pm M$ have opposite eigenvalues, $e^{i\alpha}$ and $-e^{i\alpha}$, respectively.
	
\section{Relationship between $\mathcal{H}_{\rm DSM} (\vec{k})$ and an effective Hamiltonian for ${\rm Cd}_3 {\rm As}_2$}\label{appendix:Cd3As2}

In this appendix, we show that our model $\mathcal{H}_{\rm DSM} (\boldsymbol{k})$ for a Dirac semimetal [Eq.~(\ref{eq:model_for_Dirac_semimetals})] has similar energy spectra to an effective $\boldsymbol{k}\cdot \boldsymbol{p}$ model for  ${\rm Cd}_3 {\rm As}_2$ known as a Dirac semimetal \cite{PhysRevB.88.125427, yi2014evidence, doi:10.1021/ic403163d}. 
 ${\rm Cd}_3 {\rm As}_2$ respects space groups $\# 137$ ($P4_2 /nmc1'$) and $\# 142$ ($I4_1 /acd1'$)  in room-temperature and high-temperature, respectively \cite{PhysRevB.88.125427, doi:10.1021/ic403163d}. 
 When ${\rm Cd}_3 {\rm As}_2$ has such centrosymmetric structures, 
 the effective Hamiltonian \cite{PhysRevB.88.125427} around the $\Gamma$ point is given by  
\begin{align}
	\mathcal{H}_{\Gamma}(\boldsymbol{k}) = &\epsilon_0 (\boldsymbol{k}) +\begin{pmatrix}
		M(\boldsymbol{k}) & Ak_{+} & 0 &0 \\
		Ak_{-} & -M(\boldsymbol{k}) & 0 & 0 \\
		0 & 0 & M (\boldsymbol{k}) & -A k_- \\ 
		0 & 0 & -A k_{+} & -M(\boldsymbol{k})
	\end{pmatrix}  \nonumber \\
&+O(k^2),
\end{align}
 in terms of the four basis $\ket{S_{J=\frac{1}{2}}, J_z =\frac{1}{2}}$, $\ket{P_{J=\frac{3}{2}}, J_z=\frac{3}{2}}$, $\ket{S_{J=\frac{1}{2}},J_z =-\frac{1}{2}}$, and $\ket{P_{J=\frac{3}{2}}, J_z= -\frac{3}{2}}$, 
where $\epsilon_0 (\boldsymbol{k})=C_0 +C_1 k^{2}_z +C_2(k^2_x+k^2_y)$, $k_\pm = k_x \pm ik_y$, and $M(\boldsymbol{k})=M_0-M_1 k^2_z-M_2(k^2_x+k^2_y)$ with parameters $M_0<0$, $M_1<0$, and $M_2 <0$. 
We can rewrite this Hamiltonian as 
\begin{align}
	\mathcal{H}_{\Gamma}(\boldsymbol{k})=&\epsilon_0 (\boldsymbol{k}) +A(k_x \sigma_z \tau_x - k_y \sigma_0 \tau_y)\nonumber \\
	&+M(\boldsymbol{k})\sigma_0 \tau_z+ O(k^2),
\end{align}
where $\sigma_i$ and $\tau_i$ ($i=x,y,z$) are the Pauli matrices corresponding to $J_z = \pm J$ and $J=\frac{3}{2}, \frac{1}{2}$, respectively. 
By using a unitary matrix
\begin{align}
	U_{D}=\sigma_{x} \frac{\tau_x-i\tau_y }{2}+\sigma_z \frac{\tau_x+i\tau_y}{2},
\end{align}
we can obtain the following equation:
\begin{align}
	\mathcal{H}'_{\Gamma}(\boldsymbol{k})  \equiv& U_{D} \mathcal{H}_{\Gamma}(\boldsymbol{k}) U^{\dagger}_{D} \nonumber \\
	=&\epsilon_0 (\boldsymbol{k}) +A(k_x \sigma_x \tau_x+k_y \sigma_y \tau_x)\nonumber \\
	&-M(\boldsymbol{k})\sigma_0 \tau_z+ O(k^2).
\end{align}
The energy dispersion of $\mathcal{H}'_{\Gamma}(\boldsymbol{k})$ is given by $E(\boldsymbol{k})=\epsilon_0 (\boldsymbol{k})\pm \sqrt{M(\boldsymbol{k})+A^2 k_+ k_-}$, resulting in a pair of Dirac points at $(k_x, k_y, k_z)=(0,0,\pm\sqrt{\frac{M_0}{M_1}})$.  If we set $\epsilon_0 (\boldsymbol{k})$ to be zero, $\mathcal{H}'_{\Gamma}(\boldsymbol{k})$ is identical to the expansion of $\mathcal{H}_{\rm DSM}(\boldsymbol{k})$ [Eq.~(\ref{eq:model_for_Dirac_semimetals})] to the second order $k^2$ around the $\Gamma$ point.


\begin{thebibliography}{85}%
\makeatletter
\providecommand \@ifxundefined [1]{%
 \@ifx{#1\undefined}
}%
\providecommand \@ifnum [1]{%
 \ifnum #1\expandafter \@firstoftwo
 \else \expandafter \@secondoftwo
 \fi
}%
\providecommand \@ifx [1]{%
 \ifx #1\expandafter \@firstoftwo
 \else \expandafter \@secondoftwo
 \fi
}%
\providecommand \natexlab [1]{#1}%
\providecommand \enquote  [1]{``#1''}%
\providecommand \bibnamefont  [1]{#1}%
\providecommand \bibfnamefont [1]{#1}%
\providecommand \citenamefont [1]{#1}%
\providecommand \href@noop [0]{\@secondoftwo}%
\providecommand \href [0]{\begingroup \@sanitize@url \@href}%
\providecommand \@href[1]{\@@startlink{#1}\@@href}%
\providecommand \@@href[1]{\endgroup#1\@@endlink}%
\providecommand \@sanitize@url [0]{\catcode `\\12\catcode `\$12\catcode
  `\&12\catcode `\#12\catcode `\^12\catcode `\_12\catcode `\%12\relax}%
\providecommand \@@startlink[1]{}%
\providecommand \@@endlink[0]{}%
\providecommand \url  [0]{\begingroup\@sanitize@url \@url }%
\providecommand \@url [1]{\endgroup\@href {#1}{\urlprefix }}%
\providecommand \urlprefix  [0]{URL }%
\providecommand \Eprint [0]{\href }%
\providecommand \doibase [0]{https://doi.org/}%
\providecommand \selectlanguage [0]{\@gobble}%
\providecommand \bibinfo  [0]{\@secondoftwo}%
\providecommand \bibfield  [0]{\@secondoftwo}%
\providecommand \translation [1]{[#1]}%
\providecommand \BibitemOpen [0]{}%
\providecommand \bibitemStop [0]{}%
\providecommand \bibitemNoStop [0]{.\EOS\space}%
\providecommand \EOS [0]{\spacefactor3000\relax}%
\providecommand \BibitemShut  [1]{\csname bibitem#1\endcsname}%
\let\auto@bib@innerbib\@empty
\bibitem [{\citenamefont {Sitte}\ \emph {et~al.}(2012)\citenamefont {Sitte},
  \citenamefont {Rosch}, \citenamefont {Altman},\ and\ \citenamefont
  {Fritz}}]{PhysRevLett.108.126807}%
  \BibitemOpen
  \bibfield  {author} {\bibinfo {author} {\bibfnamefont {M.}~\bibnamefont
  {Sitte}}, \bibinfo {author} {\bibfnamefont {A.}~\bibnamefont {Rosch}},
  \bibinfo {author} {\bibfnamefont {E.}~\bibnamefont {Altman}},\ and\ \bibinfo
  {author} {\bibfnamefont {L.}~\bibnamefont {Fritz}},\ }\bibfield  {title}
  {\bibinfo {title} {{Topological Insulators in Magnetic Fields: Quantum Hall
  Effect and Edge Channels with a Nonquantized $\ensuremath{\theta}$ Term}},\
  }\href {https://doi.org/10.1103/PhysRevLett.108.126807} {\bibfield  {journal}
  {\bibinfo  {journal} {Phys. Rev. Lett.}\ }\textbf {\bibinfo {volume} {108}},\
  \bibinfo {pages} {126807} (\bibinfo {year} {2012})}\BibitemShut {NoStop}%
\bibitem [{\citenamefont {Zhang}\ \emph {et~al.}(2013)\citenamefont {Zhang},
  \citenamefont {Kane},\ and\ \citenamefont {Mele}}]{PhysRevLett.110.046404}%
  \BibitemOpen
  \bibfield  {author} {\bibinfo {author} {\bibfnamefont {F.}~\bibnamefont
  {Zhang}}, \bibinfo {author} {\bibfnamefont {C.~L.}\ \bibnamefont {Kane}},\
  and\ \bibinfo {author} {\bibfnamefont {E.~J.}\ \bibnamefont {Mele}},\
  }\bibfield  {title} {\bibinfo {title} {{Surface State Magnetization and
  Chiral Edge States on Topological Insulators}},\ }\href
  {https://doi.org/10.1103/PhysRevLett.110.046404} {\bibfield  {journal}
  {\bibinfo  {journal} {Phys. Rev. Lett.}\ }\textbf {\bibinfo {volume} {110}},\
  \bibinfo {pages} {046404} (\bibinfo {year} {2013})}\BibitemShut {NoStop}%
\bibitem [{\citenamefont {Benalcazar}\ \emph
  {et~al.}(2017{\natexlab{a}})\citenamefont {Benalcazar}, \citenamefont
  {Bernevig},\ and\ \citenamefont {Hughes}}]{benalcazar2017quantized}%
  \BibitemOpen
  \bibfield  {author} {\bibinfo {author} {\bibfnamefont {W.~A.}\ \bibnamefont
  {Benalcazar}}, \bibinfo {author} {\bibfnamefont {B.~A.}\ \bibnamefont
  {Bernevig}},\ and\ \bibinfo {author} {\bibfnamefont {T.~L.}\ \bibnamefont
  {Hughes}},\ }\bibfield  {title} {\bibinfo {title} {{Quantized electric
  multipole insulators}},\ }\href {https://doi.org/10.1126/science.aah6442}
  {\bibfield  {journal} {\bibinfo  {journal} {Science}\ }\textbf {\bibinfo
  {volume} {357}},\ \bibinfo {pages} {61} (\bibinfo {year}
  {2017}{\natexlab{a}})}\BibitemShut {NoStop}%
\bibitem [{\citenamefont {Benalcazar}\ \emph
  {et~al.}(2017{\natexlab{b}})\citenamefont {Benalcazar}, \citenamefont
  {Bernevig},\ and\ \citenamefont {Hughes}}]{PhysRevB.96.245115}%
  \BibitemOpen
  \bibfield  {author} {\bibinfo {author} {\bibfnamefont {W.~A.}\ \bibnamefont
  {Benalcazar}}, \bibinfo {author} {\bibfnamefont {B.~A.}\ \bibnamefont
  {Bernevig}},\ and\ \bibinfo {author} {\bibfnamefont {T.~L.}\ \bibnamefont
  {Hughes}},\ }\bibfield  {title} {\bibinfo {title} {{Electric multipole
  moments, topological multipole moment pumping, and chiral hinge states in
  crystalline insulators}},\ }\href
  {https://doi.org/10.1103/PhysRevB.96.245115} {\bibfield  {journal} {\bibinfo
  {journal} {Phys. Rev. B}\ }\textbf {\bibinfo {volume} {96}},\ \bibinfo
  {pages} {245115} (\bibinfo {year} {2017}{\natexlab{b}})}\BibitemShut
  {NoStop}%
\bibitem [{\citenamefont {Song}\ \emph {et~al.}(2017)\citenamefont {Song},
  \citenamefont {Fang},\ and\ \citenamefont {Fang}}]{PhysRevLett.119.246402}%
  \BibitemOpen
  \bibfield  {author} {\bibinfo {author} {\bibfnamefont {Z.}~\bibnamefont
  {Song}}, \bibinfo {author} {\bibfnamefont {Z.}~\bibnamefont {Fang}},\ and\
  \bibinfo {author} {\bibfnamefont {C.}~\bibnamefont {Fang}},\ }\bibfield
  {title} {\bibinfo {title} {{$(d\ensuremath{-}2)$-Dimensional Edge States of
  Rotation Symmetry Protected Topological States}},\ }\href
  {https://doi.org/10.1103/PhysRevLett.119.246402} {\bibfield  {journal}
  {\bibinfo  {journal} {Phys. Rev. Lett.}\ }\textbf {\bibinfo {volume} {119}},\
  \bibinfo {pages} {246402} (\bibinfo {year} {2017})}\BibitemShut {NoStop}%
\bibitem [{\citenamefont {Langbehn}\ \emph {et~al.}(2017)\citenamefont
  {Langbehn}, \citenamefont {Peng}, \citenamefont {Trifunovic}, \citenamefont
  {von Oppen},\ and\ \citenamefont {Brouwer}}]{PhysRevLett.119.246401}%
  \BibitemOpen
  \bibfield  {author} {\bibinfo {author} {\bibfnamefont {J.}~\bibnamefont
  {Langbehn}}, \bibinfo {author} {\bibfnamefont {Y.}~\bibnamefont {Peng}},
  \bibinfo {author} {\bibfnamefont {L.}~\bibnamefont {Trifunovic}}, \bibinfo
  {author} {\bibfnamefont {F.}~\bibnamefont {von Oppen}},\ and\ \bibinfo
  {author} {\bibfnamefont {P.~W.}\ \bibnamefont {Brouwer}},\ }\bibfield
  {title} {\bibinfo {title} {{Reflection-Symmetric Second-Order Topological
  Insulators and Superconductors}},\ }\href
  {https://doi.org/10.1103/PhysRevLett.119.246401} {\bibfield  {journal}
  {\bibinfo  {journal} {Phys. Rev. Lett.}\ }\textbf {\bibinfo {volume} {119}},\
  \bibinfo {pages} {246401} (\bibinfo {year} {2017})}\BibitemShut {NoStop}%
\bibitem [{\citenamefont {Schindler}\ \emph
  {et~al.}(2018{\natexlab{a}})\citenamefont {Schindler}, \citenamefont {Cook},
  \citenamefont {Vergniory}, \citenamefont {Wang}, \citenamefont {Parkin},
  \citenamefont {Bernevig},\ and\ \citenamefont
  {Neupert}}]{schindler2018higher}%
  \BibitemOpen
  \bibfield  {author} {\bibinfo {author} {\bibfnamefont {F.}~\bibnamefont
  {Schindler}}, \bibinfo {author} {\bibfnamefont {A.~M.}\ \bibnamefont {Cook}},
  \bibinfo {author} {\bibfnamefont {M.~G.}\ \bibnamefont {Vergniory}}, \bibinfo
  {author} {\bibfnamefont {Z.}~\bibnamefont {Wang}}, \bibinfo {author}
  {\bibfnamefont {S.~S.}\ \bibnamefont {Parkin}}, \bibinfo {author}
  {\bibfnamefont {B.~A.}\ \bibnamefont {Bernevig}},\ and\ \bibinfo {author}
  {\bibfnamefont {T.}~\bibnamefont {Neupert}},\ }\bibfield  {title} {\bibinfo
  {title} {{Higher-order topological insulators}},\ }\href
  {https://doi.org/10.1126/sciadv.aat0346} {\bibfield  {journal} {\bibinfo
  {journal} {Sci. Adv.}\ }\textbf {\bibinfo {volume} {4}},\ \bibinfo {pages}
  {eaat0346} (\bibinfo {year} {2018}{\natexlab{a}})}\BibitemShut {NoStop}%
\bibitem [{\citenamefont {Fang}\ and\ \citenamefont
  {Fu}(2019)}]{fang2017rotation}%
  \BibitemOpen
  \bibfield  {author} {\bibinfo {author} {\bibfnamefont {C.}~\bibnamefont
  {Fang}}\ and\ \bibinfo {author} {\bibfnamefont {L.}~\bibnamefont {Fu}},\
  }\bibfield  {title} {\bibinfo {title} {{New classes of topological
  crystalline insulators having surface rotation anomaly}},\ }\href
  {https://advances.sciencemag.org/content/5/12/eaat2374.abstract} {\bibfield
  {journal} {\bibinfo  {journal} {Sci. Adv.}\ }\textbf {\bibinfo {volume}
  {5}},\ \bibinfo {pages} {eaat2374} (\bibinfo {year} {2019})}\BibitemShut
  {NoStop}%
\bibitem [{\citenamefont {Hasan}\ and\ \citenamefont
  {Kane}(2010)}]{RevModPhys.82.3045}%
  \BibitemOpen
  \bibfield  {author} {\bibinfo {author} {\bibfnamefont {M.~Z.}\ \bibnamefont
  {Hasan}}\ and\ \bibinfo {author} {\bibfnamefont {C.~L.}\ \bibnamefont
  {Kane}},\ }\bibfield  {title} {\bibinfo {title} {{Colloquium: Topological
  insulators}},\ }\href {https://doi.org/10.1103/RevModPhys.82.3045} {\bibfield
   {journal} {\bibinfo  {journal} {Rev. Mod. Phys.}\ }\textbf {\bibinfo
  {volume} {82}},\ \bibinfo {pages} {3045} (\bibinfo {year}
  {2010})}\BibitemShut {NoStop}%
\bibitem [{\citenamefont {Qi}\ and\ \citenamefont
  {Zhang}(2011)}]{RevModPhys.83.1057}%
  \BibitemOpen
  \bibfield  {author} {\bibinfo {author} {\bibfnamefont {X.-L.}\ \bibnamefont
  {Qi}}\ and\ \bibinfo {author} {\bibfnamefont {S.-C.}\ \bibnamefont {Zhang}},\
  }\bibfield  {title} {\bibinfo {title} {{Topological insulators and
  superconductors}},\ }\href {https://doi.org/10.1103/RevModPhys.83.1057}
  {\bibfield  {journal} {\bibinfo  {journal} {Rev. Mod. Phys.}\ }\textbf
  {\bibinfo {volume} {83}},\ \bibinfo {pages} {1057} (\bibinfo {year}
  {2011})}\BibitemShut {NoStop}%
\bibitem [{\citenamefont {Zhang}\ \emph {et~al.}(2020)\citenamefont {Zhang},
  \citenamefont {Lin}, \citenamefont {Wang}, \citenamefont {Xiong},
  \citenamefont {Tian}, \citenamefont {Lu}, \citenamefont {Chen},\ and\
  \citenamefont {Jiang}}]{zhang2020symmetry}%
  \BibitemOpen
  \bibfield  {author} {\bibinfo {author} {\bibfnamefont {X.}~\bibnamefont
  {Zhang}}, \bibinfo {author} {\bibfnamefont {Z.-K.}\ \bibnamefont {Lin}},
  \bibinfo {author} {\bibfnamefont {H.-X.}\ \bibnamefont {Wang}}, \bibinfo
  {author} {\bibfnamefont {Z.}~\bibnamefont {Xiong}}, \bibinfo {author}
  {\bibfnamefont {Y.}~\bibnamefont {Tian}}, \bibinfo {author} {\bibfnamefont
  {M.-H.}\ \bibnamefont {Lu}}, \bibinfo {author} {\bibfnamefont {Y.-F.}\
  \bibnamefont {Chen}},\ and\ \bibinfo {author} {\bibfnamefont {J.-H.}\
  \bibnamefont {Jiang}},\ }\bibfield  {title} {\bibinfo {title}
  {Symmetry-protected hierarchy of anomalous multipole topological band gaps in
  nonsymmorphic metacrystals},\ }\href
  {https://www.nature.com/articles/s41467-019-13861-4} {\bibfield  {journal}
  {\bibinfo  {journal} {Nat. Commun.}\ }\textbf {\bibinfo {volume} {11}},\
  \bibinfo {pages} {1} (\bibinfo {year} {2020})}\BibitemShut {NoStop}%
\bibitem [{\citenamefont {Kooi}\ \emph {et~al.}(2020)\citenamefont {Kooi},
  \citenamefont {van Miert},\ and\ \citenamefont
  {Ortix}}]{PhysRevB.102.041122}%
  \BibitemOpen
  \bibfield  {author} {\bibinfo {author} {\bibfnamefont {S.~H.}\ \bibnamefont
  {Kooi}}, \bibinfo {author} {\bibfnamefont {G.}~\bibnamefont {van Miert}},\
  and\ \bibinfo {author} {\bibfnamefont {C.}~\bibnamefont {Ortix}},\ }\bibfield
   {title} {\bibinfo {title} {Hybrid-order topology of weak topological
  insulators},\ }\href {https://doi.org/10.1103/PhysRevB.102.041122} {\bibfield
   {journal} {\bibinfo  {journal} {Phys. Rev. B}\ }\textbf {\bibinfo {volume}
  {102}},\ \bibinfo {pages} {041122(R)} (\bibinfo {year} {2020})}\BibitemShut
  {NoStop}%
\bibitem [{\citenamefont {Szumniak}\ \emph {et~al.}(2020)\citenamefont
  {Szumniak}, \citenamefont {Loss},\ and\ \citenamefont
  {Klinovaja}}]{PhysRevB.102.125126}%
  \BibitemOpen
  \bibfield  {author} {\bibinfo {author} {\bibfnamefont {P.}~\bibnamefont
  {Szumniak}}, \bibinfo {author} {\bibfnamefont {D.}~\bibnamefont {Loss}},\
  and\ \bibinfo {author} {\bibfnamefont {J.}~\bibnamefont {Klinovaja}},\
  }\bibfield  {title} {\bibinfo {title} {Hinge modes and surface states in
  second-order topological three-dimensional quantum hall systems induced by
  charge density modulation},\ }\href
  {https://doi.org/10.1103/PhysRevB.102.125126} {\bibfield  {journal} {\bibinfo
   {journal} {Phys. Rev. B}\ }\textbf {\bibinfo {volume} {102}},\ \bibinfo
  {pages} {125126} (\bibinfo {year} {2020})}\BibitemShut {NoStop}%
\bibitem [{\citenamefont {Yang}\ \emph {et~al.}(2021)\citenamefont {Yang},
  \citenamefont {Lu}, \citenamefont {Yan}, \citenamefont {Huang}, \citenamefont
  {Deng},\ and\ \citenamefont {Liu}}]{PhysRevLett.126.156801}%
  \BibitemOpen
  \bibfield  {author} {\bibinfo {author} {\bibfnamefont {Y.}~\bibnamefont
  {Yang}}, \bibinfo {author} {\bibfnamefont {J.}~\bibnamefont {Lu}}, \bibinfo
  {author} {\bibfnamefont {M.}~\bibnamefont {Yan}}, \bibinfo {author}
  {\bibfnamefont {X.}~\bibnamefont {Huang}}, \bibinfo {author} {\bibfnamefont
  {W.}~\bibnamefont {Deng}},\ and\ \bibinfo {author} {\bibfnamefont
  {Z.}~\bibnamefont {Liu}},\ }\bibfield  {title} {\bibinfo {title}
  {Hybrid-order topological insulators in a phononic crystal},\ }\href
  {https://doi.org/10.1103/PhysRevLett.126.156801} {\bibfield  {journal}
  {\bibinfo  {journal} {Phys. Rev. Lett.}\ }\textbf {\bibinfo {volume} {126}},\
  \bibinfo {pages} {156801} (\bibinfo {year} {2021})}\BibitemShut {NoStop}%
\bibitem [{\citenamefont {Murakami}(2007)}]{murakami2007phase}%
  \BibitemOpen
  \bibfield  {author} {\bibinfo {author} {\bibfnamefont {S.}~\bibnamefont
  {Murakami}},\ }\bibfield  {title} {\bibinfo {title} {{Phase transition
  between the quantum spin Hall and insulator phases in 3D: emergence of a
  topological gapless phase}},\ }\href
  {http://dx.doi.org/10.1088/1367-2630/9/9/356} {\bibfield  {journal} {\bibinfo
   {journal} {New. J. Phys.}\ }\textbf {\bibinfo {volume} {9}},\ \bibinfo
  {pages} {356} (\bibinfo {year} {2007})}\BibitemShut {NoStop}%
\bibitem [{\citenamefont {Wan}\ \emph {et~al.}(2011)\citenamefont {Wan},
  \citenamefont {Turner}, \citenamefont {Vishwanath},\ and\ \citenamefont
  {Savrasov}}]{PhysRevB.83.205101}%
  \BibitemOpen
  \bibfield  {author} {\bibinfo {author} {\bibfnamefont {X.}~\bibnamefont
  {Wan}}, \bibinfo {author} {\bibfnamefont {A.~M.}\ \bibnamefont {Turner}},
  \bibinfo {author} {\bibfnamefont {A.}~\bibnamefont {Vishwanath}},\ and\
  \bibinfo {author} {\bibfnamefont {S.~Y.}\ \bibnamefont {Savrasov}},\
  }\bibfield  {title} {\bibinfo {title} {{Topological semimetal and Fermi-arc
  surface states in the electronic structure of pyrochlore iridates}},\ }\href
  {https://doi.org/10.1103/PhysRevB.83.205101} {\bibfield  {journal} {\bibinfo
  {journal} {Phys. Rev. B}\ }\textbf {\bibinfo {volume} {83}},\ \bibinfo
  {pages} {205101} (\bibinfo {year} {2011})}\BibitemShut {NoStop}%
\bibitem [{\citenamefont {Yang}\ \emph {et~al.}(2011)\citenamefont {Yang},
  \citenamefont {Lu},\ and\ \citenamefont {Ran}}]{PhysRevB.84.075129}%
  \BibitemOpen
  \bibfield  {author} {\bibinfo {author} {\bibfnamefont {K.-Y.}\ \bibnamefont
  {Yang}}, \bibinfo {author} {\bibfnamefont {Y.-M.}\ \bibnamefont {Lu}},\ and\
  \bibinfo {author} {\bibfnamefont {Y.}~\bibnamefont {Ran}},\ }\bibfield
  {title} {\bibinfo {title} {{Quantum Hall effects in a Weyl semimetal:
  Possible application in pyrochlore iridates}},\ }\href
  {https://doi.org/10.1103/PhysRevB.84.075129} {\bibfield  {journal} {\bibinfo
  {journal} {Phys. Rev. B}\ }\textbf {\bibinfo {volume} {84}},\ \bibinfo
  {pages} {075129} (\bibinfo {year} {2011})}\BibitemShut {NoStop}%
\bibitem [{\citenamefont {Burkov}\ and\ \citenamefont
  {Balents}(2011)}]{PhysRevLett.107.127205}%
  \BibitemOpen
  \bibfield  {author} {\bibinfo {author} {\bibfnamefont {A.~A.}\ \bibnamefont
  {Burkov}}\ and\ \bibinfo {author} {\bibfnamefont {L.}~\bibnamefont
  {Balents}},\ }\bibfield  {title} {\bibinfo {title} {{Weyl Semimetal in a
  Topological Insulator Multilayer}},\ }\href
  {https://doi.org/10.1103/PhysRevLett.107.127205} {\bibfield  {journal}
  {\bibinfo  {journal} {Phys. Rev. Lett.}\ }\textbf {\bibinfo {volume} {107}},\
  \bibinfo {pages} {127205} (\bibinfo {year} {2011})}\BibitemShut {NoStop}%
\bibitem [{\citenamefont {Murakami}\ \emph {et~al.}(2017)\citenamefont
  {Murakami}, \citenamefont {Hirayama}, \citenamefont {Okugawa},\ and\
  \citenamefont {Miyake}}]{doi:10.1126/sciadv.1602680}%
  \BibitemOpen
  \bibfield  {author} {\bibinfo {author} {\bibfnamefont {S.}~\bibnamefont
  {Murakami}}, \bibinfo {author} {\bibfnamefont {M.}~\bibnamefont {Hirayama}},
  \bibinfo {author} {\bibfnamefont {R.}~\bibnamefont {Okugawa}},\ and\ \bibinfo
  {author} {\bibfnamefont {T.}~\bibnamefont {Miyake}},\ }\bibfield  {title}
  {\bibinfo {title} {{Emergence of topological semimetals in gap closing in
  semiconductors without inversion symmetry}},\ }\href
  {https://www.science.org/doi/abs/10.1126/sciadv.1602680} {\bibfield
  {journal} {\bibinfo  {journal} {Sci. Adv.}\ }\textbf {\bibinfo {volume}
  {3}},\ \bibinfo {pages} {e1602680} (\bibinfo {year} {2017})}\BibitemShut
  {NoStop}%
\bibitem [{\citenamefont {Wang}\ \emph {et~al.}(2012)\citenamefont {Wang},
  \citenamefont {Sun}, \citenamefont {Chen}, \citenamefont {Franchini},
  \citenamefont {Xu}, \citenamefont {Weng}, \citenamefont {Dai},\ and\
  \citenamefont {Fang}}]{PhysRevB.85.195320}%
  \BibitemOpen
  \bibfield  {author} {\bibinfo {author} {\bibfnamefont {Z.}~\bibnamefont
  {Wang}}, \bibinfo {author} {\bibfnamefont {Y.}~\bibnamefont {Sun}}, \bibinfo
  {author} {\bibfnamefont {X.-Q.}\ \bibnamefont {Chen}}, \bibinfo {author}
  {\bibfnamefont {C.}~\bibnamefont {Franchini}}, \bibinfo {author}
  {\bibfnamefont {G.}~\bibnamefont {Xu}}, \bibinfo {author} {\bibfnamefont
  {H.}~\bibnamefont {Weng}}, \bibinfo {author} {\bibfnamefont {X.}~\bibnamefont
  {Dai}},\ and\ \bibinfo {author} {\bibfnamefont {Z.}~\bibnamefont {Fang}},\
  }\bibfield  {title} {\bibinfo {title} {{Dirac semimetal and topological phase
  transitions in ${A}_{3}$Bi ($A=\text{Na}$, K, Rb)}},\ }\href
  {https://doi.org/10.1103/PhysRevB.85.195320} {\bibfield  {journal} {\bibinfo
  {journal} {Phys. Rev. B}\ }\textbf {\bibinfo {volume} {85}},\ \bibinfo
  {pages} {195320} (\bibinfo {year} {2012})}\BibitemShut {NoStop}%
\bibitem [{\citenamefont {Young}\ \emph {et~al.}(2012)\citenamefont {Young},
  \citenamefont {Zaheer}, \citenamefont {Teo}, \citenamefont {Kane},
  \citenamefont {Mele},\ and\ \citenamefont {Rappe}}]{PhysRevLett.108.140405}%
  \BibitemOpen
  \bibfield  {author} {\bibinfo {author} {\bibfnamefont {S.~M.}\ \bibnamefont
  {Young}}, \bibinfo {author} {\bibfnamefont {S.}~\bibnamefont {Zaheer}},
  \bibinfo {author} {\bibfnamefont {J.~C.~Y.}\ \bibnamefont {Teo}}, \bibinfo
  {author} {\bibfnamefont {C.~L.}\ \bibnamefont {Kane}}, \bibinfo {author}
  {\bibfnamefont {E.~J.}\ \bibnamefont {Mele}},\ and\ \bibinfo {author}
  {\bibfnamefont {A.~M.}\ \bibnamefont {Rappe}},\ }\bibfield  {title} {\bibinfo
  {title} {{Dirac Semimetal in Three Dimensions}},\ }\href
  {https://doi.org/10.1103/PhysRevLett.108.140405} {\bibfield  {journal}
  {\bibinfo  {journal} {Phys. Rev. Lett.}\ }\textbf {\bibinfo {volume} {108}},\
  \bibinfo {pages} {140405} (\bibinfo {year} {2012})}\BibitemShut {NoStop}%
\bibitem [{\citenamefont {Steinberg}\ \emph {et~al.}(2014)\citenamefont
  {Steinberg}, \citenamefont {Young}, \citenamefont {Zaheer}, \citenamefont
  {Kane}, \citenamefont {Mele},\ and\ \citenamefont
  {Rappe}}]{PhysRevLett.112.036403}%
  \BibitemOpen
  \bibfield  {author} {\bibinfo {author} {\bibfnamefont {J.~A.}\ \bibnamefont
  {Steinberg}}, \bibinfo {author} {\bibfnamefont {S.~M.}\ \bibnamefont
  {Young}}, \bibinfo {author} {\bibfnamefont {S.}~\bibnamefont {Zaheer}},
  \bibinfo {author} {\bibfnamefont {C.~L.}\ \bibnamefont {Kane}}, \bibinfo
  {author} {\bibfnamefont {E.~J.}\ \bibnamefont {Mele}},\ and\ \bibinfo
  {author} {\bibfnamefont {A.~M.}\ \bibnamefont {Rappe}},\ }\bibfield  {title}
  {\bibinfo {title} {{Bulk Dirac Points in Distorted Spinels}},\ }\href
  {https://doi.org/10.1103/PhysRevLett.112.036403} {\bibfield  {journal}
  {\bibinfo  {journal} {Phys. Rev. Lett.}\ }\textbf {\bibinfo {volume} {112}},\
  \bibinfo {pages} {036403} (\bibinfo {year} {2014})}\BibitemShut {NoStop}%
\bibitem [{\citenamefont {Hosur}(2012)}]{PhysRevB.86.195102}%
  \BibitemOpen
  \bibfield  {author} {\bibinfo {author} {\bibfnamefont {P.}~\bibnamefont
  {Hosur}},\ }\bibfield  {title} {\bibinfo {title} {{Friedel oscillations due
  to Fermi arcs in Weyl semimetals}},\ }\href
  {https://doi.org/10.1103/PhysRevB.86.195102} {\bibfield  {journal} {\bibinfo
  {journal} {Phys. Rev. B}\ }\textbf {\bibinfo {volume} {86}},\ \bibinfo
  {pages} {195102} (\bibinfo {year} {2012})}\BibitemShut {NoStop}%
\bibitem [{\citenamefont {Xu}\ \emph {et~al.}(2011)\citenamefont {Xu},
  \citenamefont {Weng}, \citenamefont {Wang}, \citenamefont {Dai},\ and\
  \citenamefont {Fang}}]{PhysRevLett.107.186806}%
  \BibitemOpen
  \bibfield  {author} {\bibinfo {author} {\bibfnamefont {G.}~\bibnamefont
  {Xu}}, \bibinfo {author} {\bibfnamefont {H.}~\bibnamefont {Weng}}, \bibinfo
  {author} {\bibfnamefont {Z.}~\bibnamefont {Wang}}, \bibinfo {author}
  {\bibfnamefont {X.}~\bibnamefont {Dai}},\ and\ \bibinfo {author}
  {\bibfnamefont {Z.}~\bibnamefont {Fang}},\ }\bibfield  {title} {\bibinfo
  {title} {{Chern Semimetal and the Quantized Anomalous Hall Effect in
  ${\mathrm{HgCr}}_{2}{\mathrm{Se}}_{4}$}},\ }\href
  {https://doi.org/10.1103/PhysRevLett.107.186806} {\bibfield  {journal}
  {\bibinfo  {journal} {Phys. Rev. Lett.}\ }\textbf {\bibinfo {volume} {107}},\
  \bibinfo {pages} {186806} (\bibinfo {year} {2011})}\BibitemShut {NoStop}%
\bibitem [{\citenamefont {Ojanen}(2013)}]{PhysRevB.87.245112}%
  \BibitemOpen
  \bibfield  {author} {\bibinfo {author} {\bibfnamefont {T.}~\bibnamefont
  {Ojanen}},\ }\bibfield  {title} {\bibinfo {title} {{Helical Fermi arcs and
  surface states in time-reversal invariant Weyl semimetals}},\ }\href
  {https://doi.org/10.1103/PhysRevB.87.245112} {\bibfield  {journal} {\bibinfo
  {journal} {Phys. Rev. B}\ }\textbf {\bibinfo {volume} {87}},\ \bibinfo
  {pages} {245112} (\bibinfo {year} {2013})}\BibitemShut {NoStop}%
\bibitem [{\citenamefont {Okugawa}\ and\ \citenamefont
  {Murakami}(2014)}]{PhysRevB.89.235315}%
  \BibitemOpen
  \bibfield  {author} {\bibinfo {author} {\bibfnamefont {R.}~\bibnamefont
  {Okugawa}}\ and\ \bibinfo {author} {\bibfnamefont {S.}~\bibnamefont
  {Murakami}},\ }\bibfield  {title} {\bibinfo {title} {{Dispersion of Fermi
  arcs in Weyl semimetals and their evolutions to Dirac cones}},\ }\href
  {https://doi.org/10.1103/PhysRevB.89.235315} {\bibfield  {journal} {\bibinfo
  {journal} {Phys. Rev. B}\ }\textbf {\bibinfo {volume} {89}},\ \bibinfo
  {pages} {235315} (\bibinfo {year} {2014})}\BibitemShut {NoStop}%
\bibitem [{\citenamefont {Ezawa}(2018{\natexlab{a}})}]{PhysRevLett.120.026801}%
  \BibitemOpen
  \bibfield  {author} {\bibinfo {author} {\bibfnamefont {M.}~\bibnamefont
  {Ezawa}},\ }\bibfield  {title} {\bibinfo {title} {{Higher-Order Topological
  Insulators and Semimetals on the Breathing Kagome and Pyrochlore Lattices}},\
  }\href {https://doi.org/10.1103/PhysRevLett.120.026801} {\bibfield  {journal}
  {\bibinfo  {journal} {Phys. Rev. Lett.}\ }\textbf {\bibinfo {volume} {120}},\
  \bibinfo {pages} {026801} (\bibinfo {year} {2018}{\natexlab{a}})}\BibitemShut
  {NoStop}%
\bibitem [{\citenamefont {Lin}\ and\ \citenamefont
  {Hughes}(2018)}]{PhysRevB.98.241103}%
  \BibitemOpen
  \bibfield  {author} {\bibinfo {author} {\bibfnamefont {M.}~\bibnamefont
  {Lin}}\ and\ \bibinfo {author} {\bibfnamefont {T.~L.}\ \bibnamefont
  {Hughes}},\ }\bibfield  {title} {\bibinfo {title} {Topological quadrupolar
  semimetals},\ }\href {https://doi.org/10.1103/PhysRevB.98.241103} {\bibfield
  {journal} {\bibinfo  {journal} {Phys. Rev. B}\ }\textbf {\bibinfo {volume}
  {98}},\ \bibinfo {pages} {241103(R)} (\bibinfo {year} {2018})}\BibitemShut
  {NoStop}%
\bibitem [{\citenamefont {C\ifmmode \u{a}\else \u{a}\fi{}lug\ifmmode~\u{a}\else
  \u{a}\fi{}ru}\ \emph {et~al.}(2019)\citenamefont {C\ifmmode \u{a}\else
  \u{a}\fi{}lug\ifmmode~\u{a}\else \u{a}\fi{}ru}, \citenamefont {Juri\ifmmode
  \check{c}\else \v{c}\fi{}i\ifmmode~\acute{c}\else \'{c}\fi{}},\ and\
  \citenamefont {Roy}}]{PhysRevB.99.041301}%
  \BibitemOpen
  \bibfield  {author} {\bibinfo {author} {\bibfnamefont {D.}~\bibnamefont
  {C\ifmmode \u{a}\else \u{a}\fi{}lug\ifmmode~\u{a}\else \u{a}\fi{}ru}},
  \bibinfo {author} {\bibfnamefont {V.}~\bibnamefont {Juri\ifmmode
  \check{c}\else \v{c}\fi{}i\ifmmode~\acute{c}\else \'{c}\fi{}}},\ and\
  \bibinfo {author} {\bibfnamefont {B.}~\bibnamefont {Roy}},\ }\bibfield
  {title} {\bibinfo {title} {{Higher-order topological phases: A general
  principle of construction}},\ }\href
  {https://doi.org/10.1103/PhysRevB.99.041301} {\bibfield  {journal} {\bibinfo
  {journal} {Phys. Rev. B}\ }\textbf {\bibinfo {volume} {99}},\ \bibinfo
  {pages} {041301(R)} (\bibinfo {year} {2019})}\BibitemShut {NoStop}%
\bibitem [{\citenamefont {Wieder}\ \emph {et~al.}(2020)\citenamefont {Wieder},
  \citenamefont {Wang}, \citenamefont {Cano}, \citenamefont {Dai},
  \citenamefont {Schoop}, \citenamefont {Bradlyn},\ and\ \citenamefont
  {Bernevig}}]{wieder2020strong}%
  \BibitemOpen
  \bibfield  {author} {\bibinfo {author} {\bibfnamefont {B.~J.}\ \bibnamefont
  {Wieder}}, \bibinfo {author} {\bibfnamefont {Z.}~\bibnamefont {Wang}},
  \bibinfo {author} {\bibfnamefont {J.}~\bibnamefont {Cano}}, \bibinfo {author}
  {\bibfnamefont {X.}~\bibnamefont {Dai}}, \bibinfo {author} {\bibfnamefont
  {L.~M.}\ \bibnamefont {Schoop}}, \bibinfo {author} {\bibfnamefont
  {B.}~\bibnamefont {Bradlyn}},\ and\ \bibinfo {author} {\bibfnamefont {B.~A.}\
  \bibnamefont {Bernevig}},\ }\bibfield  {title} {\bibinfo {title} {{Strong and
  fragile topological Dirac semimetals with higher-order Fermi arcs}},\ }\href
  {https://www.nature.com/articles/s41467-020-14443-5} {\bibfield  {journal}
  {\bibinfo  {journal} {Nat. Commun.}\ }\textbf {\bibinfo {volume} {11}},\
  \bibinfo {pages} {1} (\bibinfo {year} {2020})}\BibitemShut {NoStop}%
\bibitem [{\citenamefont {Li}\ \emph {et~al.}(2020)\citenamefont {Li},
  \citenamefont {Wang}, \citenamefont {Li}, \citenamefont {Zheng},
  \citenamefont {Brinkman}, \citenamefont {Yu},\ and\ \citenamefont
  {Liao}}]{PhysRevLett.124.156601}%
  \BibitemOpen
  \bibfield  {author} {\bibinfo {author} {\bibfnamefont {C.-Z.}\ \bibnamefont
  {Li}}, \bibinfo {author} {\bibfnamefont {A.-Q.}\ \bibnamefont {Wang}},
  \bibinfo {author} {\bibfnamefont {C.}~\bibnamefont {Li}}, \bibinfo {author}
  {\bibfnamefont {W.-Z.}\ \bibnamefont {Zheng}}, \bibinfo {author}
  {\bibfnamefont {A.}~\bibnamefont {Brinkman}}, \bibinfo {author}
  {\bibfnamefont {D.-P.}\ \bibnamefont {Yu}},\ and\ \bibinfo {author}
  {\bibfnamefont {Z.-M.}\ \bibnamefont {Liao}},\ }\bibfield  {title} {\bibinfo
  {title} {{Reducing Electronic Transport Dimension to Topological Hinge States
  by Increasing Geometry Size of Dirac Semimetal Josephson Junctions}},\ }\href
  {https://doi.org/10.1103/PhysRevLett.124.156601} {\bibfield  {journal}
  {\bibinfo  {journal} {Phys. Rev. Lett.}\ }\textbf {\bibinfo {volume} {124}},\
  \bibinfo {pages} {156601} (\bibinfo {year} {2020})}\BibitemShut {NoStop}%
\bibitem [{\citenamefont {Zeng}\ \emph {et~al.}(2020)\citenamefont {Zeng},
  \citenamefont {Yang},\ and\ \citenamefont {Xu}}]{PhysRevB.101.241104}%
  \BibitemOpen
  \bibfield  {author} {\bibinfo {author} {\bibfnamefont {Q.-B.}\ \bibnamefont
  {Zeng}}, \bibinfo {author} {\bibfnamefont {Y.-B.}\ \bibnamefont {Yang}},\
  and\ \bibinfo {author} {\bibfnamefont {Y.}~\bibnamefont {Xu}},\ }\bibfield
  {title} {\bibinfo {title} {{Higher-order topological insulators and
  semimetals in generalized Aubry-Andr\'e-Harper models}},\ }\href
  {https://doi.org/10.1103/PhysRevB.101.241104} {\bibfield  {journal} {\bibinfo
   {journal} {Phys. Rev. B}\ }\textbf {\bibinfo {volume} {101}},\ \bibinfo
  {pages} {241104(R)} (\bibinfo {year} {2020})}\BibitemShut {NoStop}%
\bibitem [{\citenamefont {Szab\'o}\ and\ \citenamefont
  {Roy}(2020)}]{PhysRevResearch.2.043197}%
  \BibitemOpen
  \bibfield  {author} {\bibinfo {author} {\bibfnamefont {A.~L.}\ \bibnamefont
  {Szab\'o}}\ and\ \bibinfo {author} {\bibfnamefont {B.}~\bibnamefont {Roy}},\
  }\bibfield  {title} {\bibinfo {title} {Dirty higher-order dirac semimetal:
  Quantum criticality and bulk-boundary correspondence},\ }\href
  {https://doi.org/10.1103/PhysRevResearch.2.043197} {\bibfield  {journal}
  {\bibinfo  {journal} {Phys. Rev. Research}\ }\textbf {\bibinfo {volume}
  {2}},\ \bibinfo {pages} {043197} (\bibinfo {year} {2020})}\BibitemShut
  {NoStop}%
\bibitem [{\citenamefont {Roy}(2019)}]{PhysRevResearch.1.032048}%
  \BibitemOpen
  \bibfield  {author} {\bibinfo {author} {\bibfnamefont {B.}~\bibnamefont
  {Roy}},\ }\bibfield  {title} {\bibinfo {title} {{Antiunitary symmetry
  protected higher-order topological phases}},\ }\href
  {https://doi.org/10.1103/PhysRevResearch.1.032048} {\bibfield  {journal}
  {\bibinfo  {journal} {Phys. Rev. Research}\ }\textbf {\bibinfo {volume}
  {1}},\ \bibinfo {pages} {032048} (\bibinfo {year} {2019})}\BibitemShut
  {NoStop}%
\bibitem [{\citenamefont {Ghorashi}\ \emph {et~al.}(2020)\citenamefont
  {Ghorashi}, \citenamefont {Li},\ and\ \citenamefont
  {Hughes}}]{PhysRevLett.125.266804}%
  \BibitemOpen
  \bibfield  {author} {\bibinfo {author} {\bibfnamefont {S.~A.~A.}\
  \bibnamefont {Ghorashi}}, \bibinfo {author} {\bibfnamefont {T.}~\bibnamefont
  {Li}},\ and\ \bibinfo {author} {\bibfnamefont {T.~L.}\ \bibnamefont
  {Hughes}},\ }\bibfield  {title} {\bibinfo {title} {{Higher-Order Weyl
  Semimetals}},\ }\href {https://doi.org/10.1103/PhysRevLett.125.266804}
  {\bibfield  {journal} {\bibinfo  {journal} {Phys. Rev. Lett.}\ }\textbf
  {\bibinfo {volume} {125}},\ \bibinfo {pages} {266804} (\bibinfo {year}
  {2020})}\BibitemShut {NoStop}%
\bibitem [{\citenamefont {Wang}\ \emph {et~al.}(2020)\citenamefont {Wang},
  \citenamefont {Lin}, \citenamefont {Jiang}, \citenamefont {Guo},\ and\
  \citenamefont {Jiang}}]{PhysRevLett.125.146401}%
  \BibitemOpen
  \bibfield  {author} {\bibinfo {author} {\bibfnamefont {H.-X.}\ \bibnamefont
  {Wang}}, \bibinfo {author} {\bibfnamefont {Z.-K.}\ \bibnamefont {Lin}},
  \bibinfo {author} {\bibfnamefont {B.}~\bibnamefont {Jiang}}, \bibinfo
  {author} {\bibfnamefont {G.-Y.}\ \bibnamefont {Guo}},\ and\ \bibinfo {author}
  {\bibfnamefont {J.-H.}\ \bibnamefont {Jiang}},\ }\bibfield  {title} {\bibinfo
  {title} {{Higher-Order Weyl Semimetals}},\ }\href
  {https://doi.org/10.1103/PhysRevLett.125.146401} {\bibfield  {journal}
  {\bibinfo  {journal} {Phys. Rev. Lett.}\ }\textbf {\bibinfo {volume} {125}},\
  \bibinfo {pages} {146401} (\bibinfo {year} {2020})}\BibitemShut {NoStop}%
\bibitem [{\citenamefont {Fleury}(2021)}]{fleury2021sound}%
  \BibitemOpen
  \bibfield  {author} {\bibinfo {author} {\bibfnamefont {R.}~\bibnamefont
  {Fleury}},\ }\bibfield  {title} {\bibinfo {title} {The sound of weyl
  hinges},\ }\href {https://www.nature.com/articles/s41563-021-01018-y}
  {\bibfield  {journal} {\bibinfo  {journal} {Nat. Mater.}\ }\textbf {\bibinfo
  {volume} {20}},\ \bibinfo {pages} {716} (\bibinfo {year} {2021})}\BibitemShut
  {NoStop}%
\bibitem [{\citenamefont {Luo}\ \emph {et~al.}(2021)\citenamefont {Luo},
  \citenamefont {Wang}, \citenamefont {Lin}, \citenamefont {Jiang},
  \citenamefont {Wu}, \citenamefont {Li},\ and\ \citenamefont
  {Jiang}}]{luo2021observation}%
  \BibitemOpen
  \bibfield  {author} {\bibinfo {author} {\bibfnamefont {L.}~\bibnamefont
  {Luo}}, \bibinfo {author} {\bibfnamefont {H.-X.}\ \bibnamefont {Wang}},
  \bibinfo {author} {\bibfnamefont {Z.-K.}\ \bibnamefont {Lin}}, \bibinfo
  {author} {\bibfnamefont {B.}~\bibnamefont {Jiang}}, \bibinfo {author}
  {\bibfnamefont {Y.}~\bibnamefont {Wu}}, \bibinfo {author} {\bibfnamefont
  {F.}~\bibnamefont {Li}},\ and\ \bibinfo {author} {\bibfnamefont {J.-H.}\
  \bibnamefont {Jiang}},\ }\bibfield  {title} {\bibinfo {title} {Observation of
  a phononic higher-order weyl semimetal},\ }\href
  {https://www.nature.com/articles/s41563-021-00985-6} {\bibfield  {journal}
  {\bibinfo  {journal} {Nat. Mater.}\ }\textbf {\bibinfo {volume} {20}},\
  \bibinfo {pages} {794} (\bibinfo {year} {2021})}\BibitemShut {NoStop}%
\bibitem [{\citenamefont {Wei}\ \emph {et~al.}(2021)\citenamefont {Wei},
  \citenamefont {Zhang}, \citenamefont {Deng}, \citenamefont {Lu},
  \citenamefont {Huang}, \citenamefont {Yan}, \citenamefont {Chen},
  \citenamefont {Liu},\ and\ \citenamefont {Jia}}]{wei2021higher}%
  \BibitemOpen
  \bibfield  {author} {\bibinfo {author} {\bibfnamefont {Q.}~\bibnamefont
  {Wei}}, \bibinfo {author} {\bibfnamefont {X.}~\bibnamefont {Zhang}}, \bibinfo
  {author} {\bibfnamefont {W.}~\bibnamefont {Deng}}, \bibinfo {author}
  {\bibfnamefont {J.}~\bibnamefont {Lu}}, \bibinfo {author} {\bibfnamefont
  {X.}~\bibnamefont {Huang}}, \bibinfo {author} {\bibfnamefont
  {M.}~\bibnamefont {Yan}}, \bibinfo {author} {\bibfnamefont {G.}~\bibnamefont
  {Chen}}, \bibinfo {author} {\bibfnamefont {Z.}~\bibnamefont {Liu}},\ and\
  \bibinfo {author} {\bibfnamefont {S.}~\bibnamefont {Jia}},\ }\bibfield
  {title} {\bibinfo {title} {Higher-order topological semimetal in acoustic
  crystals},\ }\href {https://www.nature.com/articles/s41563-021-00933-4}
  {\bibfield  {journal} {\bibinfo  {journal} {Nature Materials}\ }\textbf
  {\bibinfo {volume} {20}},\ \bibinfo {pages} {812} (\bibinfo {year}
  {2021})}\BibitemShut {NoStop}%
\bibitem [{\citenamefont {Zhang}\ \emph {et~al.}(2021)\citenamefont {Zhang},
  \citenamefont {Wu}, \citenamefont {Chen},\ and\ \citenamefont
  {Jiang}}]{PhysRevB.104.014203}%
  \BibitemOpen
  \bibfield  {author} {\bibinfo {author} {\bibfnamefont {Z.-Q.}\ \bibnamefont
  {Zhang}}, \bibinfo {author} {\bibfnamefont {B.-L.}\ \bibnamefont {Wu}},
  \bibinfo {author} {\bibfnamefont {C.-Z.}\ \bibnamefont {Chen}},\ and\
  \bibinfo {author} {\bibfnamefont {H.}~\bibnamefont {Jiang}},\ }\bibfield
  {title} {\bibinfo {title} {{Global phase diagram of disordered higher-order
  Weyl semimetals}},\ }\href {https://doi.org/10.1103/PhysRevB.104.014203}
  {\bibfield  {journal} {\bibinfo  {journal} {Phys. Rev. B}\ }\textbf {\bibinfo
  {volume} {104}},\ \bibinfo {pages} {014203} (\bibinfo {year}
  {2021})}\BibitemShut {NoStop}%
\bibitem [{\citenamefont {Rui}\ \emph {et~al.}(2021)\citenamefont {Rui},
  \citenamefont {Zhang}, \citenamefont {Hirschmann}, \citenamefont {Zheng},
  \citenamefont {Schnyder}, \citenamefont {Trauzettel},\ and\ \citenamefont
  {Wang}}]{PhysRevB.103.184510}%
  \BibitemOpen
  \bibfield  {author} {\bibinfo {author} {\bibfnamefont {W.~B.}\ \bibnamefont
  {Rui}}, \bibinfo {author} {\bibfnamefont {S.-B.}\ \bibnamefont {Zhang}},
  \bibinfo {author} {\bibfnamefont {M.~M.}\ \bibnamefont {Hirschmann}},
  \bibinfo {author} {\bibfnamefont {Z.}~\bibnamefont {Zheng}}, \bibinfo
  {author} {\bibfnamefont {A.~P.}\ \bibnamefont {Schnyder}}, \bibinfo {author}
  {\bibfnamefont {B.}~\bibnamefont {Trauzettel}},\ and\ \bibinfo {author}
  {\bibfnamefont {Z.~D.}\ \bibnamefont {Wang}},\ }\bibfield  {title} {\bibinfo
  {title} {{Higher-order Weyl superconductors with anisotropic Weyl-point
  connectivity}},\ }\href {https://doi.org/10.1103/PhysRevB.103.184510}
  {\bibfield  {journal} {\bibinfo  {journal} {Phys. Rev. B}\ }\textbf {\bibinfo
  {volume} {103}},\ \bibinfo {pages} {184510} (\bibinfo {year}
  {2021})}\BibitemShut {NoStop}%
\bibitem [{\citenamefont {Geier}\ \emph {et~al.}(2018)\citenamefont {Geier},
  \citenamefont {Trifunovic}, \citenamefont {Hoskam},\ and\ \citenamefont
  {Brouwer}}]{PhysRevB.97.205135}%
  \BibitemOpen
  \bibfield  {author} {\bibinfo {author} {\bibfnamefont {M.}~\bibnamefont
  {Geier}}, \bibinfo {author} {\bibfnamefont {L.}~\bibnamefont {Trifunovic}},
  \bibinfo {author} {\bibfnamefont {M.}~\bibnamefont {Hoskam}},\ and\ \bibinfo
  {author} {\bibfnamefont {P.~W.}\ \bibnamefont {Brouwer}},\ }\bibfield
  {title} {\bibinfo {title} {{Second-order topological insulators and
  superconductors with an order-two crystalline symmetry}},\ }\href
  {https://doi.org/10.1103/PhysRevB.97.205135} {\bibfield  {journal} {\bibinfo
  {journal} {Phys. Rev. B}\ }\textbf {\bibinfo {volume} {97}},\ \bibinfo
  {pages} {205135} (\bibinfo {year} {2018})}\BibitemShut {NoStop}%
\bibitem [{\citenamefont {Kunst}\ \emph {et~al.}(2018)\citenamefont {Kunst},
  \citenamefont {van Miert},\ and\ \citenamefont
  {Bergholtz}}]{PhysRevB.97.241405}%
  \BibitemOpen
  \bibfield  {author} {\bibinfo {author} {\bibfnamefont {F.~K.}\ \bibnamefont
  {Kunst}}, \bibinfo {author} {\bibfnamefont {G.}~\bibnamefont {van Miert}},\
  and\ \bibinfo {author} {\bibfnamefont {E.~J.}\ \bibnamefont {Bergholtz}},\
  }\bibfield  {title} {\bibinfo {title} {{Lattice models with exactly solvable
  topological hinge and corner states}},\ }\href
  {https://doi.org/10.1103/PhysRevB.97.241405} {\bibfield  {journal} {\bibinfo
  {journal} {Phys. Rev. B}\ }\textbf {\bibinfo {volume} {97}},\ \bibinfo
  {pages} {241405(R)} (\bibinfo {year} {2018})}\BibitemShut {NoStop}%
\bibitem [{\citenamefont {Schindler}\ \emph
  {et~al.}(2018{\natexlab{b}})\citenamefont {Schindler}, \citenamefont {Wang},
  \citenamefont {Vergniory}, \citenamefont {Cook}, \citenamefont {Murani},
  \citenamefont {Sengupta}, \citenamefont {Kasumov}, \citenamefont {Deblock},
  \citenamefont {Jeon}, \citenamefont {Drozdov}, \citenamefont {Bouchiat},
  \citenamefont {Gu\'eron}, \citenamefont {Yazdani}, \citenamefont {Bernevig},\
  and\ \citenamefont {Neupert}}]{schindler2018higherbismuth}%
  \BibitemOpen
  \bibfield  {author} {\bibinfo {author} {\bibfnamefont {F.}~\bibnamefont
  {Schindler}}, \bibinfo {author} {\bibfnamefont {Z.}~\bibnamefont {Wang}},
  \bibinfo {author} {\bibfnamefont {M.~G.}\ \bibnamefont {Vergniory}}, \bibinfo
  {author} {\bibfnamefont {A.~M.}\ \bibnamefont {Cook}}, \bibinfo {author}
  {\bibfnamefont {A.}~\bibnamefont {Murani}}, \bibinfo {author} {\bibfnamefont
  {S.}~\bibnamefont {Sengupta}}, \bibinfo {author} {\bibfnamefont {A.~Y.}\
  \bibnamefont {Kasumov}}, \bibinfo {author} {\bibfnamefont {R.}~\bibnamefont
  {Deblock}}, \bibinfo {author} {\bibfnamefont {S.}~\bibnamefont {Jeon}},
  \bibinfo {author} {\bibfnamefont {I.}~\bibnamefont {Drozdov}}, \bibinfo
  {author} {\bibfnamefont {H.}~\bibnamefont {Bouchiat}}, \bibinfo {author}
  {\bibfnamefont {S.}~\bibnamefont {Gu\'eron}}, \bibinfo {author}
  {\bibfnamefont {A.}~\bibnamefont {Yazdani}}, \bibinfo {author} {\bibfnamefont
  {B.~A.}\ \bibnamefont {Bernevig}},\ and\ \bibinfo {author} {\bibfnamefont
  {T.}~\bibnamefont {Neupert}},\ }\bibfield  {title} {\bibinfo {title}
  {{Higher-order topology in bismuth}},\ }\href
  {https://doi.org/10.1038/s41567-018-0224-7} {\bibfield  {journal} {\bibinfo
  {journal} {Nat. Phys.}\ }\textbf {\bibinfo {volume} {14}},\ \bibinfo {pages}
  {918} (\bibinfo {year} {2018}{\natexlab{b}})}\BibitemShut {NoStop}%
\bibitem [{\citenamefont {Xie}\ \emph {et~al.}(2018)\citenamefont {Xie},
  \citenamefont {Wang}, \citenamefont {Wang}, \citenamefont {Zhu},
  \citenamefont {Jiang}, \citenamefont {Lu},\ and\ \citenamefont
  {Chen}}]{PhysRevB.98.205147}%
  \BibitemOpen
  \bibfield  {author} {\bibinfo {author} {\bibfnamefont {B.-Y.}\ \bibnamefont
  {Xie}}, \bibinfo {author} {\bibfnamefont {H.-F.}\ \bibnamefont {Wang}},
  \bibinfo {author} {\bibfnamefont {H.-X.}\ \bibnamefont {Wang}}, \bibinfo
  {author} {\bibfnamefont {X.-Y.}\ \bibnamefont {Zhu}}, \bibinfo {author}
  {\bibfnamefont {J.-H.}\ \bibnamefont {Jiang}}, \bibinfo {author}
  {\bibfnamefont {M.-H.}\ \bibnamefont {Lu}},\ and\ \bibinfo {author}
  {\bibfnamefont {Y.-F.}\ \bibnamefont {Chen}},\ }\bibfield  {title} {\bibinfo
  {title} {{Second-order photonic topological insulator with corner states}},\
  }\href {https://doi.org/10.1103/PhysRevB.98.205147} {\bibfield  {journal}
  {\bibinfo  {journal} {Phys. Rev. B}\ }\textbf {\bibinfo {volume} {98}},\
  \bibinfo {pages} {205147} (\bibinfo {year} {2018})}\BibitemShut {NoStop}%
\bibitem [{\citenamefont {Serra-Garcia}\ \emph {et~al.}(2018)\citenamefont
  {Serra-Garcia}, \citenamefont {Peri}, \citenamefont {S{\"u}sstrunk},
  \citenamefont {Bilal}, \citenamefont {Larsen}, \citenamefont {Villanueva},\
  and\ \citenamefont {Huber}}]{serra2018observationnature555}%
  \BibitemOpen
  \bibfield  {author} {\bibinfo {author} {\bibfnamefont {M.}~\bibnamefont
  {Serra-Garcia}}, \bibinfo {author} {\bibfnamefont {V.}~\bibnamefont {Peri}},
  \bibinfo {author} {\bibfnamefont {R.}~\bibnamefont {S{\"u}sstrunk}}, \bibinfo
  {author} {\bibfnamefont {O.~R.}\ \bibnamefont {Bilal}}, \bibinfo {author}
  {\bibfnamefont {T.}~\bibnamefont {Larsen}}, \bibinfo {author} {\bibfnamefont
  {L.~G.}\ \bibnamefont {Villanueva}},\ and\ \bibinfo {author} {\bibfnamefont
  {S.~D.}\ \bibnamefont {Huber}},\ }\bibfield  {title} {\bibinfo {title}
  {{Observation of a phononic quadrupole topological insulator}},\ }\href
  {https://doi.org/10.1038/nature25156} {\bibfield  {journal} {\bibinfo
  {journal} {Nature}\ }\textbf {\bibinfo {volume} {555}},\ \bibinfo {pages}
  {342} (\bibinfo {year} {2018})}\BibitemShut {NoStop}%
\bibitem [{\citenamefont {Peterson}\ \emph {et~al.}(2018)\citenamefont
  {Peterson}, \citenamefont {Benalcazar}, \citenamefont {Hughes},\ and\
  \citenamefont {Bahl}}]{peterson2018quantizedNature7695}%
  \BibitemOpen
  \bibfield  {author} {\bibinfo {author} {\bibfnamefont {C.~W.}\ \bibnamefont
  {Peterson}}, \bibinfo {author} {\bibfnamefont {W.~A.}\ \bibnamefont
  {Benalcazar}}, \bibinfo {author} {\bibfnamefont {T.~L.}\ \bibnamefont
  {Hughes}},\ and\ \bibinfo {author} {\bibfnamefont {G.}~\bibnamefont {Bahl}},\
  }\bibfield  {title} {\bibinfo {title} {{A quantized microwave quadrupole
  insulator with topologically protected corner states}},\ }\href
  {https://doi.org/10.1038/nature25777} {\bibfield  {journal} {\bibinfo
  {journal} {Nature}\ }\textbf {\bibinfo {volume} {555}},\ \bibinfo {pages}
  {346} (\bibinfo {year} {2018})}\BibitemShut {NoStop}%
\bibitem [{\citenamefont {Imhof}\ \emph {et~al.}(2018)\citenamefont {Imhof},
  \citenamefont {Berger}, \citenamefont {Bayer}, \citenamefont {Brehm},
  \citenamefont {Molenkamp}, \citenamefont {Kiessling}, \citenamefont
  {Schindler}, \citenamefont {Lee}, \citenamefont {Greiter}, \citenamefont
  {Neupert},\ and\ \citenamefont {Thomale}}]{imhof2018topolectricalnatphys}%
  \BibitemOpen
  \bibfield  {author} {\bibinfo {author} {\bibfnamefont {S.}~\bibnamefont
  {Imhof}}, \bibinfo {author} {\bibfnamefont {C.}~\bibnamefont {Berger}},
  \bibinfo {author} {\bibfnamefont {F.}~\bibnamefont {Bayer}}, \bibinfo
  {author} {\bibfnamefont {J.}~\bibnamefont {Brehm}}, \bibinfo {author}
  {\bibfnamefont {L.~W.}\ \bibnamefont {Molenkamp}}, \bibinfo {author}
  {\bibfnamefont {T.}~\bibnamefont {Kiessling}}, \bibinfo {author}
  {\bibfnamefont {F.}~\bibnamefont {Schindler}}, \bibinfo {author}
  {\bibfnamefont {C.~H.}\ \bibnamefont {Lee}}, \bibinfo {author} {\bibfnamefont
  {M.}~\bibnamefont {Greiter}}, \bibinfo {author} {\bibfnamefont
  {T.}~\bibnamefont {Neupert}},\ and\ \bibinfo {author} {\bibfnamefont
  {R.}~\bibnamefont {Thomale}},\ }\bibfield  {title} {\bibinfo {title}
  {{Topolectrical-circuit realization of topological corner modes}},\ }\href
  {https://doi.org/10.1038/s41567-018-0246-1} {\bibfield  {journal} {\bibinfo
  {journal} {Nat. Phys.}\ }\textbf {\bibinfo {volume} {14}},\ \bibinfo {pages}
  {925} (\bibinfo {year} {2018})}\BibitemShut {NoStop}%
\bibitem [{\citenamefont {Peng}\ and\ \citenamefont
  {Refael}(2019)}]{PhysRevLett.123.016806}%
  \BibitemOpen
  \bibfield  {author} {\bibinfo {author} {\bibfnamefont {Y.}~\bibnamefont
  {Peng}}\ and\ \bibinfo {author} {\bibfnamefont {G.}~\bibnamefont {Refael}},\
  }\bibfield  {title} {\bibinfo {title} {Floquet second-order topological
  insulators from nonsymmorphic space-time symmetries},\ }\href
  {https://doi.org/10.1103/PhysRevLett.123.016806} {\bibfield  {journal}
  {\bibinfo  {journal} {Phys. Rev. Lett.}\ }\textbf {\bibinfo {volume} {123}},\
  \bibinfo {pages} {016806} (\bibinfo {year} {2019})}\BibitemShut {NoStop}%
\bibitem [{\citenamefont {Wang}\ \emph {et~al.}(2019)\citenamefont {Wang},
  \citenamefont {Wieder}, \citenamefont {Li}, \citenamefont {Yan},\ and\
  \citenamefont {Bernevig}}]{wang2018higher}%
  \BibitemOpen
  \bibfield  {author} {\bibinfo {author} {\bibfnamefont {Z.}~\bibnamefont
  {Wang}}, \bibinfo {author} {\bibfnamefont {B.~J.}\ \bibnamefont {Wieder}},
  \bibinfo {author} {\bibfnamefont {J.}~\bibnamefont {Li}}, \bibinfo {author}
  {\bibfnamefont {B.}~\bibnamefont {Yan}},\ and\ \bibinfo {author}
  {\bibfnamefont {B.~A.}\ \bibnamefont {Bernevig}},\ }\bibfield  {title}
  {\bibinfo {title} {{Higher-Order Topology, Monopole Nodal Lines, and the
  Origin of Large Fermi Arcs in Transition Metal Dichalcogenides
  $X{\mathrm{Te}}_{2}$ ($X=\mathrm{Mo},\mathrm{W}$)}},\ }\href
  {https://doi.org/10.1103/PhysRevLett.123.186401} {\bibfield  {journal}
  {\bibinfo  {journal} {Phys. Rev. Lett.}\ }\textbf {\bibinfo {volume} {123}},\
  \bibinfo {pages} {186401} (\bibinfo {year} {2019})}\BibitemShut {NoStop}%
\bibitem [{\citenamefont {Sheng}\ \emph {et~al.}(2019)\citenamefont {Sheng},
  \citenamefont {Chen}, \citenamefont {Liu}, \citenamefont {Chen},
  \citenamefont {Yu}, \citenamefont {Zhao},\ and\ \citenamefont
  {Yang}}]{sheng2019two}%
  \BibitemOpen
  \bibfield  {author} {\bibinfo {author} {\bibfnamefont {X.-L.}\ \bibnamefont
  {Sheng}}, \bibinfo {author} {\bibfnamefont {C.}~\bibnamefont {Chen}},
  \bibinfo {author} {\bibfnamefont {H.}~\bibnamefont {Liu}}, \bibinfo {author}
  {\bibfnamefont {Z.}~\bibnamefont {Chen}}, \bibinfo {author} {\bibfnamefont
  {Z.-M.}\ \bibnamefont {Yu}}, \bibinfo {author} {\bibfnamefont {Y.~X.}\
  \bibnamefont {Zhao}},\ and\ \bibinfo {author} {\bibfnamefont {S.~A.}\
  \bibnamefont {Yang}},\ }\bibfield  {title} {\bibinfo {title} {Two-dimensional
  second-order topological insulator in graphdiyne},\ }\href
  {https://doi.org/10.1103/PhysRevLett.123.256402} {\bibfield  {journal}
  {\bibinfo  {journal} {Phys. Rev. Lett.}\ }\textbf {\bibinfo {volume} {123}},\
  \bibinfo {pages} {256402} (\bibinfo {year} {2019})}\BibitemShut {NoStop}%
\bibitem [{\citenamefont {Agarwala}\ \emph {et~al.}(2020)\citenamefont
  {Agarwala}, \citenamefont {Juri\ifmmode \check{c}\else
  \v{c}\fi{}i\ifmmode~\acute{c}\else \'{c}\fi{}},\ and\ \citenamefont
  {Roy}}]{agarwala2019higher}%
  \BibitemOpen
  \bibfield  {author} {\bibinfo {author} {\bibfnamefont {A.}~\bibnamefont
  {Agarwala}}, \bibinfo {author} {\bibfnamefont {V.}~\bibnamefont {Juri\ifmmode
  \check{c}\else \v{c}\fi{}i\ifmmode~\acute{c}\else \'{c}\fi{}}},\ and\
  \bibinfo {author} {\bibfnamefont {B.}~\bibnamefont {Roy}},\ }\bibfield
  {title} {\bibinfo {title} {Higher-order topological insulators in amorphous
  solids},\ }\href {https://doi.org/10.1103/PhysRevResearch.2.012067}
  {\bibfield  {journal} {\bibinfo  {journal} {Phys. Rev. Research}\ }\textbf
  {\bibinfo {volume} {2}},\ \bibinfo {pages} {012067(R)} (\bibinfo {year}
  {2020})}\BibitemShut {NoStop}%
\bibitem [{\citenamefont {Chen}\ \emph
  {et~al.}(2020{\natexlab{a}})\citenamefont {Chen}, \citenamefont {Chen},
  \citenamefont {Gao}, \citenamefont {Zhou},\ and\ \citenamefont
  {Xu}}]{chen2019higher}%
  \BibitemOpen
  \bibfield  {author} {\bibinfo {author} {\bibfnamefont {R.}~\bibnamefont
  {Chen}}, \bibinfo {author} {\bibfnamefont {C.-Z.}\ \bibnamefont {Chen}},
  \bibinfo {author} {\bibfnamefont {J.-H.}\ \bibnamefont {Gao}}, \bibinfo
  {author} {\bibfnamefont {B.}~\bibnamefont {Zhou}},\ and\ \bibinfo {author}
  {\bibfnamefont {D.-H.}\ \bibnamefont {Xu}},\ }\bibfield  {title} {\bibinfo
  {title} {Higher-order topological insulators in quasicrystals},\ }\href
  {https://doi.org/10.1103/PhysRevLett.124.036803} {\bibfield  {journal}
  {\bibinfo  {journal} {Phys. Rev. Lett.}\ }\textbf {\bibinfo {volume} {124}},\
  \bibinfo {pages} {036803} (\bibinfo {year} {2020}{\natexlab{a}})}\BibitemShut
  {NoStop}%
\bibitem [{\citenamefont {Fukui}\ and\ \citenamefont
  {Hatsugai}(2018)}]{PhysRevB.98.035147}%
  \BibitemOpen
  \bibfield  {author} {\bibinfo {author} {\bibfnamefont {T.}~\bibnamefont
  {Fukui}}\ and\ \bibinfo {author} {\bibfnamefont {Y.}~\bibnamefont
  {Hatsugai}},\ }\bibfield  {title} {\bibinfo {title} {Entanglement
  polarization for the topological quadrupole phase},\ }\href
  {https://doi.org/10.1103/PhysRevB.98.035147} {\bibfield  {journal} {\bibinfo
  {journal} {Phys. Rev. B}\ }\textbf {\bibinfo {volume} {98}},\ \bibinfo
  {pages} {035147} (\bibinfo {year} {2018})}\BibitemShut {NoStop}%
\bibitem [{\citenamefont {Okugawa}\ \emph {et~al.}(2019)\citenamefont
  {Okugawa}, \citenamefont {Hayashi},\ and\ \citenamefont
  {Nakanishi}}]{okugawa2019second}%
  \BibitemOpen
  \bibfield  {author} {\bibinfo {author} {\bibfnamefont {R.}~\bibnamefont
  {Okugawa}}, \bibinfo {author} {\bibfnamefont {S.}~\bibnamefont {Hayashi}},\
  and\ \bibinfo {author} {\bibfnamefont {T.}~\bibnamefont {Nakanishi}},\
  }\bibfield  {title} {\bibinfo {title} {{Second-order topological phases
  protected by chiral symmetry}},\ }\href
  {https://doi.org/10.1103/PhysRevB.100.235302} {\bibfield  {journal} {\bibinfo
   {journal} {Phys. Rev. B}\ }\textbf {\bibinfo {volume} {100}},\ \bibinfo
  {pages} {235302} (\bibinfo {year} {2019})}\BibitemShut {NoStop}%
\bibitem [{\citenamefont {Ghosh}\ \emph {et~al.}(2020)\citenamefont {Ghosh},
  \citenamefont {Paul},\ and\ \citenamefont {Saha}}]{ghosh2019engineering}%
  \BibitemOpen
  \bibfield  {author} {\bibinfo {author} {\bibfnamefont {A.~K.}\ \bibnamefont
  {Ghosh}}, \bibinfo {author} {\bibfnamefont {G.~C.}\ \bibnamefont {Paul}},\
  and\ \bibinfo {author} {\bibfnamefont {A.}~\bibnamefont {Saha}},\ }\bibfield
  {title} {\bibinfo {title} {{Higher order topological insulator via periodic
  driving}},\ }\href {https://doi.org/10.1103/PhysRevB.101.235403} {\bibfield
  {journal} {\bibinfo  {journal} {Phys. Rev. B}\ }\textbf {\bibinfo {volume}
  {101}},\ \bibinfo {pages} {235403} (\bibinfo {year} {2020})}\BibitemShut
  {NoStop}%
\bibitem [{\citenamefont {Hirayama}\ \emph {et~al.}(2020)\citenamefont
  {Hirayama}, \citenamefont {Takahashi}, \citenamefont {Matsuishi},
  \citenamefont {Hosono},\ and\ \citenamefont {Murakami}}]{hirayama2020higher}%
  \BibitemOpen
  \bibfield  {author} {\bibinfo {author} {\bibfnamefont {M.}~\bibnamefont
  {Hirayama}}, \bibinfo {author} {\bibfnamefont {R.}~\bibnamefont {Takahashi}},
  \bibinfo {author} {\bibfnamefont {S.}~\bibnamefont {Matsuishi}}, \bibinfo
  {author} {\bibfnamefont {H.}~\bibnamefont {Hosono}},\ and\ \bibinfo {author}
  {\bibfnamefont {S.}~\bibnamefont {Murakami}},\ }\bibfield  {title} {\bibinfo
  {title} {{Higher-order topological crystalline insulating phase and quantized
  hinge charge in topological electride apatite}},\ }\href
  {https://doi.org/10.1103/PhysRevResearch.2.043131} {\bibfield  {journal}
  {\bibinfo  {journal} {Phys. Rev. Research}\ }\textbf {\bibinfo {volume}
  {2}},\ \bibinfo {pages} {043131} (\bibinfo {year} {2020})}\BibitemShut
  {NoStop}%
\bibitem [{\citenamefont {Chen}\ \emph
  {et~al.}(2020{\natexlab{b}})\citenamefont {Chen}, \citenamefont {Song},
  \citenamefont {Zhao}, \citenamefont {Chen}, \citenamefont {Yu}, \citenamefont
  {Sheng},\ and\ \citenamefont {Yang}}]{chen2020universal}%
  \BibitemOpen
  \bibfield  {author} {\bibinfo {author} {\bibfnamefont {C.}~\bibnamefont
  {Chen}}, \bibinfo {author} {\bibfnamefont {Z.}~\bibnamefont {Song}}, \bibinfo
  {author} {\bibfnamefont {J.-Z.}\ \bibnamefont {Zhao}}, \bibinfo {author}
  {\bibfnamefont {Z.}~\bibnamefont {Chen}}, \bibinfo {author} {\bibfnamefont
  {Z.-M.}\ \bibnamefont {Yu}}, \bibinfo {author} {\bibfnamefont {X.-L.}\
  \bibnamefont {Sheng}},\ and\ \bibinfo {author} {\bibfnamefont {S.~A.}\
  \bibnamefont {Yang}},\ }\bibfield  {title} {\bibinfo {title} {{Universal
  Approach to Magnetic Second-Order Topological Insulator}},\ }\href
  {https://doi.org/10.1103/PhysRevLett.125.056402} {\bibfield  {journal}
  {\bibinfo  {journal} {Phys. Rev. Lett.}\ }\textbf {\bibinfo {volume} {125}},\
  \bibinfo {pages} {056402} (\bibinfo {year} {2020}{\natexlab{b}})}\BibitemShut
  {NoStop}%
\bibitem [{\citenamefont {Klitzing}\ \emph {et~al.}(1980)\citenamefont
  {Klitzing}, \citenamefont {Dorda},\ and\ \citenamefont
  {Pepper}}]{PhysRevLett.45.494}%
  \BibitemOpen
  \bibfield  {author} {\bibinfo {author} {\bibfnamefont {K.~v.}\ \bibnamefont
  {Klitzing}}, \bibinfo {author} {\bibfnamefont {G.}~\bibnamefont {Dorda}},\
  and\ \bibinfo {author} {\bibfnamefont {M.}~\bibnamefont {Pepper}},\
  }\bibfield  {title} {\bibinfo {title} {{New Method for High-Accuracy
  Determination of the Fine-Structure Constant Based on Quantized Hall
  Resistance}},\ }\href {https://doi.org/10.1103/PhysRevLett.45.494} {\bibfield
   {journal} {\bibinfo  {journal} {Phys. Rev. Lett.}\ }\textbf {\bibinfo
  {volume} {45}},\ \bibinfo {pages} {494} (\bibinfo {year} {1980})}\BibitemShut
  {NoStop}%
\bibitem [{\citenamefont {Kruthoff}\ \emph {et~al.}(2017)\citenamefont
  {Kruthoff}, \citenamefont {de~Boer}, \citenamefont {van Wezel}, \citenamefont
  {Kane},\ and\ \citenamefont {Slager}}]{PhysRevX.7.041069}%
  \BibitemOpen
  \bibfield  {author} {\bibinfo {author} {\bibfnamefont {J.}~\bibnamefont
  {Kruthoff}}, \bibinfo {author} {\bibfnamefont {J.}~\bibnamefont {de~Boer}},
  \bibinfo {author} {\bibfnamefont {J.}~\bibnamefont {van Wezel}}, \bibinfo
  {author} {\bibfnamefont {C.~L.}\ \bibnamefont {Kane}},\ and\ \bibinfo
  {author} {\bibfnamefont {R.-J.}\ \bibnamefont {Slager}},\ }\bibfield  {title}
  {\bibinfo {title} {{Topological Classification of Crystalline Insulators
  through Band Structure Combinatorics}},\ }\href
  {https://doi.org/10.1103/PhysRevX.7.041069} {\bibfield  {journal} {\bibinfo
  {journal} {Phys. Rev. X}\ }\textbf {\bibinfo {volume} {7}},\ \bibinfo {pages}
  {041069} (\bibinfo {year} {2017})}\BibitemShut {NoStop}%
\bibitem [{\citenamefont {Po}\ \emph {et~al.}(2017)\citenamefont {Po},
  \citenamefont {Vishwanath},\ and\ \citenamefont {Watanabe}}]{po2017symmetry}%
  \BibitemOpen
  \bibfield  {author} {\bibinfo {author} {\bibfnamefont {H.~C.}\ \bibnamefont
  {Po}}, \bibinfo {author} {\bibfnamefont {A.}~\bibnamefont {Vishwanath}},\
  and\ \bibinfo {author} {\bibfnamefont {H.}~\bibnamefont {Watanabe}},\
  }\bibfield  {title} {\bibinfo {title} {{Symmetry-based indicators of band
  topology in the 230 space groups}},\ }\href
  {https://doi.org/10.1038/s41467-017-00133-2} {\bibfield  {journal} {\bibinfo
  {journal} {Nat. Commun.}\ }\textbf {\bibinfo {volume} {8}},\ \bibinfo {pages}
  {50} (\bibinfo {year} {2017})}\BibitemShut {NoStop}%
\bibitem [{\citenamefont {Bradlyn}\ \emph {et~al.}(2017)\citenamefont
  {Bradlyn}, \citenamefont {Elcoro}, \citenamefont {Cano}, \citenamefont
  {Vergniory}, \citenamefont {Wang}, \citenamefont {Felser}, \citenamefont
  {Aroyo},\ and\ \citenamefont {Bernevig}}]{bradlyn2017topological}%
  \BibitemOpen
  \bibfield  {author} {\bibinfo {author} {\bibfnamefont {B.}~\bibnamefont
  {Bradlyn}}, \bibinfo {author} {\bibfnamefont {L.}~\bibnamefont {Elcoro}},
  \bibinfo {author} {\bibfnamefont {J.}~\bibnamefont {Cano}}, \bibinfo {author}
  {\bibfnamefont {M.}~\bibnamefont {Vergniory}}, \bibinfo {author}
  {\bibfnamefont {Z.}~\bibnamefont {Wang}}, \bibinfo {author} {\bibfnamefont
  {C.}~\bibnamefont {Felser}}, \bibinfo {author} {\bibfnamefont {M.~I.}\
  \bibnamefont {Aroyo}},\ and\ \bibinfo {author} {\bibfnamefont {B.~A.}\
  \bibnamefont {Bernevig}},\ }\bibfield  {title} {\bibinfo {title} {Topological
  quantum chemistry},\ }\href {https://www.nature.com/articles/nature23268}
  {\bibfield  {journal} {\bibinfo  {journal} {Nature}\ }\textbf {\bibinfo
  {volume} {547}},\ \bibinfo {pages} {298} (\bibinfo {year}
  {2017})}\BibitemShut {NoStop}%
\bibitem [{\citenamefont {Watanabe}\ \emph {et~al.}(2018)\citenamefont
  {Watanabe}, \citenamefont {Po},\ and\ \citenamefont
  {Vishwanath}}]{watanabe2018structure}%
  \BibitemOpen
  \bibfield  {author} {\bibinfo {author} {\bibfnamefont {H.}~\bibnamefont
  {Watanabe}}, \bibinfo {author} {\bibfnamefont {H.~C.}\ \bibnamefont {Po}},\
  and\ \bibinfo {author} {\bibfnamefont {A.}~\bibnamefont {Vishwanath}},\
  }\bibfield  {title} {\bibinfo {title} {Structure and topology of band
  structures in the 1651 magnetic space groups},\ }\href
  {https://advances.sciencemag.org/content/4/8/eaat8685} {\bibfield  {journal}
  {\bibinfo  {journal} {Sci. Adv.}\ }\textbf {\bibinfo {volume} {4}},\ \bibinfo
  {pages} {eaat8685} (\bibinfo {year} {2018})}\BibitemShut {NoStop}%
\bibitem [{\citenamefont {Song}\ \emph {et~al.}(2018)\citenamefont {Song},
  \citenamefont {Zhang}, \citenamefont {Fang},\ and\ \citenamefont
  {Fang}}]{song2018quantitative}%
  \BibitemOpen
  \bibfield  {author} {\bibinfo {author} {\bibfnamefont {Z.}~\bibnamefont
  {Song}}, \bibinfo {author} {\bibfnamefont {T.}~\bibnamefont {Zhang}},
  \bibinfo {author} {\bibfnamefont {Z.}~\bibnamefont {Fang}},\ and\ \bibinfo
  {author} {\bibfnamefont {C.}~\bibnamefont {Fang}},\ }\bibfield  {title}
  {\bibinfo {title} {{Quantitative mappings between symmetry and topology in
  solids}},\ }\href {https://www.nature.com/articles/s41467-018-06010-w}
  {\bibfield  {journal} {\bibinfo  {journal} {Nat. Commun}\ }\textbf {\bibinfo
  {volume} {9}},\ \bibinfo {pages} {1} (\bibinfo {year} {2018})}\BibitemShut
  {NoStop}%
\bibitem [{\citenamefont {Ono}\ and\ \citenamefont
  {Watanabe}(2018)}]{PhysRevB.98.115150}%
  \BibitemOpen
  \bibfield  {author} {\bibinfo {author} {\bibfnamefont {S.}~\bibnamefont
  {Ono}}\ and\ \bibinfo {author} {\bibfnamefont {H.}~\bibnamefont {Watanabe}},\
  }\bibfield  {title} {\bibinfo {title} {{Unified understanding of symmetry
  indicators for all internal symmetry classes}},\ }\href
  {https://doi.org/10.1103/PhysRevB.98.115150} {\bibfield  {journal} {\bibinfo
  {journal} {Phys. Rev. B}\ }\textbf {\bibinfo {volume} {98}},\ \bibinfo
  {pages} {115150} (\bibinfo {year} {2018})}\BibitemShut {NoStop}%
\bibitem [{\citenamefont {Khalaf}(2018)}]{PhysRevB.97.205136}%
  \BibitemOpen
  \bibfield  {author} {\bibinfo {author} {\bibfnamefont {E.}~\bibnamefont
  {Khalaf}},\ }\bibfield  {title} {\bibinfo {title} {{Higher-order topological
  insulators and superconductors protected by inversion symmetry}},\ }\href
  {https://doi.org/10.1103/PhysRevB.97.205136} {\bibfield  {journal} {\bibinfo
  {journal} {Phys. Rev. B}\ }\textbf {\bibinfo {volume} {97}},\ \bibinfo
  {pages} {205136} (\bibinfo {year} {2018})}\BibitemShut {NoStop}%
\bibitem [{\citenamefont {Matsugatani}\ and\ \citenamefont
  {Watanabe}(2018)}]{PhysRevB.98.205129}%
  \BibitemOpen
  \bibfield  {author} {\bibinfo {author} {\bibfnamefont {A.}~\bibnamefont
  {Matsugatani}}\ and\ \bibinfo {author} {\bibfnamefont {H.}~\bibnamefont
  {Watanabe}},\ }\bibfield  {title} {\bibinfo {title} {{Connecting higher-order
  topological insulators to lower-dimensional topological insulators}},\ }\href
  {https://doi.org/10.1103/PhysRevB.98.205129} {\bibfield  {journal} {\bibinfo
  {journal} {Phys. Rev. B}\ }\textbf {\bibinfo {volume} {98}},\ \bibinfo
  {pages} {205129} (\bibinfo {year} {2018})}\BibitemShut {NoStop}%
\bibitem [{\citenamefont {Xu}\ \emph {et~al.}(2019)\citenamefont {Xu},
  \citenamefont {Song}, \citenamefont {Wang}, \citenamefont {Weng},\ and\
  \citenamefont {Dai}}]{PhysRevLett.122.256402}%
  \BibitemOpen
  \bibfield  {author} {\bibinfo {author} {\bibfnamefont {Y.}~\bibnamefont
  {Xu}}, \bibinfo {author} {\bibfnamefont {Z.}~\bibnamefont {Song}}, \bibinfo
  {author} {\bibfnamefont {Z.}~\bibnamefont {Wang}}, \bibinfo {author}
  {\bibfnamefont {H.}~\bibnamefont {Weng}},\ and\ \bibinfo {author}
  {\bibfnamefont {X.}~\bibnamefont {Dai}},\ }\bibfield  {title} {\bibinfo
  {title} {{Higher-Order Topology of the Axion Insulator
  ${\mathrm{EuIn}}_{2}{\mathrm{As}}_{2}$}},\ }\href
  {https://doi.org/10.1103/PhysRevLett.122.256402} {\bibfield  {journal}
  {\bibinfo  {journal} {Phys. Rev. Lett.}\ }\textbf {\bibinfo {volume} {122}},\
  \bibinfo {pages} {256402} (\bibinfo {year} {2019})}\BibitemShut {NoStop}%
\bibitem [{\citenamefont {Tanaka}\ \emph
  {et~al.}(2020{\natexlab{a}})\citenamefont {Tanaka}, \citenamefont
  {Takahashi}, \citenamefont {Zhang},\ and\ \citenamefont
  {Murakami}}]{PhysRevResearch.2.043274}%
  \BibitemOpen
  \bibfield  {author} {\bibinfo {author} {\bibfnamefont {Y.}~\bibnamefont
  {Tanaka}}, \bibinfo {author} {\bibfnamefont {R.}~\bibnamefont {Takahashi}},
  \bibinfo {author} {\bibfnamefont {T.}~\bibnamefont {Zhang}},\ and\ \bibinfo
  {author} {\bibfnamefont {S.}~\bibnamefont {Murakami}},\ }\bibfield  {title}
  {\bibinfo {title} {{Theory of inversion-${\mathbb{Z}}_{4}$ protected
  topological chiral hinge states and its applications to layered
  antiferromagnets}},\ }\href
  {https://doi.org/10.1103/PhysRevResearch.2.043274} {\bibfield  {journal}
  {\bibinfo  {journal} {Phys. Rev. Research}\ }\textbf {\bibinfo {volume}
  {2}},\ \bibinfo {pages} {043274} (\bibinfo {year}
  {2020}{\natexlab{a}})}\BibitemShut {NoStop}%
\bibitem [{\citenamefont {Khalaf}\ \emph {et~al.}(2018)\citenamefont {Khalaf},
  \citenamefont {Po}, \citenamefont {Vishwanath},\ and\ \citenamefont
  {Watanabe}}]{PhysRevX.8.031070}%
  \BibitemOpen
  \bibfield  {author} {\bibinfo {author} {\bibfnamefont {E.}~\bibnamefont
  {Khalaf}}, \bibinfo {author} {\bibfnamefont {H.~C.}\ \bibnamefont {Po}},
  \bibinfo {author} {\bibfnamefont {A.}~\bibnamefont {Vishwanath}},\ and\
  \bibinfo {author} {\bibfnamefont {H.}~\bibnamefont {Watanabe}},\ }\bibfield
  {title} {\bibinfo {title} {{Symmetry Indicators and Anomalous Surface States
  of Topological Crystalline Insulators}},\ }\href
  {https://doi.org/10.1103/PhysRevX.8.031070} {\bibfield  {journal} {\bibinfo
  {journal} {Phys. Rev. X}\ }\textbf {\bibinfo {volume} {8}},\ \bibinfo {pages}
  {031070} (\bibinfo {year} {2018})}\BibitemShut {NoStop}%
\bibitem [{\citenamefont {Tanaka}\ \emph
  {et~al.}(2020{\natexlab{b}})\citenamefont {Tanaka}, \citenamefont
  {Takahashi},\ and\ \citenamefont {Murakami}}]{PhysRevB.101.115120}%
  \BibitemOpen
  \bibfield  {author} {\bibinfo {author} {\bibfnamefont {Y.}~\bibnamefont
  {Tanaka}}, \bibinfo {author} {\bibfnamefont {R.}~\bibnamefont {Takahashi}},\
  and\ \bibinfo {author} {\bibfnamefont {S.}~\bibnamefont {Murakami}},\
  }\bibfield  {title} {\bibinfo {title} {{Appearance of hinge states in
  second-order topological insulators via the cutting procedure}},\ }\href
  {https://doi.org/10.1103/PhysRevB.101.115120} {\bibfield  {journal} {\bibinfo
   {journal} {Phys. Rev. B}\ }\textbf {\bibinfo {volume} {101}},\ \bibinfo
  {pages} {115120} (\bibinfo {year} {2020}{\natexlab{b}})}\BibitemShut
  {NoStop}%
\bibitem [{\citenamefont {Takahashi}\ \emph {et~al.}(2020)\citenamefont
  {Takahashi}, \citenamefont {Tanaka},\ and\ \citenamefont
  {Murakami}}]{PhysRevResearch.2.013300}%
  \BibitemOpen
  \bibfield  {author} {\bibinfo {author} {\bibfnamefont {R.}~\bibnamefont
  {Takahashi}}, \bibinfo {author} {\bibfnamefont {Y.}~\bibnamefont {Tanaka}},\
  and\ \bibinfo {author} {\bibfnamefont {S.}~\bibnamefont {Murakami}},\
  }\bibfield  {title} {\bibinfo {title} {{Bulk-edge and bulk-hinge
  correspondence in inversion-symmetric insulators}},\ }\href
  {https://doi.org/10.1103/PhysRevResearch.2.013300} {\bibfield  {journal}
  {\bibinfo  {journal} {Phys. Rev. Research}\ }\textbf {\bibinfo {volume}
  {2}},\ \bibinfo {pages} {013300} (\bibinfo {year} {2020})}\BibitemShut
  {NoStop}%
\bibitem [{\citenamefont {Ezawa}(2018{\natexlab{b}})}]{PhysRevB.97.155305}%
  \BibitemOpen
  \bibfield  {author} {\bibinfo {author} {\bibfnamefont {M.}~\bibnamefont
  {Ezawa}},\ }\bibfield  {title} {\bibinfo {title} {{Magnetic second-order
  topological insulators and semimetals}},\ }\href
  {https://doi.org/10.1103/PhysRevB.97.155305} {\bibfield  {journal} {\bibinfo
  {journal} {Phys. Rev. B}\ }\textbf {\bibinfo {volume} {97}},\ \bibinfo
  {pages} {155305} (\bibinfo {year} {2018}{\natexlab{b}})}\BibitemShut
  {NoStop}%
\bibitem [{\citenamefont {van Miert}\ and\ \citenamefont
  {Ortix}(2018)}]{PhysRevB.98.081110}%
  \BibitemOpen
  \bibfield  {author} {\bibinfo {author} {\bibfnamefont {G.}~\bibnamefont {van
  Miert}}\ and\ \bibinfo {author} {\bibfnamefont {C.}~\bibnamefont {Ortix}},\
  }\bibfield  {title} {\bibinfo {title} {{Higher-order topological insulators
  protected by inversion and rotoinversion symmetries}},\ }\href
  {https://doi.org/10.1103/PhysRevB.98.081110} {\bibfield  {journal} {\bibinfo
  {journal} {Phys. Rev. B}\ }\textbf {\bibinfo {volume} {98}},\ \bibinfo
  {pages} {081110(R)} (\bibinfo {year} {2018})}\BibitemShut {NoStop}%
\bibitem [{\citenamefont {Du}\ \emph {et~al.}(2015)\citenamefont {Du},
  \citenamefont {Wan}, \citenamefont {Wang}, \citenamefont {Sheng},
  \citenamefont {Duan},\ and\ \citenamefont {Wan}}]{du2015dirac}%
  \BibitemOpen
  \bibfield  {author} {\bibinfo {author} {\bibfnamefont {Y.}~\bibnamefont
  {Du}}, \bibinfo {author} {\bibfnamefont {B.}~\bibnamefont {Wan}}, \bibinfo
  {author} {\bibfnamefont {D.}~\bibnamefont {Wang}}, \bibinfo {author}
  {\bibfnamefont {L.}~\bibnamefont {Sheng}}, \bibinfo {author} {\bibfnamefont
  {C.-G.}\ \bibnamefont {Duan}},\ and\ \bibinfo {author} {\bibfnamefont
  {X.}~\bibnamefont {Wan}},\ }\bibfield  {title} {\bibinfo {title} {{Dirac and
  Weyl Semimetal in XYBi (X= Ba, Eu; Y= Cu, Ag and Au)}},\ }\href
  {https://www.nature.com/articles/srep14423} {\bibfield  {journal} {\bibinfo
  {journal} {Sci. Rep.}\ }\textbf {\bibinfo {volume} {5}},\ \bibinfo {pages}
  {1} (\bibinfo {year} {2015})}\BibitemShut {NoStop}%
\bibitem [{Pyt()}]{PythTB}%
  \BibitemOpen
  \href@noop {} {\bibinfo {title} {Python tight binding open-source package}},\
  \bibinfo {howpublished}
  {\url{http://physics.rutgers.edu/pythtb/}}\BibitemShut {NoStop}%
\bibitem [{\citenamefont {Elcoro}\ \emph {et~al.}(2021)\citenamefont {Elcoro},
  \citenamefont {Wieder}, \citenamefont {Song}, \citenamefont {Xu},
  \citenamefont {Bradlyn},\ and\ \citenamefont
  {Bernevig}}]{elcoro2021magnetic}%
  \BibitemOpen
  \bibfield  {author} {\bibinfo {author} {\bibfnamefont {L.}~\bibnamefont
  {Elcoro}}, \bibinfo {author} {\bibfnamefont {B.~J.}\ \bibnamefont {Wieder}},
  \bibinfo {author} {\bibfnamefont {Z.}~\bibnamefont {Song}}, \bibinfo {author}
  {\bibfnamefont {Y.}~\bibnamefont {Xu}}, \bibinfo {author} {\bibfnamefont
  {B.}~\bibnamefont {Bradlyn}},\ and\ \bibinfo {author} {\bibfnamefont {B.~A.}\
  \bibnamefont {Bernevig}},\ }\bibfield  {title} {\bibinfo {title} {{Magnetic
  topological quantum chemistry}},\ }\href
  {https://www.nature.com/articles/s41467-021-26241-8} {\bibfield  {journal}
  {\bibinfo  {journal} {Nat. Commun.}\ }\textbf {\bibinfo {volume} {12}},\
  \bibinfo {pages} {1} (\bibinfo {year} {2021})}\BibitemShut {NoStop}%
\bibitem [{\citenamefont {Peng}\ \emph {et~al.}(2021)\citenamefont {Peng},
  \citenamefont {Jiang}, \citenamefont {Fang}, \citenamefont {Weng},\ and\
  \citenamefont {Fang}}]{peng2021topological}%
  \BibitemOpen
  \bibfield  {author} {\bibinfo {author} {\bibfnamefont {B.}~\bibnamefont
  {Peng}}, \bibinfo {author} {\bibfnamefont {Y.}~\bibnamefont {Jiang}},
  \bibinfo {author} {\bibfnamefont {Z.}~\bibnamefont {Fang}}, \bibinfo {author}
  {\bibfnamefont {H.}~\bibnamefont {Weng}},\ and\ \bibinfo {author}
  {\bibfnamefont {C.}~\bibnamefont {Fang}},\ }\bibfield  {title} {\bibinfo
  {title} {Topological classification and diagnosis in magnetically ordered
  electronic materials},\ }\href {https://arxiv.org/abs/2102.12645} {\bibfield
  {journal} {\bibinfo  {journal} {arXiv:2102.12645}\ } (\bibinfo {year}
  {2021})}\BibitemShut {NoStop}%
\bibitem [{\citenamefont {Wang}\ \emph {et~al.}(2013)\citenamefont {Wang},
  \citenamefont {Weng}, \citenamefont {Wu}, \citenamefont {Dai},\ and\
  \citenamefont {Fang}}]{PhysRevB.88.125427}%
  \BibitemOpen
  \bibfield  {author} {\bibinfo {author} {\bibfnamefont {Z.}~\bibnamefont
  {Wang}}, \bibinfo {author} {\bibfnamefont {H.}~\bibnamefont {Weng}}, \bibinfo
  {author} {\bibfnamefont {Q.}~\bibnamefont {Wu}}, \bibinfo {author}
  {\bibfnamefont {X.}~\bibnamefont {Dai}},\ and\ \bibinfo {author}
  {\bibfnamefont {Z.}~\bibnamefont {Fang}},\ }\bibfield  {title} {\bibinfo
  {title} {{Three-dimensional Dirac semimetal and quantum transport in
  Cd${}_{3}$As${}_{2}$}},\ }\href {https://doi.org/10.1103/PhysRevB.88.125427}
  {\bibfield  {journal} {\bibinfo  {journal} {Phys. Rev. B}\ }\textbf {\bibinfo
  {volume} {88}},\ \bibinfo {pages} {125427} (\bibinfo {year}
  {2013})}\BibitemShut {NoStop}%
\bibitem [{\citenamefont {Yi}\ \emph {et~al.}(2014)\citenamefont {Yi},
  \citenamefont {Wang}, \citenamefont {Chen}, \citenamefont {Shi},
  \citenamefont {Feng}, \citenamefont {Liang}, \citenamefont {Xie},
  \citenamefont {He}, \citenamefont {He}, \citenamefont {Peng} \emph
  {et~al.}}]{yi2014evidence}%
  \BibitemOpen
  \bibfield  {author} {\bibinfo {author} {\bibfnamefont {H.}~\bibnamefont
  {Yi}}, \bibinfo {author} {\bibfnamefont {Z.}~\bibnamefont {Wang}}, \bibinfo
  {author} {\bibfnamefont {C.}~\bibnamefont {Chen}}, \bibinfo {author}
  {\bibfnamefont {Y.}~\bibnamefont {Shi}}, \bibinfo {author} {\bibfnamefont
  {Y.}~\bibnamefont {Feng}}, \bibinfo {author} {\bibfnamefont {A.}~\bibnamefont
  {Liang}}, \bibinfo {author} {\bibfnamefont {Z.}~\bibnamefont {Xie}}, \bibinfo
  {author} {\bibfnamefont {S.}~\bibnamefont {He}}, \bibinfo {author}
  {\bibfnamefont {J.}~\bibnamefont {He}}, \bibinfo {author} {\bibfnamefont
  {Y.}~\bibnamefont {Peng}}, \emph {et~al.},\ }\bibfield  {title} {\bibinfo
  {title} {{Evidence of topological surface state in three-dimensional Dirac
  semimetal ${\mathrm{Cd}}_{3}{\mathrm{As}}_{2}$}},\ }\href
  {https://www.nature.com/articles/srep06106} {\bibfield  {journal} {\bibinfo
  {journal} {Sci. Rep.}\ }\textbf {\bibinfo {volume} {4}},\ \bibinfo {pages}
  {1} (\bibinfo {year} {2014})}\BibitemShut {NoStop}%
\bibitem [{\citenamefont {Ali}\ \emph {et~al.}(2014)\citenamefont {Ali},
  \citenamefont {Gibson}, \citenamefont {Jeon}, \citenamefont {Zhou},
  \citenamefont {Yazdani},\ and\ \citenamefont {Cava}}]{doi:10.1021/ic403163d}%
  \BibitemOpen
  \bibfield  {author} {\bibinfo {author} {\bibfnamefont {M.~N.}\ \bibnamefont
  {Ali}}, \bibinfo {author} {\bibfnamefont {Q.}~\bibnamefont {Gibson}},
  \bibinfo {author} {\bibfnamefont {S.}~\bibnamefont {Jeon}}, \bibinfo {author}
  {\bibfnamefont {B.~B.}\ \bibnamefont {Zhou}}, \bibinfo {author}
  {\bibfnamefont {A.}~\bibnamefont {Yazdani}},\ and\ \bibinfo {author}
  {\bibfnamefont {R.~J.}\ \bibnamefont {Cava}},\ }\bibfield  {title} {\bibinfo
  {title} {{The Crystal and Electronic Structures of
  ${\mathrm{Cd}}_{3}{\mathrm{As}}_{2}$, the Three-Dimensional Electronic
  Analogue of Graphene}},\ }\href {https://doi.org/10.1021/ic403163d}
  {\bibfield  {journal} {\bibinfo  {journal} {Inorganic Chem.}\ }\textbf
  {\bibinfo {volume} {53}},\ \bibinfo {pages} {4062} (\bibinfo {year}
  {2014})}\BibitemShut {NoStop}%
\bibitem [{\citenamefont {Liu}\ \emph {et~al.}(2014)\citenamefont {Liu},
  \citenamefont {Jiang}, \citenamefont {Zhou}, \citenamefont {Wang},
  \citenamefont {Zhang}, \citenamefont {Weng}, \citenamefont {Prabhakaran},
  \citenamefont {Mo}, \citenamefont {Peng}, \citenamefont {Dudin} \emph
  {et~al.}}]{liu2014stable}%
  \BibitemOpen
  \bibfield  {author} {\bibinfo {author} {\bibfnamefont {Z.}~\bibnamefont
  {Liu}}, \bibinfo {author} {\bibfnamefont {J.}~\bibnamefont {Jiang}}, \bibinfo
  {author} {\bibfnamefont {B.}~\bibnamefont {Zhou}}, \bibinfo {author}
  {\bibfnamefont {Z.}~\bibnamefont {Wang}}, \bibinfo {author} {\bibfnamefont
  {Y.}~\bibnamefont {Zhang}}, \bibinfo {author} {\bibfnamefont
  {H.}~\bibnamefont {Weng}}, \bibinfo {author} {\bibfnamefont {D.}~\bibnamefont
  {Prabhakaran}}, \bibinfo {author} {\bibfnamefont {S.~K.}\ \bibnamefont {Mo}},
  \bibinfo {author} {\bibfnamefont {H.}~\bibnamefont {Peng}}, \bibinfo {author}
  {\bibfnamefont {P.}~\bibnamefont {Dudin}}, \emph {et~al.},\ }\bibfield
  {title} {\bibinfo {title} {{A stable three-dimensional topological Dirac
  semimetal Cd${}_{3}$As${}_{2}$}},\ }\href
  {https://www.nature.com/articles/nmat3990} {\bibfield  {journal} {\bibinfo
  {journal} {Nature materials}\ }\textbf {\bibinfo {volume} {13}},\ \bibinfo
  {pages} {677} (\bibinfo {year} {2014})}\BibitemShut {NoStop}%
\bibitem [{\citenamefont {Po}\ \emph {et~al.}(2018)\citenamefont {Po},
  \citenamefont {Watanabe},\ and\ \citenamefont
  {Vishwanath}}]{PhysRevLett.121.126402}%
  \BibitemOpen
  \bibfield  {author} {\bibinfo {author} {\bibfnamefont {H.~C.}\ \bibnamefont
  {Po}}, \bibinfo {author} {\bibfnamefont {H.}~\bibnamefont {Watanabe}},\ and\
  \bibinfo {author} {\bibfnamefont {A.}~\bibnamefont {Vishwanath}},\ }\bibfield
   {title} {\bibinfo {title} {{Fragile Topology and Wannier Obstructions}},\
  }\href {https://doi.org/10.1103/PhysRevLett.121.126402} {\bibfield  {journal}
  {\bibinfo  {journal} {Phys. Rev. Lett.}\ }\textbf {\bibinfo {volume} {121}},\
  \bibinfo {pages} {126402} (\bibinfo {year} {2018})}\BibitemShut {NoStop}%
\bibitem [{\citenamefont {Altland}\ and\ \citenamefont
  {Zirnbauer}(1997)}]{PhysRevB.55.1142}%
  \BibitemOpen
  \bibfield  {author} {\bibinfo {author} {\bibfnamefont {A.}~\bibnamefont
  {Altland}}\ and\ \bibinfo {author} {\bibfnamefont {M.~R.}\ \bibnamefont
  {Zirnbauer}},\ }\bibfield  {title} {\bibinfo {title} {{Nonstandard symmetry
  classes in mesoscopic normal-superconducting hybrid structures}},\ }\href
  {https://doi.org/10.1103/PhysRevB.55.1142} {\bibfield  {journal} {\bibinfo
  {journal} {Phys. Rev. B}\ }\textbf {\bibinfo {volume} {55}},\ \bibinfo
  {pages} {1142} (\bibinfo {year} {1997})}\BibitemShut {NoStop}%
\bibitem [{\citenamefont {Fang}\ \emph {et~al.}(2012)\citenamefont {Fang},
  \citenamefont {Gilbert},\ and\ \citenamefont
  {Bernevig}}]{PhysRevB.86.115112}%
  \BibitemOpen
  \bibfield  {author} {\bibinfo {author} {\bibfnamefont {C.}~\bibnamefont
  {Fang}}, \bibinfo {author} {\bibfnamefont {M.~J.}\ \bibnamefont {Gilbert}},\
  and\ \bibinfo {author} {\bibfnamefont {B.~A.}\ \bibnamefont {Bernevig}},\
  }\bibfield  {title} {\bibinfo {title} {{Bulk topological invariants in
  noninteracting point group symmetric insulators}},\ }\href
  {https://doi.org/10.1103/PhysRevB.86.115112} {\bibfield  {journal} {\bibinfo
  {journal} {Phys. Rev. B}\ }\textbf {\bibinfo {volume} {86}},\ \bibinfo
  {pages} {115112} (\bibinfo {year} {2012})}\BibitemShut {NoStop}%
\end{thebibliography}
\end{document}